\documentclass[12pt,american]{curt_thesis}
\usepackage{times}
\usepackage[T1]{fontenc}
\usepackage[latin1]{inputenc}
\usepackage{geometry}
\usepackage{subfigure}
\usepackage{amsmath}
\usepackage{makeidx}
\makeindex
\usepackage{graphicx}
\usepackage{setspace}
\usepackage{amssymb}

\makeatletter

\let\SF@@footnote\footnote
\def\footnote{\ifx\protect\@typeset@protect
    \expandafter\SF@@footnote
  \else
    \expandafter\SF@gobble@opt
  \fi
}
\expandafter\def\csname SF@gobble@opt \endcsname{\@ifnextchar[
  \SF@gobble@twobracket
  \@gobble
}
\edef\SF@gobble@opt{\noexpand\protect
  \expandafter\noexpand\csname SF@gobble@opt \endcsname}
\def\SF@gobble@twobracket[#1]#2{}
\providecommand{\tabularnewline}{\\}

\usepackage{color} 
\definecolor{dgray}{rgb}{0.3,0.3,0.3}
\definecolor{lgray}{rgb}{0.5,0.5,0.5}

\usepackage{afterpage}

\usepackage{fancyhdr}
\renewcommand{\tableofcontents}{%
\afterpage{\lhead{TABLE OF CONTENTS - CONTINUED}}
\section*{\contentsname
\@mkboth{%
\MakeUppercase\contentsname}{\MakeUppercase\contentsname}}%
\@starttoc{toc}%
\afterpage{\lhead{}}
}
\renewcommand\listoffigures{%
\afterpage{\lhead{LIST OF FIGURES - CONTINUED}}
\section*{\listfigurename}%
\@mkboth{\MakeUppercase\listfigurename}%
              {\MakeUppercase\listfigurename}%
\@starttoc{lof}%
\afterpage{\lhead{}}
}

\renewcommand*\l@chapter[2]{%
  \addpenalty{-\@highpenalty}%
  \vskip 1.0em \@plus\p@
  \begingroup
    \advance\hangindent 1.5em
    \@dottedtocline{0}{0pt}{20pt}{ \bfseries #1}{\bfseries #2}
  \endgroup
  \penalty\@highpenalty}
\renewcommand*\l@figure[2]{%
  \@dottedtocline{1}{0pt}{30pt}{\textsc{Figure} #1}{#2}}
\renewcommand*\l@table[2]{%
  \@dottedtocline{1}{0pt}{30pt}{\textsc{Table} #1}{#2}}

\renewcommand*\l@part[2]{%
  \addpenalty{-\@highpenalty}%
  \vskip 1.0em \@plus\p@
  \begingroup
    \advance\hangindent 1.5em
    \@dottedtocline{0}{0pt}{20pt}{\large \bfseries #1}{\large \bfseries #2}
  \endgroup
  \penalty\@highpenalty}

\usepackage{babel}
\makeatother
\begin{document}
\ifx\href\undefined\else\hypersetup{linktocpage=true}\fi 

\title{Nuclear Magnetic Resonance with the Distant Dipolar Field}

\author{Curtis Andrew Corum}

\degreeaward{DOCTOR OF PHILOSOPHY}

\copyyear{2005}

\pubnum{UA doesn't use this...}

\maketitle

\chapter*{APPROVAL FORM}

\begin{center}THE UNIVERSITY OF ARIZONA GRADUATE COLLEGE\end{center}
\medskip{}

As members of the Dissertation Committee, we certify that we have
read the dissertation prepared by Curtis Andrew Corum entitled {}``Nuclear
Magnetic Resonance with the Distant Dipolar Field'' and recommend
that it be accepted as fulfilling the dissertation requirement for
the Degree of Doctor of Philosophy
\medskip{}

\_\_\_\_\_\_\_\_\_\_\_\_\_\_\_\_\_\_\_\_\hfill{}Date: 12/2/2004

Arthur F. Gmitro

\_\_\_\_\_\_\_\_\_\_\_\_\_\_\_\_\_\_\_\_\hfill{}Date: 12/2/2004

Harrison H. Barrett

\_\_\_\_\_\_\_\_\_\_\_\_\_\_\_\_\_\_\_\_\hfill{}Date: 12/2/2004

Theodore Trouard

\_\_\_\_\_\_\_\_\_\_\_\_\_\_\_\_\_\_\_\_\hfill{}Date: 12/2/2004

Jean-Phillipe Galons
\medskip{}

Final approval and acceptance of this dissertation is contingent upon
the candidate's submission of the final copies of the dissertation
to the Graduate College. 

I hereby certify that I have read this dissertation prepared under
my direction and recommend that it be accepted as fulfilling the dissertation
requirement.
\medskip{}

\_\_\_\_\_\_\_\_\_\_\_\_\_\_\_\_\_\_\_\_\hfill{}Date: 12/2/2004

Arthur F. Gmitro

\chapter*{STATEMENT BY AUTHOR}

~~~~~This dissertation has been submitted in partial fulfillment
of requirements for an advanced degree at The University of Arizona
and is deposited in the University Library to be made available to
borrowers under rules of the Library.

Brief quotations from this dissertation are allowable without special
permission, provided that accurate acknowledgment of source is made.
Requests for permission for extended quotation from or reproduction
of this manuscript in whole or in part may be granted by the copyright
holder. 

\bigskip{}
\begin{flushright}SIGNED: \_\_\_\_Curtis Andrew Corum\_\_\_\_\end{flushright}

\chapter*{ACKNOWLEDGMENTS}

Any undertaking as arduous as obtaining one's doctor of philosophy
requires the help, support, and understanding of many people. 

I wish first of all to thank my adviser Arthur F. Gmitro, whose willingness
to take on a student wanting to study a new, exciting, and somewhat
obscure area of magnetic resonance has lead to this work. Without
his early risk taking, confidence, continued support, and direction
this work would never have been accomplished. I'd also like to acknowledge
his excellent teaching, and his part in introducing me to the field
of Magnetic Resonance.

Thanks to my committee members Professors Harrison H. Barrett, Theodore
P. Trouard, and Jean-Phillipe Galons.

First I'd like to thank Harry for his support early in my graduate
work at the University of Arizona, his excellent teaching for all
his courses on imaging science and mathematics, and the privilege
to read early drafts of his recent book {}``Foundations of Image
Science''.

Ted deserves thanks for his generosity in all things, personal, professional,
and for letting me have a desk in his lab...

I cannot thank J. P. enough for all his help in learning about NMR
and MRI, the research game, and the dreaded Bruker programming environment....

Constantin Job deserves thanks putting up with all the ups and downs
associated with hardware support in the Biological Magnetic Resonance
Facility.

There are numerous others to thank for numerous reasons, and not enough
space to do so properly here.

\clearpage

\begin{dedication}
I dedicate this dissertation to my wife Katrina and our daughter Marianna.

Thanks for all your love and support! And for putting up with a husband
and dad in grad school...
\end{dedication}
\clearpage \renewcommand{\contentsname}{TABLE OF CONTENTS}

\tableofcontents

\clearpage\addcontentsline{toc}{chapter}{LIST OF FIGURES} \renewcommand{\listfigurename}{LIST OF FIGURES}

\listoffigures

\clearpage\addcontentsline{toc}{chapter}{LIST OF TABLES} \renewcommand{\listtablename}{LIST OF TABLES}

\listoftables

\chapter*{{\large \begin{center}ABSTRACT\end{center}}\addcontentsline{toc}{chapter}{ABSTRACT}}

Distant dipolar field (DDF)-based nuclear magnetic resonance is an
active research area with many fundamental properties still not well
understood. Already several intriguing applications have developed,
like HOMOGENIZED and IDEAL spectroscopy, that allow high resolution
spectra to be obtained in inhomogeneous fields, such as in-vivo. The
theoretical and experimental research in this thesis concentrates
on the fundamental signal properties of DDF-based sequences in the
presence of relaxation ($T_{1}$ and $T_{2}$) and diffusion. A general
introduction to magnetic resonance phenomenon is followed by a more
in depth introduction to the DDF and its effects. A novel analytical
signal equation has been developed to describe the effects of $T_{2}$
relaxation and diffusing spatially modulated longitudinal spins during
the signal build period of an HOMOGENIZED cross peak. Diffusion of
the longitudinal spins results in a lengthening of the effective dipolar
demagnetization time, delaying the re-phasing of coupled anti-phase
states in the quantum picture. In the classical picture the unwinding
rate of spatially twisted magnetization is no longer constant, but
decays exponentially with time. The expression is experimentally verified
for the HOMOGENIZED spectrum of 100mM TSP in $H_{2}O$ at 4.7T. Equations
have also been developed for the case of multiple repetition steady
state 1d and 2d spectroscopic sequences with incomplete magnetization
recovery, leading to spatially varying longitudinal magnetization.
Experimental verification has been accomplished by imaging the profile.
The equations should be found generally applicable for those interested
in DDF-based spectroscopy and imaging.

\part{The Basics\label{par:The-Basics}}

\chapter{INTRODUCTION}

\section{Motivation and Setting}

Distant dipolar field\index{distant dipolar field} (DDF\index{DDF})
based nuclear magnetic resonance is a relatively new area of research.
It utilizes what had been thought of as the negligible interaction
between macroscopic groups of spins in a liquid. This is in contrast
to microscopic interactions which contribute to relaxation effects.

The macroscopic or {}``distant'' dipolar field now becomes a new
tool, added to the already overflowing toolbox of physical and physiological
effects utilized in magnetic resonance spectroscopy\index{magnetic resonance spectroscopy}
(MRS\index{MRS}) and imaging\index{magnetic resonance imaging} (MRI\index{MRI}).
It offers many exciting possibilities such novel contrast imaging,
motion insensitivity \cite{KRC+03}, and mesoscale (below the size
of single voxel) spatial frequency selectivity.

One of the most intriguing features, at least for in-vivo spectroscopy,
is insensitivity to $B_{0}$ inhomogeneity. This was demonstrated
by Warren et al. in the HOMOGENIZED \index{HOMOGENIZED}2d spectroscopy
sequence \cite{VLW96}.

The work presented in this thesis was motivated by trying to apply
HOMOGENIZED to an NMR compatible bioreactor system \cite{GGM+93}.
This system has practical limits for line-widths obtainable in localized
spectroscopy, which HOMOGENIZED could potentially overcome. As the
work progressed it became obvious that there were still fundamental
issues not well understood for HOMOGENIZED and DDF in general. The
work then shifted to understanding such fundamental issues as signal
dependence on $T_{1}$, $T_{2}$, and diffusion, as well as the fundamental
nature and spatial origin of the signal.

\section{Prehistory of NMR}

Semantics and the lens of hindsight make any historical and even scientific
historical fact open to interpretation. The field of nuclear magnetic
resonance (NMR) is generally said to originate with the announcement
by I. I. Rabi\index{Rabi, I. I.} et al.\cite{RZM+38,RMK+38} of a
new {}``resonant'' technique for measuring the magnetic moment of
nuclei in a molecular beam passing through a magnetic field. This
became quickly established as a powerful technique for the measurement
of magnetic properties of nuclei. Subsequently E. Purcell\index{Purcell, Edward}
et al.\cite{PTP46} looked at resonant absorption of radio-frequency
energy in protons in semi-solid paraffin. Nearly simultaneously F.
Bloch\index{Bloch, Felix} et al.\cite{BHP46} reported resonant ''induction''
in liquid water. These successes were preceded by earlier efforts
in the Netherlands and in Russia\cite{Pfe99}. The importance of NMR
was highlighted by the awarding of the Nobel Prize\index{Nobel Prize}
for Physics to Rabi in 1944, and to Bloch and Purcell in 1952.

The infant technique of NMR in liquids and solids quickly established
itself as a useful probe of numerous physical properties of nuclei,
atoms and molecules in solution and solids. Over the years it has
developed from a technique of experimental physics to one of experimental
chemistry, to a routine analytical tool in chemistry and to some degree
solid state physics and materials science. It then branched into radiology/medical
imaging as Magnetic Resonance Imaging (MRI, originally called nuclear
magnetic resonance imaging, the unpopular term {}``nuclear'' being
dropped).

\chapter{NUCLEAR MAGNETISM}

\section{The NMR Phenomenon}

The matter that surrounds us is composed of atoms and molecules, arranged
as atomic or molecular gas mixtures (the atmosphere), liquid mixtures
or solutions (the ocean, lakes, tap water, gasoline, urine), liquid
crystals, solids (rocks, metals, glasses, etc.), plasmas (consisting
of partly or wholly ionized atoms and molecules) and more complicated
suspensions, composites, and living systems. All atoms in all these
states of matter contain a nucleus and some of these nuclei (those
with an odd number of protons or neutrons) possess a net spin and
a magnetic moment \cite[section 1.3.3, pp 12-15]{Lev01}. The magnitude
of the proton and other nuclear magnetic moments has been measured
to great accuracy thanks to the resonant atomic beam experiments of
Rabi et al. \cite{RMK+38} and followers. The origin of the nuclear
spin and magnetic moment is the domain of subatomic physics, specifically
quantum chromodynamics, and is still an active theoretical \cite{HRG04}
and experimental \cite{SBH+04} research topic.

\section{Susceptibility and Magnetization}

A material has macroscopic magnetic properties determined by its magnetic
susceptibility (see reference\index{magnetic susceptibility} \cite{DHK03}
and appendix \ref{sec:Equilibrium-Magnetization}). The {}``DC''
susceptibility $\chi$ determines the equilibrium magnetization of
a sample when placed in an external field. It is a classical dimensionless
quantity that represents the average tendency of the individual magnetic
dipole moments to align due to a magnetic field. It is a function
of sample composition, phase (gas, liquid, solid, plasma), and temperature
(see appendix \ref{sec:Equilibrium-Magnetization}). The total {}``DC''
susceptibility can be broken up into two components, electronic and
nuclear. The electronic susceptibility usually dominates. In fact
for $^{1}H$ in $H_{2}O$ at room temperature, $\frac{\chi_{n}}{\chi_{e}}\sim10^{-5}$.
We have

\begin{equation}
\chi=\chi_{e}+\chi_{n}\label{eq:chi_total}\end{equation}
and

\begin{equation}
\vec{M}_{0}=\frac{\chi}{\mu_{0}}\,\vec{B}_{0}.\label{eq:m_chi_b}\end{equation}

$\chi$ is in general a tensor quantity, and can be nonlinear (saturation
for ferromagnetic materials) and include history effects. For water
and many (but not all) biological materials, $\chi$ can be considered
a constant scalar quantity, in which case the direction of net magnetization
is parallel to the field. $\mu_{0}$ is the {}``permeability of free
space'' needed for the SI system of units.

We can break up the magnetization into two components, electronic
and nuclear, based on the susceptibility component that gives rise
to the magnetization. We can further break up the nuclear component
into contributions from different types of nuclei. We write this as

\begin{equation}
M_{0}=M_{0e}+M_{0n}=\frac{(\chi_{e}+\chi_{n})}{\mu_{0}}B_{0}\label{eq:M_total}\end{equation}
and

\begin{equation}
M_{0n}=M_{0n1}+M_{0n2}...=\frac{(\chi_{n1}+\chi_{n2}...)}{\mu_{0}}B_{0}.\label{eq:M_nuc}\end{equation}

The main effect of electronic magnetization in NMR is to cause inhomogeneous
broadening of the resonance spectrum (see section \ref{sec:Field-Inhomogeneity}
and reference \cite{YH94}) and the chemical shift (see section \ref{sec:Chemical-Shift}).
We will drop the {}``n'' from $M_{0n}$ from now on and use $M_{0}$
to denote the equilibrium nuclear magnetization, and $\chi$ to denote
nuclear susceptibility.

At room temperature (298K) the $\chi$ of pure 55.56M $^{1}H_{2}O$
due to the two $^{1}H$ protons is $\chi=4.07\times10^{-9}$. The
corresponding $M_{0}$ at 9.4T is $M_{0}=.0305\frac{A}{m}$.

\section{Precession}

\begin{figure}
\begin{center}\includegraphics[%
  width=0.50\columnwidth,
  keepaspectratio]{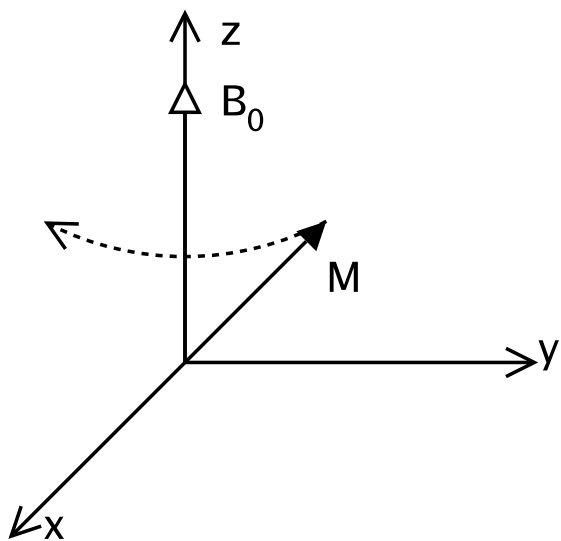}\end{center}

\caption{\label{cap:Precession}Precession when $\gamma>0$.}
\end{figure}
It is a well established fact that a magnetic dipole with moment perturbed
from alignment with an external magnetic field will precess (Figure
\ref{cap:Precession}). This is the underlying physical basis for
NMR. Precession\index{Precession} is due to the torque $\vec{\mu}\times\vec{B_{0}}$
acting on the non-zero angular momentum of the nucleus \cite[eq (7)]{Blo46}.
The rate of precession is determined by the magnetogyric ratio\index{magnetogyric ratio}
(often called the gyromagnetic ratio\index{gyromagnetic ratio}),
denoted by $\gamma$. This is the ratio (for a given nucleus) of magnetic
moment to (spin) angular momentum, where $\hbar\vec{I}$ is the angular
momentum of the nucleus. We write the magnetic moment in terms of
$\gamma$ as\begin{equation}
\vec{\mu}\equiv\gamma\,\hbar\,\vec{I}.\label{eq:gamma}\end{equation}
 Table \ref{cap:Some-common-Nuclei} in Appendix \ref{cha:NMR-Data}
shows $\gamma$ for various common nuclei.

Mathematically we can express the precession the angular momentum
for an ensemble of nuclei ($\vec{L}=\frac{1}{V}\sum_{n=1}^{N}\vec{l}_{n}$)
by a differential equation, the torque being equal to the time rate
of change of angular momentum as

\begin{equation}
\frac{d\,\vec{L}}{d\, t}=\vec{M}\times\vec{B}_{0}.\label{eq:angular_momentum}\end{equation}
 We can put this in the more useful form (since $\gamma\,\vec{L}=\vec{M}$)
\begin{equation}
\frac{d\,\vec{M}}{d\, t}=\gamma\vec{M}\times\vec{B}_{0}.\label{eq:change_in_M}\end{equation}
We note that when non-zero, the change in $\vec{M}$, $\frac{d\,\vec{M}}{d\, t}$,
is always orthogonal to $\vec{M}$ as well as $\vec{B}_{0}$. This
results in the circular {}``precession'' about $\vec{B}_{0}$.

The solution to equation \ref{eq:change_in_M} is best carried out
in spherical coordinates, with $\vec{B}_{0}$ oriented along the $z$
polar axis. Then we have\begin{equation}
\frac{d\,\vec{M}}{d\, t}=-\gamma\, M\, B_{0}sin(\theta)\,\hat{\phi}.\label{eq:dM/dt_spherical}\end{equation}
 Since the tip of $\vec{M}$ must traverse a {}``distance'' $2\pi\, cos(\theta)\, M$
to make a full revolution, this corresponds to rotation about $\hat{z}$
at a rate $\frac{d\,\phi}{d\, t}=\omega_{0}=-\gamma\, B_{0}$ at a
constant $\theta$. Note that in figure \ref{cap:Precession} the
sense of rotation is left handed or clockwise about $\hat{z}$. This
is because most nuclei of interest have a positive magnetogyric ratio%
\footnote{A caution to the reader: This sign convention is not always followed
in the literature. For a discussion of the sign convention followed
in this dissertation, see Appendix \ref{cha:The-Levitt-Sign-Conventions}
and references \cite{Lev97,Lev01}.%
} , $\gamma>0$, although some nuclei posses $\gamma<0$. 

In Cartesian coordinates the solution becomes\begin{equation}
\vec{M}(t)=M_{0}[sin(\theta_{0})\, cos(\phi_{0}+\omega_{0}t)\,\hat{x}+sin(\theta_{0})\, sin(\phi_{0}+\omega_{0}t)\,\hat{y}+cos(\theta_{0})\,\hat{z}],\label{eq:M_of_t}\end{equation}
where $M_{0}$, $\theta_{0}$ and $\phi_{0}$ determine the initial
magnitude and orientation of $\vec{M}$.

The frequency of precession \begin{equation}
f=\frac{\omega_{0}}{2\pi}=-\frac{\gamma\, B_{0}}{2\pi}\label{eq:larmor_requency}\end{equation}
 is called the Larmor frequency\index{Larmor frequency}.

\section{Longitudinal and Transverse Components\label{sec:Longitudinal-and-Transverse}}

\begin{figure}
\begin{center}\includegraphics[%
  width=0.50\columnwidth,
  keepaspectratio]{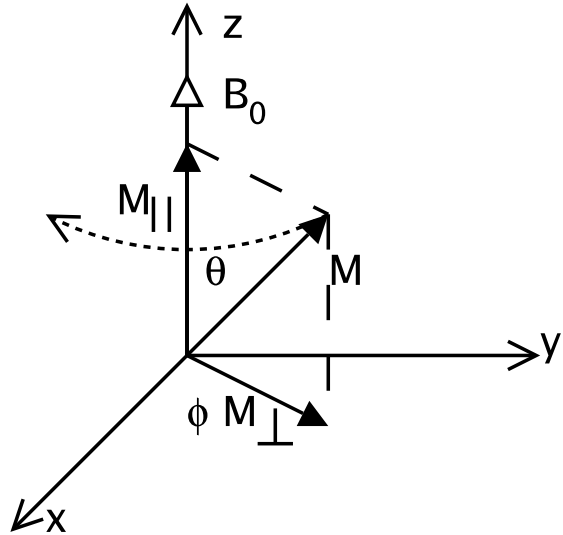}\end{center}

\caption{\label{cap:Components}Longitudinal and Transverse components of
$\vec{M}$.}
\end{figure}
It is helpful to distinguish between longitudinal and transverse components
of the magnetization (Figure \ref{cap:Components}). The longitudinal
(oriented $\Vert$ to $\vec{B}_{0}$) component does not precess,
while the transverse (oriented $\bot$ to $\vec{B}_{0}$) does precess.
The longitudinal and transverse components of magnetization relax
differently. We will discuss relaxation properties in chapter \ref{cha:Relaxation}.
Distinguishing between the longitudinal and transverse components
of the magnetization will also be useful later when we discuss the
distant dipolar field in Part \ref{par:DDF}. The components are defined
as\begin{equation}
\vec{M}=\vec{M}_{\Vert}+\vec{M}_{\bot},\label{eq:long_and_tran}\end{equation}
\begin{equation}
\vec{M}_{\Vert}\equiv M_{0}\, cos(\theta_{0})\,\hat{z},\label{eq:long}\end{equation}
and\begin{equation}
\vec{M}_{\bot}\equiv M_{0}[sin(\theta_{0})\, cos(\omega_{0}t+\phi_{0})\,\hat{x}+sin(\theta_{0})\, sin(\omega_{0}t+\phi_{0})\,\hat{y}].\label{eq:tran}\end{equation}

We can introduce an even further convenience, denoting the $\hat{x}$
component as the real part and the $\hat{y}$ component as the imaginary
part of a complex scalar value, written as\begin{equation}
\vec{M}_{\bot}\equiv Re(M_{\bot})\,\hat{x}+Im(M_{\bot})\,\hat{y}.\label{eq:M_tran_vector}\end{equation}
This gives us the form

\begin{equation}
M_{\bot}(t)=M_{0}sin(\theta_{0})\, e^{i\,(\omega_{0}t+\phi_{0})}.\label{eq:complex_tran}\end{equation}
Note that $\omega_{0}<0$ corresponds to clockwise or left-handed
precession about $\hat{z}$ for nuclei with $\gamma>0$.

The longitudinal magnetization is always real and can be written as
a real scalar\[
M_{\parallel}=M_{0}\, cos(\theta_{0}).\]

\section{Rotating Frame}

\begin{figure}
\begin{center}\includegraphics[%
  width=0.50\columnwidth,
  keepaspectratio]{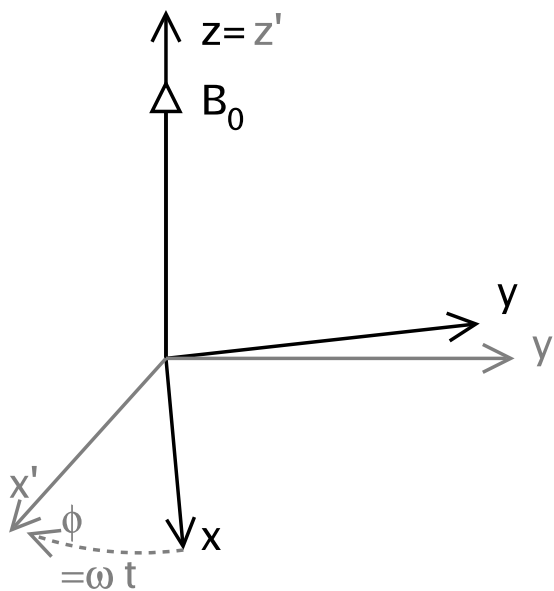}\end{center}

\caption{\label{cap:Rotating-frame}Rotating frame, $\omega_{0}<0$ and $t>0$.}
\end{figure}
Another helpful concept is the rotating frame \cite{RRS54}. We construct
another Cartesian coordinate system, whose $z'$ axis coincides with
the laboratory frame $z$. The $x'$ and $y'$ axis rotates with angular
frequency $\omega$. We show this in figure \ref{cap:Rotating-frame}.
In the rotating frame, if $\omega=\omega_{0}$, the magnetization
will appear to stand still. The coordinate transformations are

\begin{equation}
\hat{x}'=\hat{x}\, cos(\omega t)+\hat{y}\, sin(\omega t),\label{eq:x_prime}\end{equation}
\begin{equation}
\hat{y}'=\hat{y}\, cos(\omega t)-\hat{x}\, sin(\omega t)\label{eq:y_prime}\end{equation}
and

\begin{equation}
\hat{z}'=\hat{z}.\label{eq:z_prime}\end{equation}

We can also define,\begin{equation}
\Delta\omega_{0}\equiv\omega_{0}-\omega,\label{eq:delta_omega}\end{equation}
the angular frequency with which magnetization will precess in the
rotating frame. This is sometimes called the resonance offset.

\begin{figure}
\begin{center}\includegraphics[%
  width=0.50\columnwidth,
  keepaspectratio]{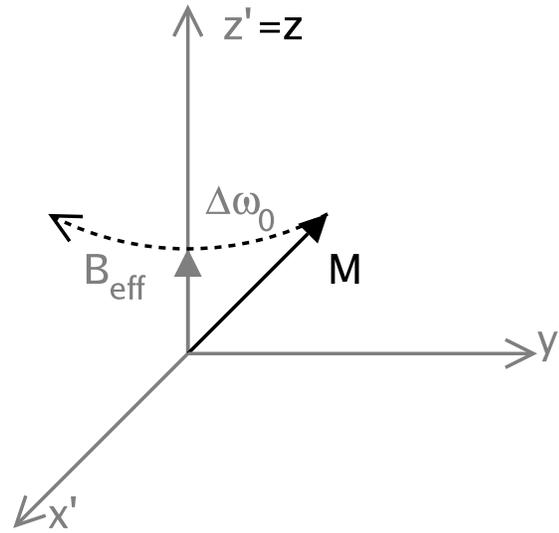}\end{center}

\caption{\label{cap:Beff}Effective field $\vec{B}_{eff}$ and resonance offset
$\Delta\omega_{0}$ in the rotating frame.}
\end{figure}

Related to equation (\ref{eq:delta_omega}) is the effective field
$\vec{B}_{eff}$. This is a fictitious field (see figure \ref{cap:Beff})
in the rotating frame such that

\begin{equation}
\vec{B}_{eff}=\frac{-\Delta\omega_{0}}{\gamma}\hat{z}.\label{eq:Beff}\end{equation}
By substituting (\ref{eq:delta_omega}) into (\ref{eq:Beff}) we get
the relation

\begin{equation}
\vec{B}_{eff}=(B_{0}+\frac{\omega}{\gamma})\,\hat{z}.\label{eq:Beff_again}\end{equation}
Note that $B_{eff}=0$ when $\omega=\omega_{0}$. The rotating frame
and effective field are extremely useful tools in understanding the
dynamics of NMR and MRI experiments. The effective field can also
include contributions from an applied radio frequency field (RF) discussed
in section \ref{sec:RF-Field}.

\chapter{OSCILLATING FIELD EFFECTS}

\section{RF Field\index{RF field}\label{sec:RF-Field}}

\begin{figure}
\begin{center}\includegraphics[%
  width=0.35\columnwidth,
  keepaspectratio]{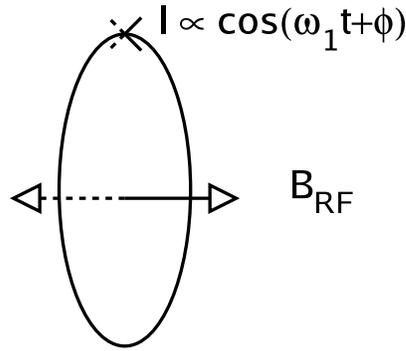}\end{center}

\caption{\label{cap:RF-magneitc-field}RF magnetic field $\vec{B}_{RF}$ due
to a oscillating current in a conducting loop {}``transmit coil.''}
\end{figure}
 The {}``resonance'' in NMR and MRI refers to the response of nuclei
to an applied oscillating magnetic field, called an {}``RF field''
or {}``RF pulse''. For commonly achievable fields and nuclei the
Larmor frequency falls within the 1-1000MHz frequency range, hence
the term Radio Frequency or RF.

In general, the magnetic (and electric) field properties in an NMR
experiment depend on the specific geometry of the RF coil, and sometimes
the geometry and absorption properties of the sample. We will consider
an idealized case of uniform RF fields and no absorption. The term
{}``$B_{1}$ inhomogeneity''\index{B1 inhomogeneity} refers to
the situation where the RF coil produces more RF magnetic field at
one location than another. Some coils are designed with a homogeneous
RF field in mind, such a solenoids or birdcages \cite{DES99,DEH99,HOK+00}.
Others such as surface coils are not, and may require $B_{1}$ insensitive
{}``adiabatic'' pulses\index{adiabatic pulse} \cite{GD01} for
experiments sensitive to $B_{1}$ inhomogeneity.

We represent an applied RF magnetic field by its components. The magnetic
field $B_{RF}$ at the center of a current loop (called the transmit
coil\index{transmit coil}) carrying an alternating current $I$ is
perpendicular to the axis of the loop as in figure \ref{cap:RF-magneitc-field}.
We must break the field $B_{RF}$ into its counter-rotating components.
The nucleus will only respond to a field rotating with the same sense
and angular frequency near its own Larmor frequency%
\footnote{There is an effect due to the counter-rotating component, causing
a minute shift in the resonance frequency while the pulse is on \cite{BS40}.%
}. Mathematically we have\begin{equation}
\vec{B}_{RF}=A\, cos(\omega_{1}t+\phi_{1})\,\hat{x},\label{eq:Brf}\end{equation}
\begin{equation}
\vec{B}_{RF}=\vec{B}_{1}+\vec{B'}_{1},\label{eq:Brf_components}\end{equation}
\begin{equation}
\vec{B}_{1}=\frac{A}{2}\, cos(\omega_{1}t+\phi_{1})\,\hat{x}+\frac{A}{2}\, sin(\omega_{1}t+\phi_{1})\,\hat{y},\label{eq:B1}\end{equation}

\begin{equation}
\vec{B'}_{1}=\frac{A}{2}\, cos(\omega_{1}t+\phi_{1})\,\hat{x}-\frac{A}{2}\, sin(\omega_{1}t+\phi_{1})\,\hat{y}.\label{eq:B'1}\end{equation}
In the complex notation introduced in section \ref{sec:Longitudinal-and-Transverse}
equation \ref{eq:complex_tran} we can write\begin{equation}
B_{1}=\frac{A}{2}\, e^{i\,(\omega_{1}\, t+\phi_{1})},\label{eq:B1_complex}\end{equation}

\begin{equation}
B'_{1}=\frac{A}{2}\, e^{i\,(-\omega_{1}\, t+\phi_{1})}.\label{eq:B'1_complex}\end{equation}

\begin{figure}
\begin{center}\includegraphics[%
  width=0.50\columnwidth,
  keepaspectratio]{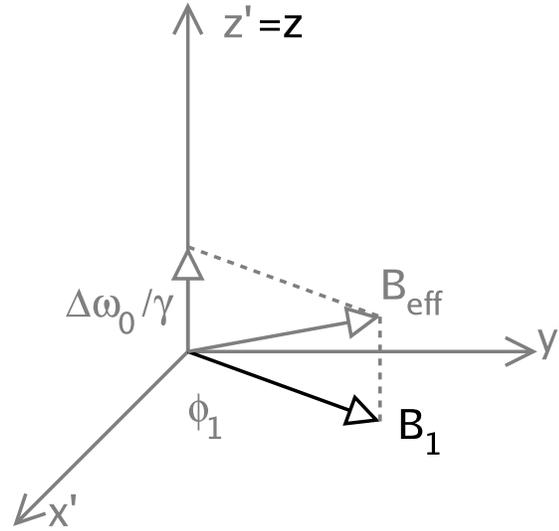}\end{center}

\caption{\label{cap:Beff_Total}Effective field $\vec{B}_{eff}$ for $\omega_{1}=\omega$.}
\end{figure}
The component $B'_{1}$ will in general have negligible effect on
the system and can be ignored. Some coils produce a rotating field
rather than a linear oscillating field, in which case no $B'_{1}$
component is produced. An advantage of these coils is efficiency of
utilization of RF power from the transmitter. In general the RF field
amplitude $A$ is a function of time. The RF field can be turned on
for periods of time, hence the term RF pulse.

An RF field with $\omega_{1}=\omega$ has a particularly simple representation
in the rotating frame: it is a constant field that does not move.
One can then add this $\vec{B}_{1}$ component to make a total $\vec{B}_{eff}$
in the rotating frame. If $\omega_{1}\neq\omega$ then the transverse
component of $\vec{B}_{eff}$ (which is $\vec{B}_{1}$) will rotate
with angular frequency $\Delta\omega_{1}=\omega_{1}-\omega$.

We can sum up these relations for $\vec{B}_{eff}$ in the rotating
frame as\begin{equation}
\vec{B}_{eff}=\frac{-\Delta\omega_{0}}{\gamma}\,\hat{z}+Re(B_{1}\, e^{i\,(\Delta\omega_{1}\, t+\phi_{1})})\,\hat{x}'+Im(B_{1}\, e^{i\,(\Delta\omega_{1}\, t+\phi_{1})})\,\hat{y}'.\label{eq:Beff_total}\end{equation}
$\phi_{1}$ is the {}``phase'' of the RF field, $\phi_{1}=0$ corresponding
to $B_{1}$ initially oriented along $\hat{x}'$ and $\phi_{1}=\frac{\pi}{2}$
corresponding to \textbf{$B_{1}$} initially oriented along $\hat{y}'$.
If the RF resonance offset $\Delta\omega_{1}=0$, then $\vec{B}_{eff}$
is constant in the rotating frame.

\section{RF Pulse\index{RF pulse}\label{sec:RF-Pulse}}

Radio Frequency (RF) pulses are the principal workhorses of NMR and
MRI. Magnetization precesses about the effective field $\vec{B}_{eff}$
in the rotating frame. For $\Delta\omega_{1}=0$, $\vec{B}_{eff}=\vec{B}_{1}$
and lies in the transverse plane. Turning on or off, or varying the
amplitude $B_{1}$ of the RF field by applying an {}``RF Pulse''
is the principal activity in any NMR experiment.

\subsection{90° Pulse\index{90 degree pulse}}

\begin{figure}
\begin{center}\includegraphics[%
  width=0.50\columnwidth,
  keepaspectratio]{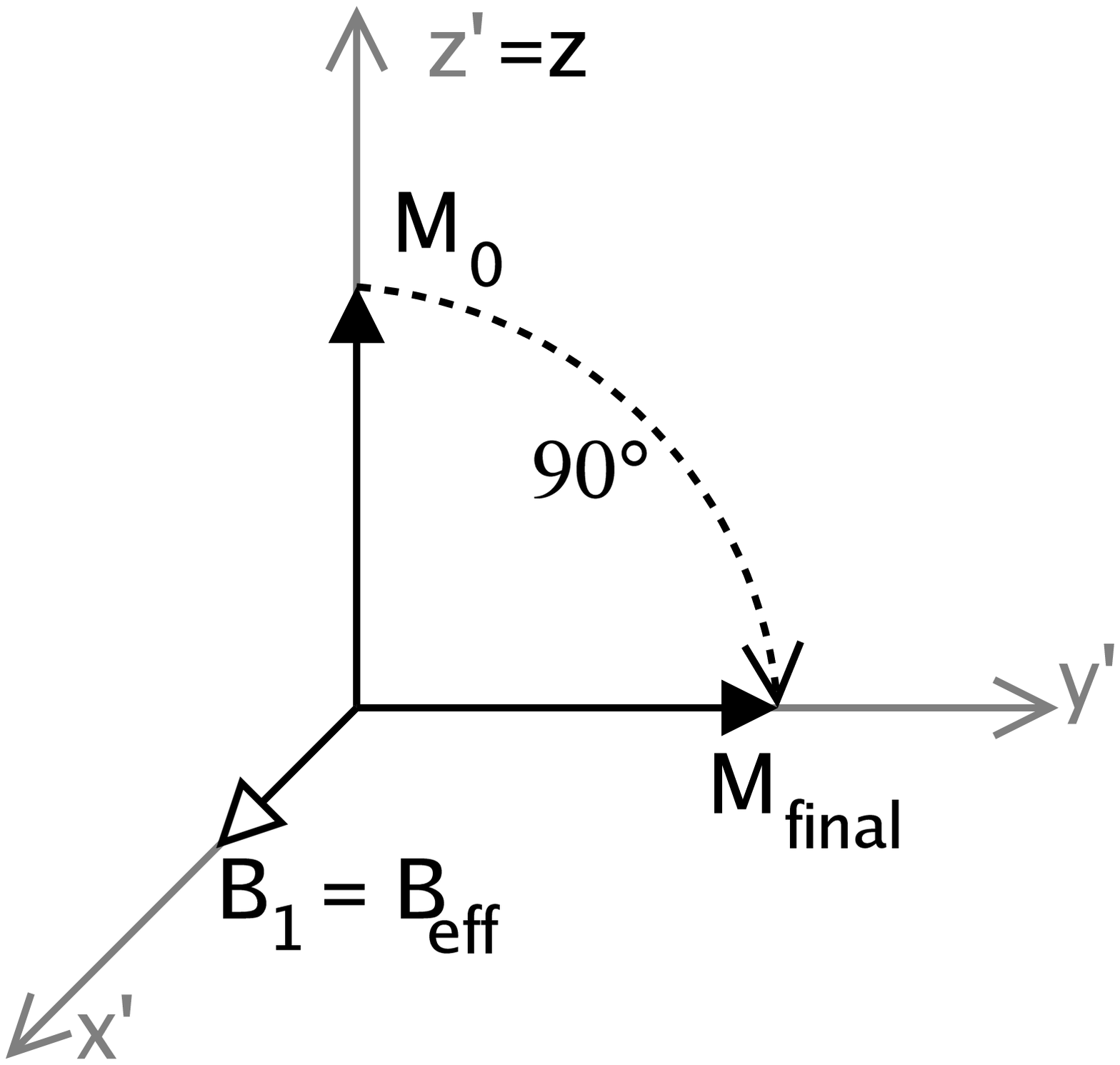}\end{center}

\caption{$90^{\circ}$ RF Pulse\label{cap:pulse90d}}
\end{figure}
Figure \ref{cap:pulse90d} shows an RF pulse that moves $\vec{M}$
from its equilibrium position $\vec{M}_{0}$ aligned with the $z$
axis into the transverse plane such that $\gamma\, B_{1}\tau=\frac{\pi}{2}$.
$\tau$ is the duration of the pulse. A more complicated but equivalent
form is\begin{equation}
\gamma\int_{0}^{\tau}\, B_{1}(t)\, dt=\frac{\pi}{2},\label{eq:90_integral}\end{equation}
which allows for the amplitude of $B_{1}$ and hence the precession
rate of $\vec{M}$ about the $B_{1}$ field to vary in time. A 90°
RF pulse acting on equilibrium magnetization is often called an excitation
pulse\index{excitation pulse}.

\subsection{180° Pulse\index{180 degree pulse}}

\begin{figure}
\begin{center}\includegraphics[%
  width=0.50\columnwidth,
  keepaspectratio]{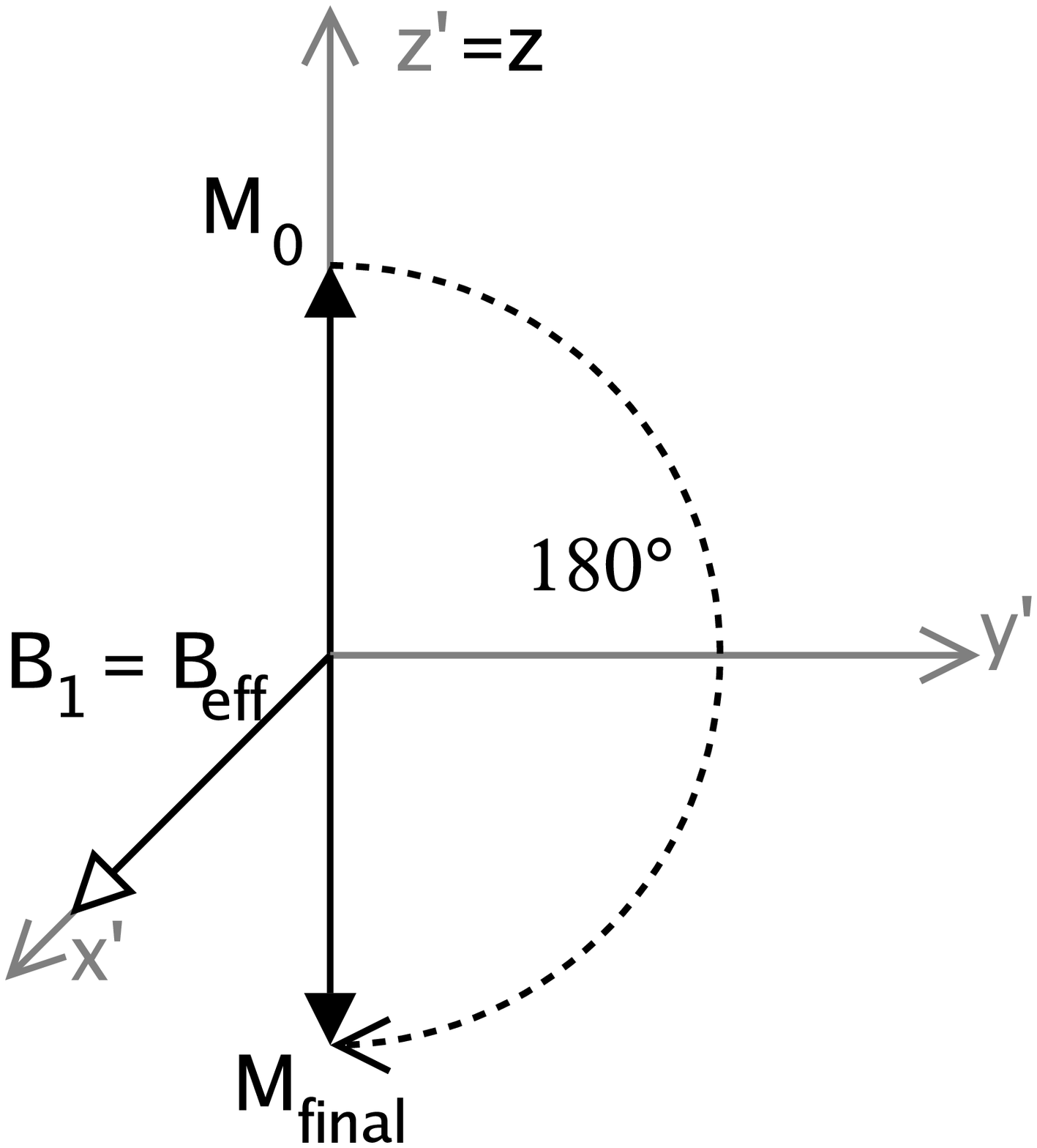}\end{center}

\caption{$180^{\circ}$ RF Pulse\label{cap:180d_pulse}}
\end{figure}
A 180° pulse inverts the magnetization from its equilibrium value.
It has twice the {}``area,'' as defined in equation \ref{eq:90_integral},
as a 90° pulse. A 180° pulse acting on equilibrium magnetization is
often called an inversion pulse\index{inversion pulse}.

\subsection{Off-Resonance Pulse\label{sub:Off-Resonance-Pulse}}

\begin{figure}
\begin{center}\includegraphics[%
  width=0.50\columnwidth,
  keepaspectratio]{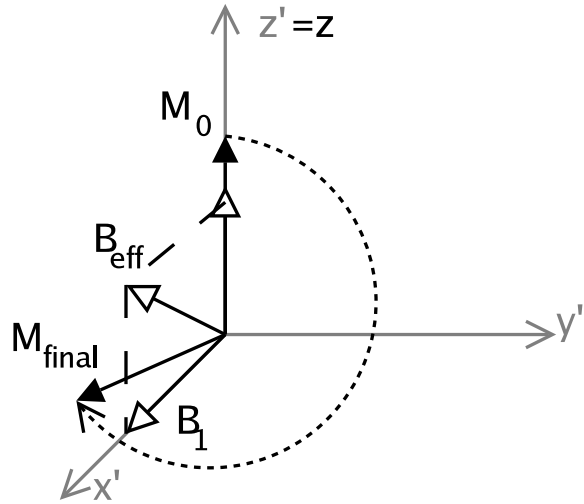}\end{center}

\caption{Same $180^{\circ}$ RF Pulse as figure \ref{cap:180d_pulse}, but
off-resonance\label{cap:180d_pulse_off}.}
\end{figure}
In the prior examples we have assumed that the RF field $B_{1}$ is
on resonance. If this is not true $B_{eff}$ will not lie in the transverse
plane. The effect of an off-resonance $B_{1}$ field is almost always
to reduce the total rotation angle compared to one on-resonance for
a given pulse. This can be seen as follows. We will assume the same
constant magnitude of $B_{1}$ field and duration as in figure \ref{cap:180d_pulse}.
A pulse with constant \textbf{$B_{1}$} is also known as a {}``hard
pulse''. On-resonance the pulse is a $180^{\circ}$pulse. Consider
$\frac{\gamma\, B_{1}}{2\pi}=500Hz$. This form is a convenient measure
of $B_{1}$ amplitude and is often shortened to $B_{1}=500Hz$. In
this case for a $180^{\circ}$pulse we need $\tau=1ms$. If the nuclei
of interest are $500Hz$ off resonance we have the following situation
seen in figure \ref{cap:180d_pulse_off}. The pulse gives less than
$90^{\circ}$ of rotation, and has a phase offset as well.

\section{Pulse Bandwidth\index{bandwidth (of and RF pulse)}}

\begin{figure}
\begin{center}\includegraphics[%
  width=5in,
  keepaspectratio]{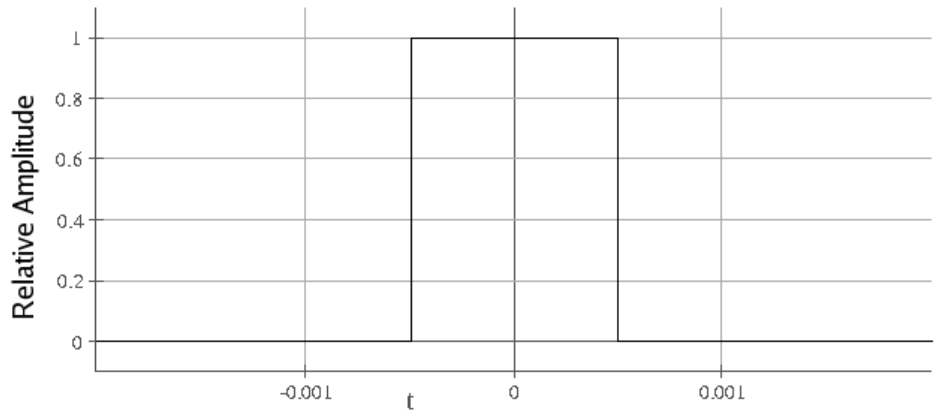}\end{center}

\caption{\label{cap:Hard Pulse}Hard pulse envelope, duration is $1ms$.}
\end{figure}
\begin{figure}
\begin{center}\includegraphics[%
  width=5in,
  keepaspectratio]{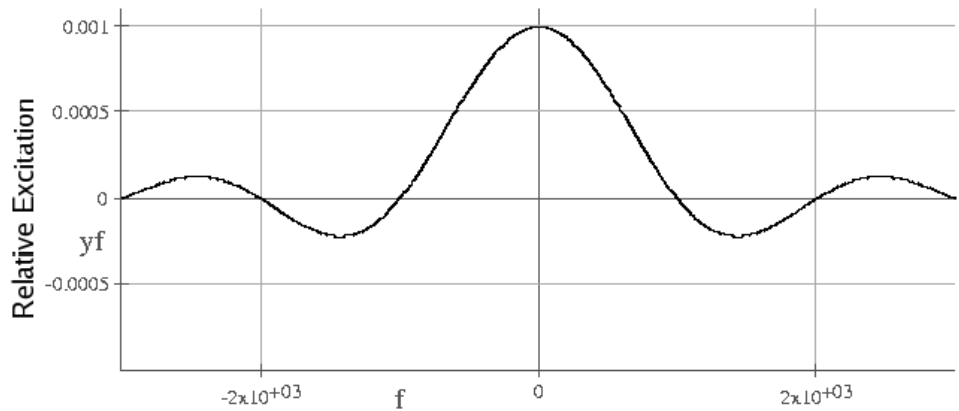}\end{center}

\caption[Fourier transform (approximate excitation profile) of the hard pulse
in figure \ref{cap:Hard Pulse}.]{\label{cap:Hard Pulse Profile}Fourier transform (approximate excitation
profile) of the hard pulse in figure \ref{cap:Hard Pulse}. The bandwidth
is approximately $1200Hz$.}
\end{figure}
 The {}``bandwidth'' of the pulse is defined as the total frequency
range for which the rotation angle is above half the on-resonance
value. In section \ref{sub:Off-Resonance-Pulse} the pulse has a bandwidth
of about $1200Hz$ or $1.2kHz$. Bandwidth in inversely proportional
to $\tau$ and depends on the shape of the pulse.

To find the bandwidth of a pulse (see figure \ref{cap:Hard Pulse})
one needs to solve the Bloch equations for the specific pulse shape
for a number of resonance offsets. One can also perform an experiment
to determine the performance of the pulse for excitation (or inversion),
this is called the excitation (or inversion) profile. The Fourier
transform of the RF pulse envelope (see figure \ref{cap:Hard Pulse Profile})
gives a good approximation to the excitation profile. The excitation
profile shows the relative rotation angle achieved versus the resonance
offset. The approximate pulse bandwidth is the full-width-half-max
of this approximate excitation profile.

\section{Free Induction Decay\index{Free Induction Decay}\index{FID}}

\begin{figure}
\begin{center}\includegraphics[%
  width=0.35\columnwidth,
  keepaspectratio]{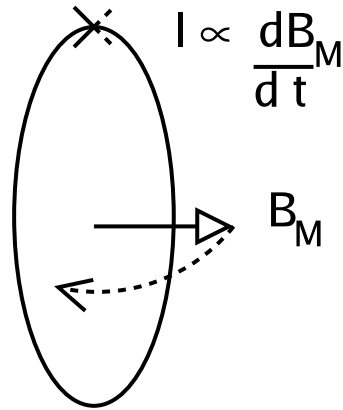}\end{center}

\caption{\label{cap:Detecting}Detecting precessing magnetization by induced
current in receiver coil.}
\end{figure}
 Following a 90° pulse, the magnetization is entirely in the transverse
plane, and continues to precess in the transverse plane in the laboratory
frame (since the RF field is zero after the pulse). In the rotating
frame the magnetization will precess according to its resonance offset
$\Delta\omega_{0}$. 

The magnetic field associated with the precessing magnetization can
be detected by its ability to induce a current in a nearby placed
coil, called the receiver coil\index{reciever coil}. The transmit
and receiver coils can be the same or different. The current induced
in the receiver coil is amplified, mixed with a local oscillator down
to the audio frequency range, and digitized.

\begin{figure}
\includegraphics[%
  bb=0bp 580bp 647bp 979bp,
  clip,
  width=5in,
  keepaspectratio]{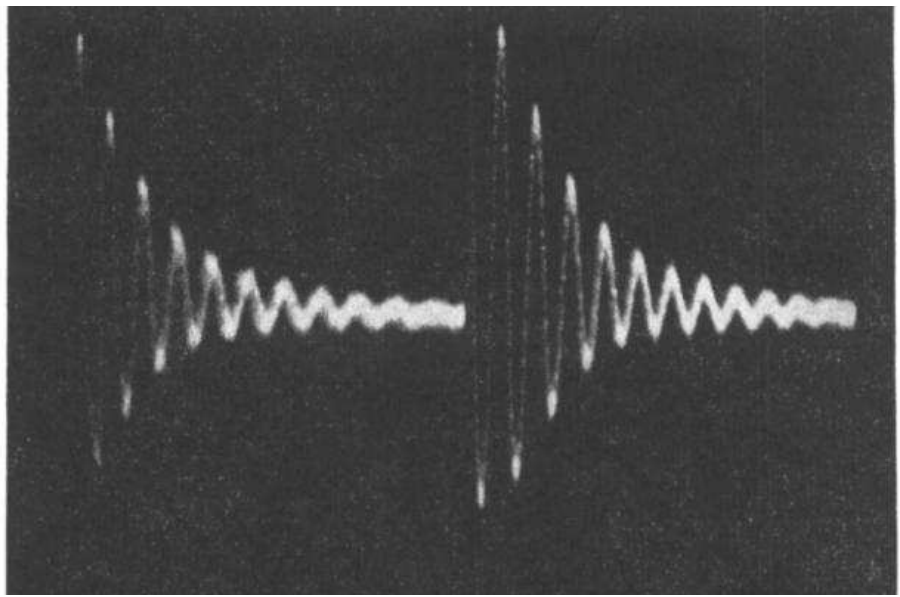}

\caption[Free induction decay (FID) for two excitations displayed on an analog
oscilloscope.]{\label{cap:FID}Free induction decay (FID) for two excitations displayed
on an analog oscilloscope. From reference \cite{Hah50}, Copyright
1950 by The American Physical Society, used with permission.}
\end{figure}

One can equate the local oscillator frequency of the receiver with
the frequency of rotation of the rotating frame. The output of the
mixer will then oscillate at the frequency of the resonance offset.
Shown in figure \ref{cap:FID} is an example oscilloscope trace from
an early pulsed NMR experiment. 

The signal is called the {}``Free Induction Decay'' or FID. The
{}``Decay'' comes from relaxation processes, which we will discuss
in section \ref{cha:Relaxation}.

\subsection{Quadrature Detection\index{quadrature detection}}

Note that there will be an ambiguity as to the sign of the offset
$\Delta\omega_{0}$ unless more information is obtained. This is achieved
by quadrature detection. The idea is to get information about both
the real and imaginary components of the precessing magnetization.
This can be done in several ways. Originally it was done in an analog
manner by having two reference oscillators (or on oscillator and a
phase shifter) and demodulating two signals, the phase of one shifted
by $90^{\circ}$ with respect to the other \cite[sec. 6.4]{RG71,Hou78}.
In digital systems it can be done in a number of ways by oversampling
and digital signal processing.

\subsection{NMR Spectrum\index{spectrum}\index{nmr spectrum}}

\begin{figure}
\subfigure[Complex FID.]{\includegraphics[%
  width=0.75\columnwidth,
  keepaspectratio]{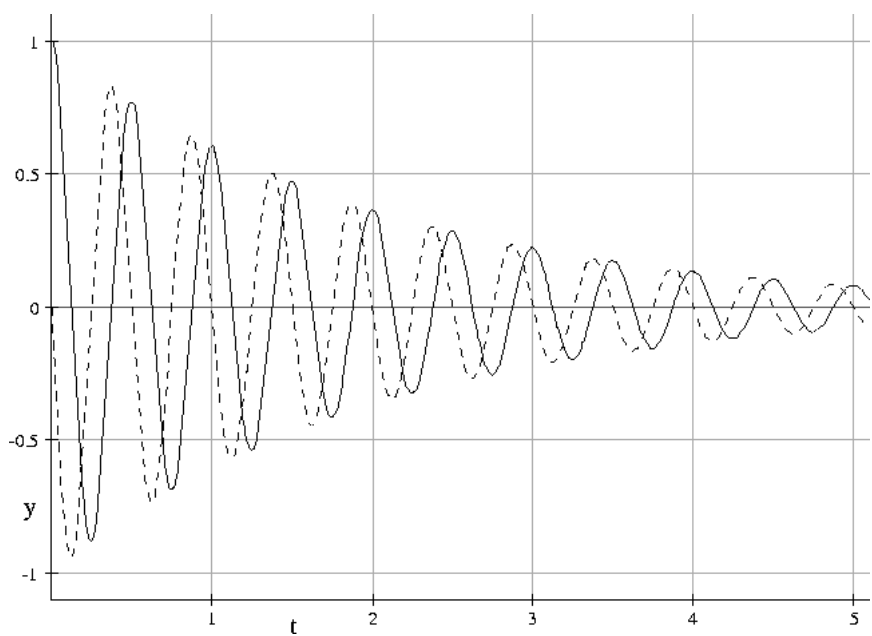}}

\subfigure[Spectrum]{\includegraphics[%
  width=0.75\columnwidth,
  keepaspectratio]{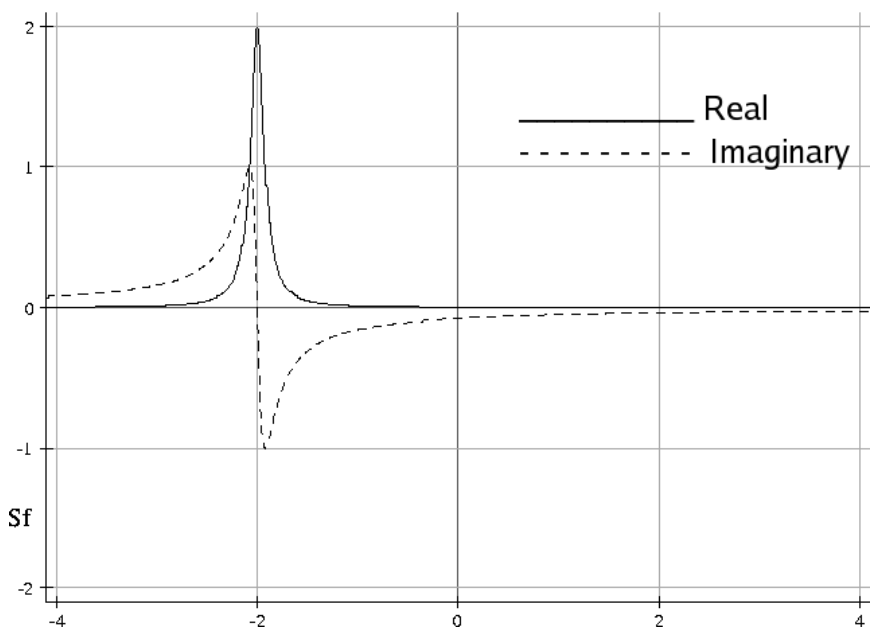}}

\caption[(a) Real and Imaginary part of the FID. (b) Complex Fourier Transform
of the FID.]{\label{cap:FIDandSPECTRUM}(a) Real and Imaginary part of the FID.
(b) Complex Fourier Transform of the FID yields the complex NMR spectrum.
Usually only the real part of the spectrum, after phase correction,
is shown.}
\end{figure}
The complex Fourier transform of the FID yields the NMR Spectrum \cite{EA66,WEH71}.
See figure \ref{cap:FIDandSPECTRUM} for a simple simulated example.
For the real part of the spectrum to be Lorentzian it is often necessary
to phase correct the spectrum \cite[sec. 5.1]{RK75,LF80}. Originally
NMR spectra were not obtained in this way, rather the RF frequency
(or $B_{0}$ field strength at constant RF frequency) was swept across
the range of interest. These so-called Absorption/Induction methods
have been shown to yield equivalent information to the Fourier method
\cite{EAB+74}, however the Fourier method has many signal-to-noise
and speed of measurement advantages and is almost universally used
in modern NMR spectrometers. The principal activity in NMR spectroscopy
is the identification of peaks of differing chemical shifts (see section
\ref{sec:Chemical-Shift}). Many other parameters can also be measured
such as relaxation rates (chapter \ref{cha:Relaxation}) and diffusion
(chapter \ref{cha:Diffusion}).

\chapter{\label{cha:Relaxation}RELAXATION\index{relaxation}}

Relaxation is the name given to processes in which magnetization decays
or returns to equilibrium. There are two principal processes of interest,
still named by their original designations and symbols \cite{BPP48}.

\section{Longitudinal Relaxation\index{longitudinal relaxation}, $T_{1}$\index{T1}}

\begin{figure}
\includegraphics[%
  width=5.5in,
  keepaspectratio]{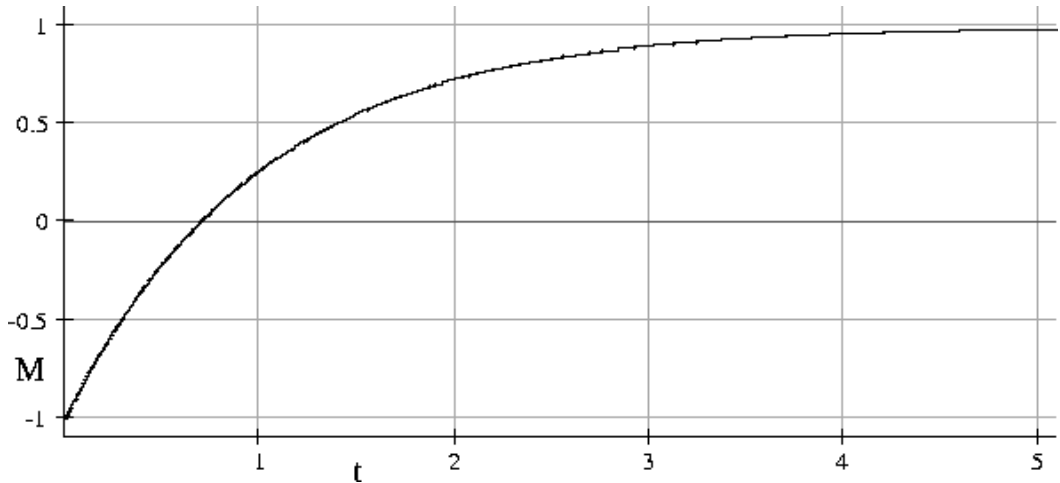}

\caption[Recovery of inverted magnetization by $T_{1}$ relaxation.]{\label{cap:Recovery}Recovery of inverted magnetization by $T_{1}$
relaxation as described by equation \ref{eq:Recovery}. $T_{1}=1$
unit.}
\end{figure}

Longitudinal relaxation, also called spin-lattice relaxation\index{spin-lattice relaxation},
using symbol $T_{1}$, describes the time scale at which magnetization
returns to thermal equilibrium, $M_{0}$, after being perturbed away
from equilibrium, such as by an RF pulse. Its effects are described
by the following equation for the recovery of the longitudinal magnetization

\begin{equation}
M_{\Vert}(t)=M_{0}-(M_{0}-M_{\Vert initial})\, e^{-t/T_{1}},\label{eq:Recovery}\end{equation}
which is the solution to the differential equation\begin{equation}
\frac{dM_{\parallel}}{dt}=\frac{M_{0}-M_{\parallel}}{T_{1}}.\label{eq:T1_de}\end{equation}

An important approximation is that $M_{\Vert}(t)\approx M_{0}$ after
a period $t=5\times T_{1}$. This can also be seen in Figure \ref{cap:Recovery}.

The term spin-lattice relaxation refers to transfer of energy from
the nuclear spins composing the macroscopic magnetization to the {}``lattice,''
a catch-all term referring to all other possible energy levels in
the system. The details of spin-lattice relaxation are beyond the
scope of this dissertation. Suffice it to say that in liquids, the
main mechanism of longitudinal relaxation is RF fields from nearby
spins causing stimulated transitions so that equilibrium is attained.
Spontaneous emission\index{spontaneous emission} processes at NMR
frequencies are entirely negligible \cite{HB97}. Information can
be found in references \cite{BPP48,Dic51,WB53,Blo56,Bloch57,Abr99,Gol01,Lev01}.

\subsection{Repetition and Recovery\index{recovery (of longitudinal magnetization)}\label{sub:Repetition-and-Recovery}}

\begin{figure}
\includegraphics[%
  width=5.5in,
  keepaspectratio]{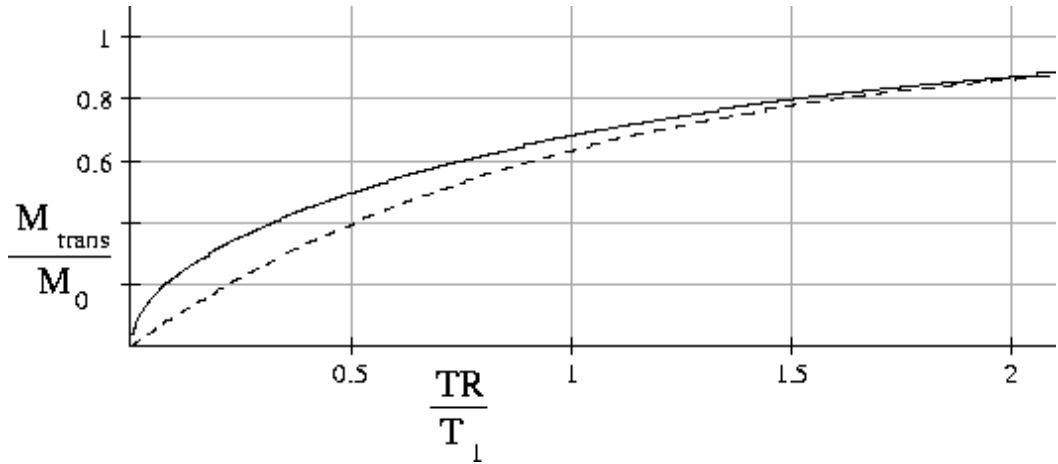}

\caption[Transverse magnetization obtained in the steady state by exciting
at the Ernst angle.]{\label{cap:Ernst_vs_90}Transverse magnetization obtained in the
steady state by exciting at the Ernst angle $\theta_{E}=arccos(e^{-\frac{TR}{T_{1}}})$
-------- vs. exciting at $90^{\circ}$ -~-~-~-. }
\end{figure}
In many NMR and MRI experiments the system is re-exited before full
relaxation (before waiting $5\times T_{1}$) has occurred. Often this
is to speed up the total time necessary to make an image in MRI or
to acquire a 2d NMR spectrum. The time between multiple excitations
is called the {}``repetition time\index{repetition time}'' and
denoted by $TR$\index{TR}. There is an optimum RF excitation pulse
to maximize the signal given a specific $TR$ and $T_{1}$ which is
called the Ernst angle\index{Ernst angle} \cite[p. 155]{Nis96}.
To find the Ernst angle we find the steady state longitudinal magnetization
after a large number of repetitions. We solve the equation\begin{equation}
M_{\Vert SS}=M_{0}-[M_{0}-M_{\Vert SS}cos(\theta)]\, e^{-t/T_{1}}\label{eq:Recovery}\end{equation}
formed by substituting $M_{\Vert}=M_{\Vert SS}$ and $M_{\Vert initial}=M_{\Vert SS}cos(\theta)$
into equation \ref{eq:Recovery}. The solution is

\begin{equation}
M_{\Vert SS}=M_{0}\frac{e^{\frac{TR}{T_{1}}}-1}{e^{\frac{TR}{T_{1}}}-cos(\theta)}.\label{eq:MperpSS}\end{equation}
The transverse magnetization immediately after excitation will be\begin{equation}
M_{\bot SS}=M_{\Vert SS}\, sin(\theta).\label{eq:MtranSS}\end{equation}
We can then find the excitation angle at which the transverse magnetization
becomes maximum, consistent with the steady-state longitudinal magnetization.
We set the result equal to zero, i.e.\begin{equation}
\frac{\partial M_{\bot SS}}{\partial\theta}=0=M_{0}\frac{e^{\frac{TR}{T_{1}}}(e^{\frac{TR}{T_{1}}}-1)\, cos(\theta)-(e^{\frac{TR}{T_{1}}}-1)}{[e^{\frac{TR}{T_{1}}}-cos(\theta)]^{2}},\label{eq:MtranSS_max}\end{equation}
which has the solution\begin{equation}
cos(\theta_{E})=e^{-\frac{TR}{T_{1}}}.\label{eq:Ernst_angle}\end{equation}
When $TR\geq5\times T_{1}$ we have $cos(\theta_{E})\approx1$ and
$\theta_{E}\approx90^{\circ}$ as expected. Figure \ref{cap:Ernst_vs_90}
shows a comparison of the signal using the Ernst angle vs. using $90^{\circ}$
as a function of $\frac{TR}{T_{1}}$.

\section{Transverse Relaxation\index{transverse relaxation}, $T_{2}$\index{T2}}

\begin{figure}
\includegraphics[%
  width=5.5in,
  keepaspectratio]{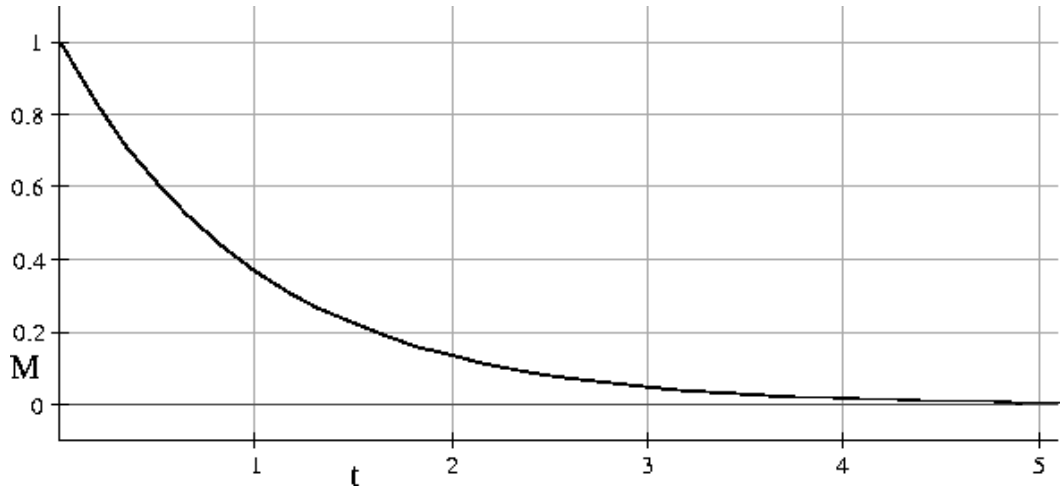}

\caption[$T_{2}$ decay of transverse magnetization.]{\label{cap:T2_decay}$T_{2}$ decay of transverse magnetization
as described by equation \ref{eq:T2_decay}. $T_{2}=1$ unit.}
\end{figure}

Transverse relaxation refers to the decay of transverse magnetization
with time. It is called spin-spin relaxation\index{spin-spin relaxation}
and is designated by the symbol $T_{2}$. Phenomenologically it can
be described by the equation\begin{equation}
M_{\bot}(t)=M_{\bot initial}\, e^{-t/T_{2}}\label{eq:T2_decay}\end{equation}
in which the initial transverse magnetization decays exponentially
with time. This is the solution to the differential equation\begin{equation}
\frac{dM_{\perp}}{dt}=-\frac{M_{\perp}}{T_{2}}.\label{eq:T2_de}\end{equation}

Figure \ref{cap:T2_decay} shows the $T_{2}$ decay curve where $T_{2}$
is one time unit in magnitude.

The term spin-spin relaxation originates from the mechanism whereby
the field from other nuclei and nearby molecules, atoms, or ions is
a random function of time, and causes a slight change in phase of
a given nuclear moment's precession. These random phase variations
accumulate over time, causing a reduction in the net macroscopic transverse
magnetization.

The details of transverse relaxation mechanisms are beyond the scope
of this dissertation; the reader is referred to references \cite{BPP48,Dic51,Blo56,Bloem56,Bloch57,Abr99,Lev01}.

\section{Field Inhomogeneity\index{field inhomogeneity}, $T_{2}^{\dagger}$\index{T2 dagger}
and $T_{2}^{*}$\index{T2 star}\label{sec:Field-Inhomogeneity}}

There is another decay process analogous to transverse $T_{2}$ relaxation.
It is due to variations in the local magnetic field, but over macroscopic
distances and in a temporally deterministic (temporally non-random)
manner. Variation in the applied field $B_{0}$ is usually called
{}``$B_{0}$ inhomogeneity''\index{B0 inhomogeneity} while susceptibility
induced variations go by the name {}``sample inhomogeneity\index{sample inhomogeneity}''
or {}``susceptibility gradients\index{susceptibility gradients}.''

The accumulated random phase variations are assumed to cause an exponential
decay $T_{2}^{\dagger}$ %
\footnote{This assumption does not always hold, often the decay is Gaussian,
or the product of Gaussian and exponential terms \cite{YH94}, \cite[sec. 20.4.1 pp. 602-603]{HBT+99}.%
}. We combine the microscopic $T_{2}$ and macroscopic $T_{2}^{\dagger}$
decay process into one decay constant $T_{2}^{*}$ with

\begin{equation}
\frac{1}{T_{2}^{*}}=\frac{1}{T_{2}}+\frac{1}{T_{2}^{\dagger}}.\label{eq:T2star_T2_T2dag}\end{equation}
The transverse magnetization will then be described by the equation\begin{equation}
M_{\bot}(t)=M_{\bot initial}\, e^{-t/T_{2}^{*}}.\label{eq:T2star_decay}\end{equation}
We will talk about $B_{0}$ and sample inhomogeneity more in section
\ref{sec:Shimming}.

\section{Chemical Shift\label{sec:Chemical-Shift}}

In addition to the applied field, field inhomogeneity, and susceptibility
fields, each nuclear spin experiences a {}``local field.'' This
is due to the field of electrons and nuclei in the rest of the molecule
containing it, and fields from nearby molecules. This field is constantly
changing due to translational, vibrational, and rotational motion.
It is the fluctuating component of this local field that leads to
relaxation \cite{BPP48}. The time average component leads to a shift
in the Larmor frequency, called the chemical shift \cite{Hah50b}. 

There are two components of the shift, a dominant field proportional
shift (due to diamagnetic effects), and another usually smaller absolute
shift due to {}``J-coupling'' through bonds to other paramagnetic
nuclei in the molecule \cite{HM51,GMS51,HM52}. 

The field proportionality constant of the field dependent chemical
shift is often denoted by the symbol $\sigma$, and can be thought
of as the normalized resonance offset relative to a {}``reference''
Larmor frequency $\omega$. This is written\begin{equation}
\sigma\equiv\frac{\omega_{0}-\omega}{\omega}.\label{eq:sigma}\end{equation}
$\sigma$ is dimensionless and is almost always reported in units
of $10^{-6}$ or {}``parts per million'' (ppm).

The chemical shift gives information about local chemical bond geometry
and average motion. It is the principal parameter of interest for
determining chemical structure using modern NMR spectroscopy \cite{Arn56}.
A more detailed discussion of the origin of chemical shift can be
found in \cite[section 7.7]{Lev01}.

\chapter{SPIN-ECHO}

The spin-echo\index{spin-echo} is another key concept of NMR and
MRI. First demonstrated by E. L. Hahn\index{Hahn, E. L.} \cite{Hah50b},
spin-echoes continue to be utilized in many NMR and MRI experiments.
The spin-echo is a way of refocusing (or re-phasing) the effects of
temporally static field inhomogeneities (see section \ref{sec:Field-Inhomogeneity}).

\begin{figure}
\includegraphics[%
  width=5.5in,
  keepaspectratio]{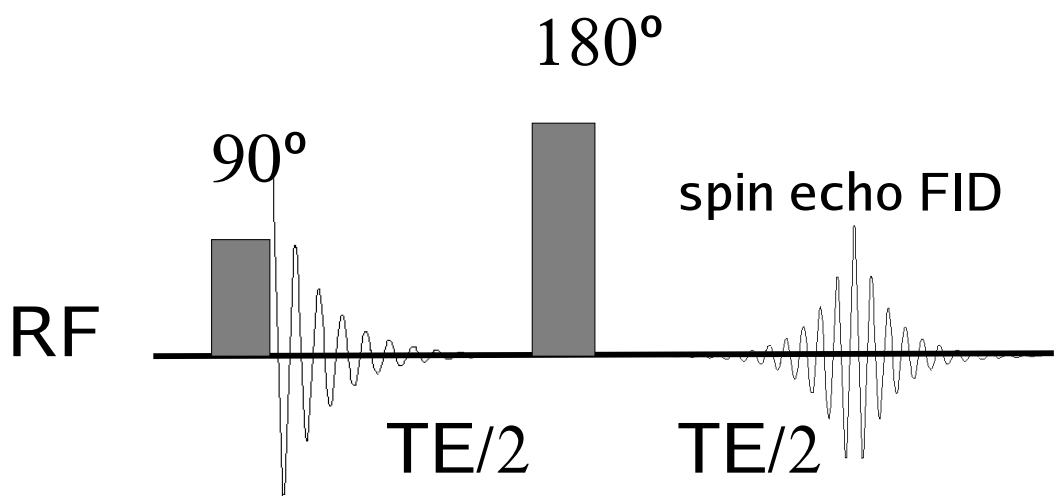}

\caption{\label{cap:Spin-Echo}Spin-echo pulse sequence.}
\end{figure}

A spin-echo consists of a 90° pulse to excite transverse magnetization
followed by a 180° pulse, shown in figure \ref{cap:Spin-Echo}. The
effect of the 180° pulse is to invert the phase of the transverse
magnetization. Any phase acquired due to field inhomogeneities or
gradients (see section \ref{cha:Gradients}) during the $TE/2$ time
period before the 180° pulse is canceled by the phase acquired during
the $TE/2$ time period after the 180° pulse. 

The envelope of a spin-echo free induction decay (FID)\index{Free Induction Decay, FID}
(not counting off resonance oscillation, such as chemical shift) is\begin{equation}
M_{\bot}(t)=M_{\bot initial}e^{-t/T_{2}}e^{-|t-TE|/T_{2}^{\dagger}}.\label{eq:spin_echo_envelope}\end{equation}
 $TE$\index{TE} is called the echo-time\index{echo-time}. Note
that the build (before $\frac{TE}{2}$) and decay (after $\frac{TE}{2}$)
sides of the FID are not symmetric in the presence of $T_{2}$ decay.

It is possible to use multiple 180° pulses (spaced $TE$ apart) and
refocus multiple echoes. This is sometimes called a CP sequence after
Carr and Purcell \cite{CP54}, who originally used such a sequence
to examine the effects of diffusion\index{diffusion} (see section
\ref{cha:Diffusion}) and $T_{2}$ relaxation. Modification of the
phase of the RF pulses (where the inversion pulses are shifted by
90° in phase) is called CPMG\index{CPMG} sequence, from Carr-Purcell-Meiboom-Gill\index{Carr-Purcell-Meiboom-Gill}
\cite{MG58}. A CPMG sequence has the desirable property of being
less sensitive to pulse amplitude errors than a CP sequence, especially
for the even echoes. This latter property is sometimes called {}``even
echo re-phasing.''

\chapter{GRADIENTS\label{cha:Gradients}}

Magnetic field gradients\index{gradients} are a useful tool in NMR
spectroscopy for destroying unwanted signals and introducing diffusion
weighting (see chapter \ref{cha:Diffusion}). Gradients are required
for MRI. 

A gradient is produced by a secondary set of magnetic coils, designed
so that the field varies linearly with position along the direction
of the gradient \cite{LCB+03}. An $x$ gradient field can be represented
by the equation\begin{equation}
\Delta\vec{B}(x)=G_{x}x\,\hat{z}.\label{eq:gradient}\end{equation}
Note that the direction of the gradient refers to the direction along
which the gradient strength varies, not the direction of the field.
MRI instruments usually possess three gradient coils, to produce orthogonal
$x$, $y$, and $z$ gradients. These can be linearly combined into
an arbitrary gradient direction $\hat{s}$.

Applying a gradient causes the magnetization to twist into a helix
along the direction of the gradient. The longer the gradient is applied,
the more twisted the transverse magnetization becomes. The resulting
NMR signal, when the magnetization is in a twisted state, is greatly
reduced when there are many twists across the sample. This is sometimes
called {}``crushing\index{crushing}'' or {}``spoiling\index{spoiling}''
the transverse magnetization.

\section{Pulsed Gradients}

\begin{figure}
\includegraphics[%
  scale=0.75]{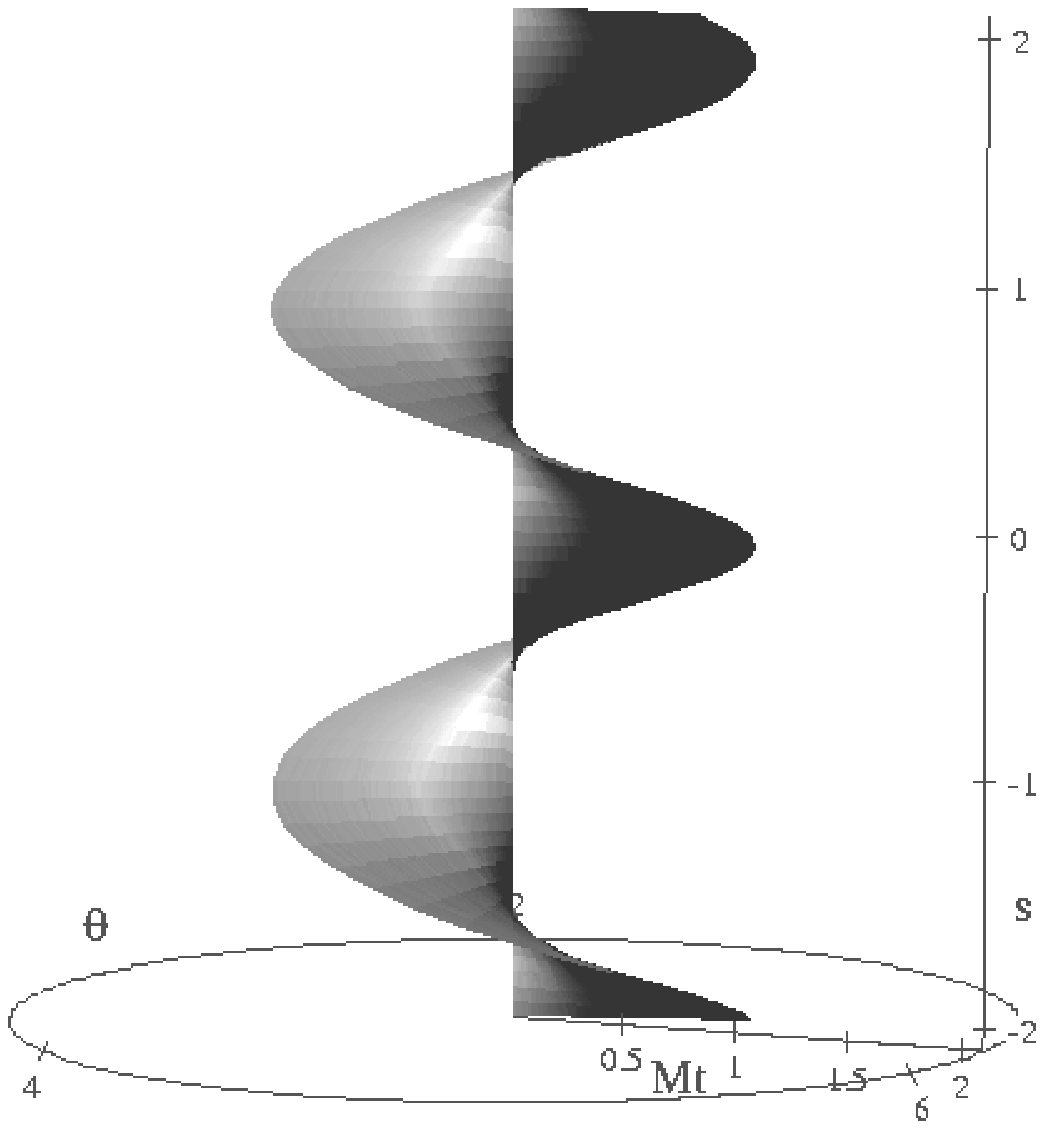}\hfill{}\includegraphics[%
  scale=0.75]{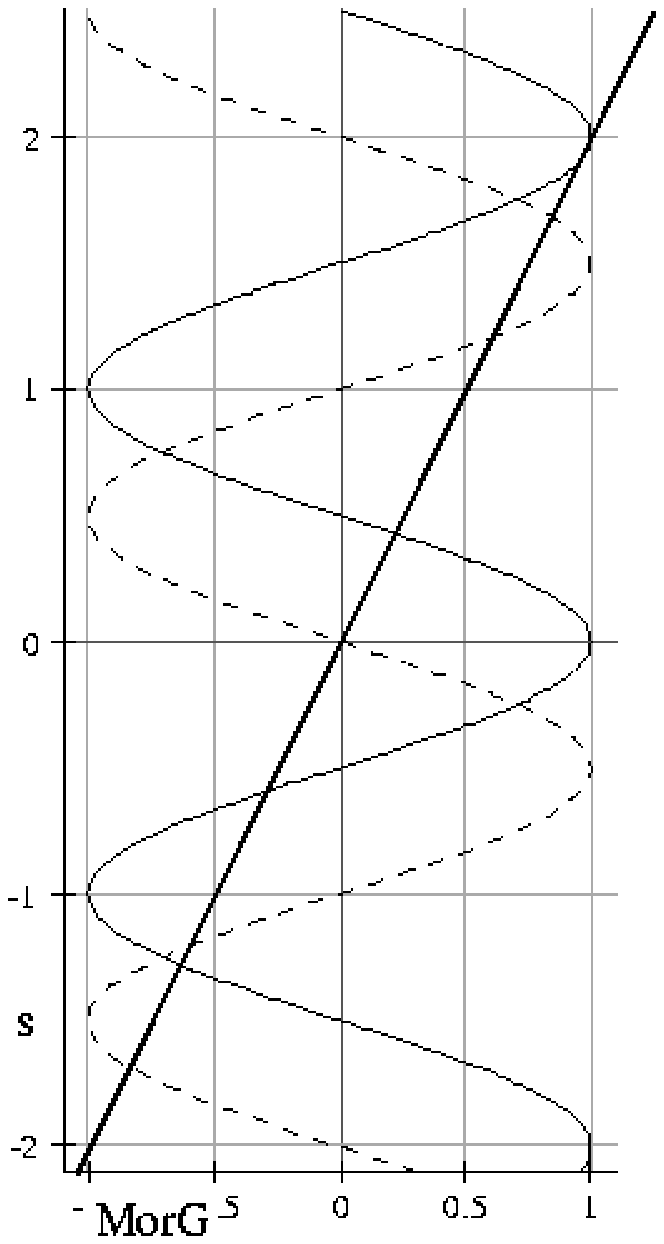}

\caption[Transverse magnetization helix after gradient pulse along arbitrary
direction $\hat{s}$.]{\label{cap:Staircase}Transverse magnetization helix after gradient
pulse along arbitrary direction $\hat{s}$. The Gradient $G_{s}$
is shown as heavy line in the component graph, $M_{x}$ is a normal
line and $M_{y}$ is dashed.}
\end{figure}

Gradient hardware is designed so that it can deliver pulses, much
like the RF coil and transmitter discussed earlier. In modern instruments
the gradient amplitude can be controlled digitally so that the amplitude
of the gradient can be made a function of time. Figure \ref{cap:Staircase}
shows the transverse magnetization\index{transverse magnetization}
along an arbitrary gradient direction $\hat{s}$ after a gradient
pulse.

\subsection{Pulse Sequence}

A series of RF and gradient pulses, interspersed with delays and acquisition
periods, is called a pulse sequence\index{pulse sequence}. We have
already seen an example (without gradients) in figure\ref{cap:Spin-Echo}.

\section{Secular Approximation of Quasi-static Fields\index{Secular Approximation}}

In the presence of a large applied magnetic field, small additional
static (or slowly varying) fields such as gradients can be treated
as a perturbation%
\footnote{also other small fields due to susceptibility and inhomogeneity%
}. We can look at the effect of a gradient on the Larmor frequency\index{Larmor frequency}
first with the gradient field $\Delta\vec{B}(x)=G_{x}x\,\hat{z}$
oriented in the same direction as the applied field $\vec{B}_{0}$=$B_{0}\hat{z}$,
and then with the gradient field oriented orthogonally $\Delta\vec{B}(x)=G_{x}x\,\hat{y}$.

When the gradient field is oriented parallel to $\vec{B}_{0}$ we
have\begin{equation}
\vec{B}=B_{0}\hat{z}+G_{x}x\,\hat{z}.\label{eq:B_plus_G}\end{equation}
The field magnitude is\begin{equation}
B=B_{0}+G_{x}x,\label{eq:B_plus_G_magnitude}\end{equation}
causing a first order change in the Larmor frequency%
\footnote{For the sign convention used in this thesis, see Appendix \ref{cap:Levitt_sign_convention}
and references \cite[section 2.5, page 30]{Lev01} or \cite{Lev97}.%
}\begin{equation}
\Delta f=\frac{-\gamma}{2\pi}(G_{x}x).\label{eq:Delta_f}\end{equation}
When the gradient field is oriented orthogonal to the large applied
field we have\begin{equation}
\vec{B}=B_{0}\hat{z}+G_{x}x\,\hat{y}.\label{eq:B_plus_G_orthogonal}\end{equation}
The field magnitude is then\begin{equation}
B=\sqrt{B_{0}^{2}+(G_{x}x)^{2}},\label{eq:B_plus_G_orth_mag}\end{equation}
which we can expand in a Taylor's series to\begin{equation}
B\approx B_{0}+\frac{(G_{x}x)^{2}}{2\, B_{0}}'\label{eq:B_plus_G_orth_taylor's}\end{equation}
which yields\begin{equation}
\Delta f\approx\frac{-\gamma}{4\pi}\frac{(G_{x}x)^{2}}{B_{0}}.\label{eq:Delta_f_orth}\end{equation}
If we set values for $B_{0}=4.7T$ and $G_{x}=200\frac{mT}{m}$ with
$x=1cm$ in the parallel case we have $|\Delta f|=85.15kHz$ and in
the orthogonal case $|\Delta f|=18.12Hz$, which is more than 3 orders
of magnitude smaller. Most susceptibility gradients and inhomogeneities
are much smaller than $200\frac{mT}{m}$, and if their field orientations
are not along $\vec{B}_{0}$, they can safely be ignored.

The above approximation of ignoring field components perpendicular
to the static field is called the secular approximation or taking
the secular component of the field. We will address the secular component
of fields that include a rotating component (are rapidly varying)
in section \ref{sec:Secular-Dipolar-Field} and appendix \ref{sec:Secular-Component-Derived}.

\chapter{DIFFUSION\label{cha:Diffusion}\index{diffusion}}

In many NMR and most MRI experiments the sample of interest is a liquid
or composed of liquids in biological compartments. There are many
fortuitous properties of a liquid sample that make NMR easier than
on a solid sample. When the nucleus of interest is in a liquid, the
random motion of molecules causes an averaging effect on the fields
due to other nearby nuclei and molecules. This contributes to so called
{}``motional narrowing\index{motional narrowing}'' giving liquids
much narrower spectral lines than solids. For details see references
\cite[sec. X]{BPP48} \cite{PS55} \cite[ch. 15]{Lev01} \cite[ch. X]{Abr99}
and \cite[sec. 5.12]{Sli90}.

We can think of a molecule in a liquid as taking a {}``random walk\index{random walk}''
in three dimensions. Assuming no macroscopic flow (or convection),
the motion will be mainly due to thermal kinetic energy and collisions
with other molecules. If the sample has no barriers, and is a normal
liquid (not a liquid crystal), the motion will be isotropic, meaning
that motion in any direction is equally probable.

\section{Fick's Laws}

Diffusion of a scalar field $c(\vec{r},\, t)$ can be described by
Fick's first law \cite{Pri97}\begin{equation}
\vec{J}(\vec{r},\, t)=-\mathbf{D}\nabla c(\vec{r},\, t).\label{eq:Ficks_1st_law}\end{equation}
 $\vec{J}$ is the flux of a given substance (or field), $c(\vec{r},\, t)$,
is the concentration (or field amplitude), and $\mathbf{D}$ is the
diffusion tensor (discussed in section \ref{sub:Diffusion-in-3d}).
In simple terms this equation says that there is a {}``flow'' from
high concentration to low concentration. Heat flow obeys a diffusion
equation and so does the motion of molecules in a liquid (if there
is no macroscopic flow or convection). We can combine this with the
continuity equation\begin{equation}
\frac{\partial c(\vec{r},\, t)}{\partial t}=-\nabla\cdot\vec{J}(\vec{r},\, t).\label{eq:continuity}\end{equation}
Equation \ref{eq:continuity} says that the time rate of change in
concentration must be equal to the divergence of the flux (what goes
into a small volume either goes out or increases the concentration).
Fick's second law, also called the diffusion equation, is therefore\begin{equation}
\frac{\partial c(\vec{r},\, t)}{\partial t}=\nabla\cdot\mathbf{D}\nabla c(\vec{r},\, t).\label{eq:Ficks_2nd_law}\end{equation}
The diffusion equation reduces to\begin{equation}
\frac{\partial c(\vec{r},\, t)}{\partial t}=D\nabla^{2}c(\vec{r},\, t),\label{eq:Ficks_2nd_law_isotropic}\end{equation}
where D is a constant, for isotropic diffusion.

\subsection{Diffusion in 1d}

We will first consider diffusion in one dimension. The probability
that a molecule $n$ will be found a distance $x$ from its starting
point is given by\begin{equation}
P(x_{n},\, t)=\frac{e^{-\frac{x_{n}^{2}}{4\, D_{x}t}}}{\sqrt{4\,\pi\, D_{x}t}}.\label{eq:diffusion_pdf}\end{equation}

Equation \ref{eq:diffusion_pdf} says that the probability is normally
distributed (as expected from a large number of random collisions
and motions), with the variance $2\, D_{x}t$ increasing linearly
with time.

\begin{figure}
\includegraphics[%
  width=1.0\columnwidth,
  keepaspectratio]{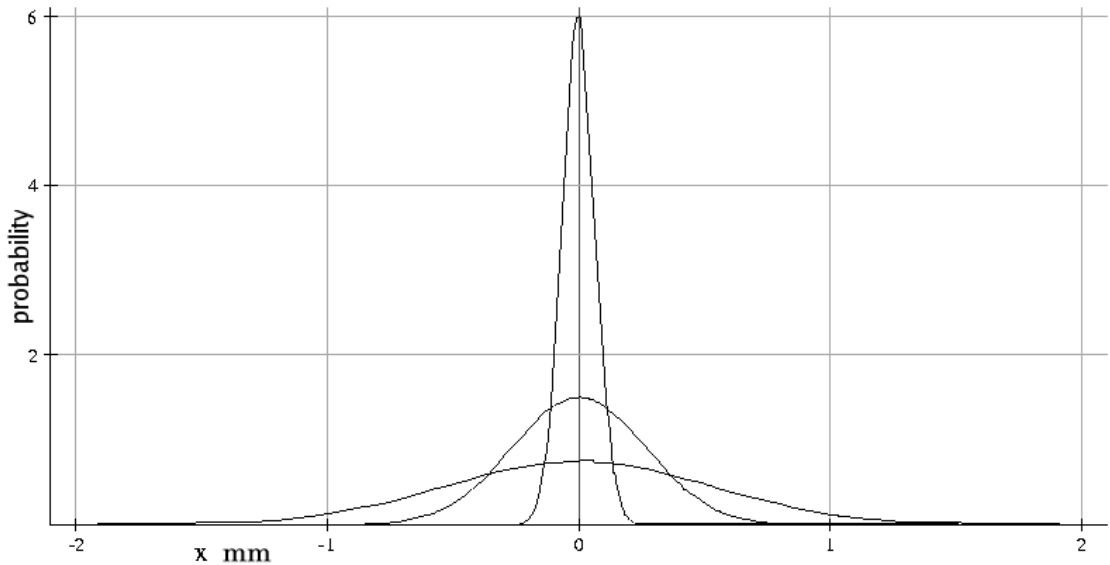}

\caption{\label{cap:Diffusion}Diffusion of water in water (self-diffusion)
at 1s, 10s, and 100s. }
\end{figure}

It also is the solution to the 1d diffusion equation\begin{equation}
\frac{\partial P(x_{n},\, t)}{\partial t}=D_{x}\,\frac{\partial^{2}P(x_{n},\, t)}{\partial x^{2}}.\label{eq:diffusion}\end{equation}
$D_{x}$ is the diffusion coefficient and has units of $[\frac{m^{2}}{s}]$.
At room temperature the diffusion coefficient of water (in water)
is $2.2\times10^{-9}\frac{m^{2}}{s}$. An example is shown in figure
\ref{cap:Diffusion}. Notice that at 100 seconds there is still only
a small probability of finding the molecule greater than 1mm from
its starting point. This is why stirring is much more effective than
diffusion for mixing at short times.

When we have a large number of identical molecules we can think of
molecules starting in a small region distributing themselves into
a larger region. We cannot predict where an individual molecule will
go, but we do know on average how they will be distributed. This is
also called an ergodic average\index{egodic average}.

\subsection{Diffusion in 3d\label{sub:Diffusion-in-3d}}

We can treat the problem of diffusion in three dimensions separably,
that is as three one dimensional problems. In this case we can consider
the possibility of the diffusion coefficients in each direction being
different. This is not the case for most pure liquids, but is often
the case in biological tissues where barriers and restriction in compartments
cause the {}``apparent diffusion coefficient'' to depend on direction.
In general, the apparent diffusion in a biological sample coefficient
could be more complicated, depending on the exact direction of interest.

In the 3d case the probability distribution is\begin{equation}
P(x_{n},\, y_{n},\, z_{n},\, t)=\frac{e^{-\frac{x_{n}^{2}}{4\, D_{x}t}}}{\sqrt{4\,\pi\, D_{x}t}}\frac{e^{-\frac{y_{n}^{2}}{4\, D_{x}t}}}{\sqrt{4\,\pi\, D_{y}t}}\frac{e^{-\frac{z_{n}^{2}}{4\, D_{x}t}}}{\sqrt{4\,\pi\, D_{z}t}}.\label{eq:diffusion_pdf_3d}\end{equation}
A useful extension is to allow the axes, while still orthogonal, to
be rotated in an arbitrary direction. This leads to the so called
{}``diffusion tensor''\cite{BML94},\begin{equation}
\mathbf{D}=\left(\begin{array}{ccc}
D_{xx} & D_{xy} & D_{xz}\\
D_{yx} & D_{yy} & D_{yz}\\
D_{zx} & D_{zy} & D_{zz}\end{array}\right).\label{eq:D}\end{equation}
We define the reciprocal diffusion tensor as\begin{equation}
\mathbf{D}_{rec}=\left(\begin{array}{ccc}
D_{xx}^{-1} & D_{xy}^{-1} & D_{xz}^{-1}\\
D_{yx}^{-1} & D_{yy}^{-1} & D_{yz}^{-1}\\
D_{zx}^{-1} & D_{zy}^{-1} & D_{zz}^{-1}\end{array}\right).\label{eq:Drec}\end{equation}
The probability distribution (dropping the $n$ subscript) becomes\begin{equation}
P(\vec{r},\, t)=\frac{e^{-\frac{\vec{r}^{T}\mathbf{D}_{rec}\vec{r}}{4\, t}}}{\sqrt{(4\,\pi\, t)^{3}\left|\mathbf{D}\right|}},\label{eq:diffusion_pdf_3d_tensor}\end{equation}
where $T$ denotes the transpose operation and $\left|\mathbf{D}\right|$
is the determinant. When diffusion is isotropic $\mathbf{D}$ becomes
a scalar $D$ and equation \ref{eq:diffusion_pdf_3d_tensor} becomes

\begin{equation}
P(r,\, t)=\frac{e^{-\frac{r^{2}}{4\, D\, t}}}{\sqrt{(4\,\pi\, D\, t)^{3}}}.\label{eq:diffusion_pdf_3d_tensor}\end{equation}
Both solutions obey the differential equation,\begin{equation}
\frac{\partial P(\vec{r},\, t)}{\partial t}=\nabla\cdot\mathbf{D}\nabla P(\vec{r},\, t),\label{eq:anisotropic_diffusion_3d}\end{equation}
which reduces to\begin{equation}
\frac{\partial P(r,\, t)}{\partial t}=D\,\nabla^{2}P(r,\, t)\label{eq:diffusion_3d}\end{equation}
for the isotropic case.

\section{Self-Diffusion in water}

In NMR and MRI we are often interested in self-diffusion\index{self-diffusion}
of water. This is the diffusion of water molecules in a solution that
is composed of other water molecules. In order for this diffusion
to be detected we must {}``label'' the water molecules in some manner.
The most convenient way to label the water molecules is by using a
gradient or RF pulse to change the amplitude or orientation of the
nuclear magnetization of the $^{1}H$ molecules. We can now talk about
the diffusion of the magnetization itself.

Since magnetization is a vector quantity we have to modify the diffusion
equation to operate on a vector field. This is to say that diffusion
operates on each of the components of the magnetization. The equation
for magnetization with (isotropic) diffusion in the rotating frame
is \begin{equation}
\frac{\partial\vec{M}}{\partial t}=D\nabla^{2}\vec{M}.\label{eq:Bloch_vec_diffusion}\end{equation}

\subsection{Diffusion Weighting\index{diffusion weighting} with Gradients\label{sub:Diffusion_Weighting}}

Application of gradients during an NMR or MRI experiment can cause
additional attenuation of the signal when there is significant diffusion.
In early experiments \cite{CP54} this was recognized as a confounding
factor in measuring $T_{2}$. Later, NMR and MRI measurement of the
diffusion properties of solutions and biological samples developed
into a rich subfield in itself \cite{ST65,BML94,Pri97,Pri98}.

When no gradients are present, diffusion will not explicitly affect
the NMR signal%
\footnote{There is however a link between diffusion, $T_{1}$ and $T_{2}$,
see {}``BPP'' Bloembergen et al. \cite{BPP48}%
}. When a gradient is applied, the phase of spins in the transverse
plane is altered as a function of position along the direction of
the gradient (there is also dependence on gradient strength and the
duration). If there is diffusion along the gradient direction, then
spins labeled with one phase will move into regions of spins having
a different phase. This causes a net reduction in the macroscopic
transverse magnetization, and detected achievable signal. A pulse
sequence where the signal responds in a known manner to diffusion
is called {}``diffusion weighted\index{diffusion weighted}.'' It
is also possible to have diffusion weighting due to diffusing longitudinal
magnetization.

We show examples and signal equations of sequences with diffusion
weighting in section \ref{sec:Stejskal-Tanner-Sequence} and \ref{sec:Stimulated Echo}.

\chapter{BLOCH EQUATIONS\label{cha:Bloch-Equations}}

The Bloch%
\footnote{sometimes called Bloch-Torrey (for tipped coordinates) or Bloch-Redfield
equations%
} equations are a set of coupled differential equations that describe
the behavior of the macroscopic magnetization \cite[ch. III. sec. II.]{Blo46,Abr99}.
The equations can account for the effects of precession, relaxation,
field inhomogeneity, and RF pulses that we have already seen in previous
sections. If one considers the magnetization as a function of space
as well as time, we can include the effects of gradients and diffusion
\cite{Tor56,Pri97}.

\section{Vector Bloch Equation}

The vector Bloch equation in the notation introduced in the previous
sections is\begin{equation}
\frac{d\vec{M}}{dt}=\gamma\,\vec{M}\times\vec{B}+\frac{(\vec{M}_{0}-\vec{M}_{\Vert})}{T_{1}}-\frac{\vec{M}_{\perp}}{T_{2}}+\nabla\cdot\mathbf{D}\nabla\vec{M}.\label{eq:Bloch_eq_vector}\end{equation}

$\vec{B}$ is assumed to include all applied fields as well as the
field $\bigtriangleup\vec{B}$ due to $\vec{B}_{0}$ inhomogeneity
and susceptibility effects. All fields could be written as functions
of $\vec{r}$ if we wish to capture inhomogeneity and gradient effects.
$\vec{B}_{1}$and $\vec{G}_{s}$are also functions of $t$ as determined
by the pulse sequence. We write all this as\begin{equation}
\vec{B}=\vec{B}_{0}+\vec{B}_{1}+\vec{G}_{s}s+\bigtriangleup\vec{B}.\label{eq:B_total}\end{equation}
In general $T_{1}$, $T_{2}$ and $\mathbf{D}$ could be functions
of $\vec{r}$ as well. We can transform to the rotating frame by replacing
$\vec{B}_{0}$ with $\frac{\Delta\omega_{0}}{\gamma}\hat{z}$ and
make sure that the frequency of $\vec{B}_{1}$ is offset accordingly.

\section{Longitudinal and Transverse Bloch Equations}

We can break the single vector equation into its longitudinal and
transverse components. We will use the complex notation for the transverse
components. The equation for the longitudinal component is\begin{equation}
\frac{dM_{\parallel}}{dt}=\gamma\,[\vec{M}\times\vec{B}]_{\parallel}+\frac{(M_{0}-M_{\parallel})}{T_{1}}+\nabla\cdot\mathbf{D}\nabla M_{\parallel}.\label{eq:Bloch_eq_long}\end{equation}
Noting that $B_{\perp}=B_{1}$, the term $\gamma\,[\vec{M}\times\vec{B}]_{\parallel}$can
be expanded (see appendix \ref{sec:Cross-Product-with-Ml-and-Mt})
to yield,

\begin{equation}
\frac{dM_{\parallel}}{dt}=\frac{i\gamma}{2}(M_{\perp}B_{1}^{*}-M_{\perp}^{*}B_{1})+\frac{(M_{0}-M_{\parallel})}{T_{1}}+\nabla\cdot\mathbf{D}\nabla M_{\parallel}.\label{eq:Bloch_eq_long_expand}\end{equation}
For the transverse component we have\begin{equation}
\frac{dM_{\perp}}{dt}=\gamma\,[\vec{M}\times\vec{B}]_{\perp}-\frac{M_{\perp}}{T_{2}}+\nabla\cdot\mathbf{D}\nabla M_{\perp},\label{eq:Bloch_eq_tran}\end{equation}
and on expanding the cross product\begin{equation}
\frac{dM_{\perp}}{dt}=i\,\gamma\,(M_{\parallel}B_{1}-M_{\perp}B_{\parallel})-\frac{M_{\perp}}{T_{2}}+\nabla\cdot\mathbf{D}\nabla M_{\perp}\label{eq:Bloch_eq_expand_tran}\end{equation}
with 

\[
B_{\parallel}=B_{0}+G\, s+\bigtriangleup B.\]

One replaces $B_{0}$ with $\frac{-\Delta\omega_{0}}{\gamma}$ in
the rotating frame. Note that in the above equations only the $B_{1}$
RF field couples the transverse and longitudinal magnetization. We
will see in part \ref{par:DDF} that there is another process called
{}``radiation dampening'' that can achieve this as well.

\chapter{SHIMMING\label{sec:Shimming}}

Most NMR and MRI experiments rely on having a constant large applied
magnetic field over the volume of the sample. The NMR signal is the
average of the magnetization from each small volume element of the
sample. If the applied field varies over the sample, the magnetization
from different regions of the sample will get out of phase. This leads
to reduction in the overall signal from the sample and broadening
of spectral lines. We discussed this effect in section \ref{sec:Field-Inhomogeneity}.

The $B_{0}$ field in an NMR spectrometer or MRI system is created
by a large magnet, in most cases a superconducting electromagnet \cite{MR87,Wil89,Fon95}.
Magnets designed for NMR and MRI have very stringent requirements
for homogeneity. In high resolution spectroscopy it is often desired
to get homogeneity of the order of 0.1Hz in a field of 600MHz over
a 1cm diameter volume. This is less than one part per billion. Homogeneity
requirements in imaging are much less stringent, but typically are
required over much larger volumes. One would usually like to achieve
10Hz over a 20cm diameter volume at a field strength of 1.5T, or approximately
0.1 parts per million (ppm). 

Because of imperfections in the magnet, inherent in the design, due
to manufacturing tolerances, or changes with age and use, all NMR
and MRI magnets have additional smaller magnets called shims to adjust
the homogeneity \cite{RH84,CH90}. Often there are two or three sets
of shims, {}``steel,'' {}``superconducting,'' and {}``room temperature.''
Steel shims are adjusted as part of the charging procedure after the
main magnet is brought up to field. They consist of either a set of
steel slugs or bands that are placed with the help of field mapping
and fitting software. Superconducting shims are secondary coils wound
within the cryostat. They are adjusted by altering their currents
after the magnet is charged and stabilized, and can also be adjusted
as part of maintenance.

In addition to imperfections in the magnet, shims compensate for susceptibility-induced
fields, which vary from sample to sample (or patient to patient in
MRI). Room temperature shims are electromagnetic coils. They are adjusted
on a per sample basis. Often this is by means of an automated {}``pre-scan''
procedure in clinical imaging. Often, in order to achieve narrow line-widths
in NMR spectroscopy, manual shimming is necessary, which can be time
consuming for the less experienced user.

\chapter{EXAMPLE PULSE SEQUENCES}

In the following sections the Bloch equations are used to solve for
the magnetization and signal for pulse sequences relevant to this
dissertation.

\section{Stejskal-Tanner Sequence\label{sec:Stejskal-Tanner-Sequence}}

\begin{figure}
\includegraphics[%
  width=1.0\columnwidth,
  keepaspectratio]{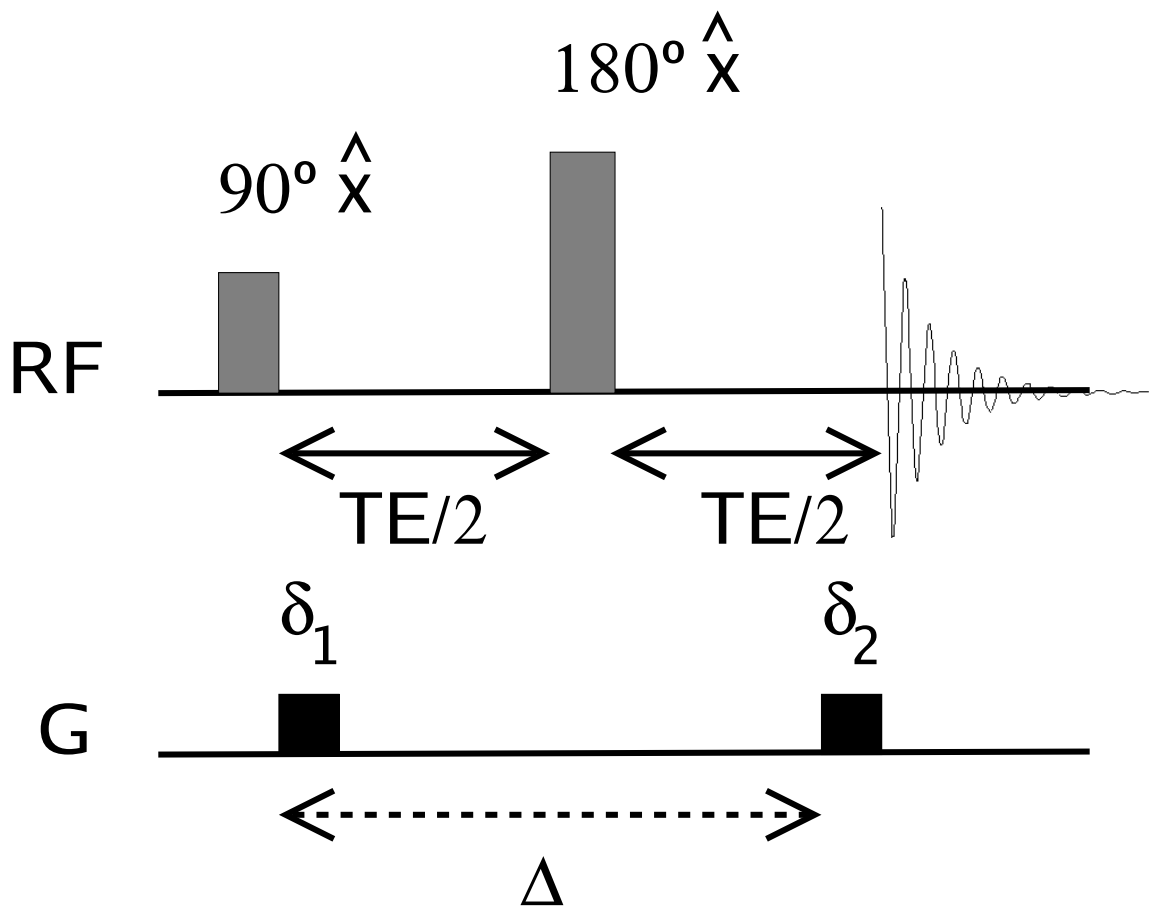}

\caption[Stejskal-Tanner sequence.]{\label{cap:st_ps}Stejskal-Tanner sequence. The large time interval
$\Delta$ is the time from the start of the first gradient pulse to
the start of the second gradient pulse. The RF pulse durations are
assumed to be negligible in this analysis.}
\end{figure}
First we will look at a simple spin-echo sequence with two pulsed
gradients shown in figure \ref{cap:st_ps}. This sequence was introduced
by Stejskal and Tanner \cite{ST65} and is often called Stejskal-Tanner
(ST) sequence or a {}``pulsed-gradient spin-echo\index{pulsed-gradient spin-echo}''
sequence.

\subsection{Initial Magnetization}

We will start with fully relaxed longitudinal magnetization

\begin{equation}
M_{\parallel}^{0}=M_{0},\label{eq:Ml_init_st}\end{equation}
which implies zero transverse magnetization

\begin{equation}
M_{\perp}^{0}=0.\label{eq:Mt_init_st}\end{equation}
We have denoted the longitudinal and transverse magnetization with
the superscript $0$ to designate the initial condition.

\subsection{Excitation Pulse}

The first RF pulse is a $90^{\circ}$ pulse. This excites all of the
magnetization into the transverse plane. We use the superscript $90$
to denote the magnetization state after the $90^{\circ}$ pulse. The
phase of the pulse is $\phi_{90}=0$ (denoted by $\hat{x}$ in the
rotating frame) so we end up with our transverse magnetization along
$\hat{y}$ (or the imaginary direction) after the pulse. We have,
therefore,\[
M_{\parallel}^{90}=0,\]
\[
M_{\perp}^{90}=i\, M_{0}.\]

The durations of both RF pulses in the sequence are assumed to be
negligible compared to the gradient durations $\delta$ and the echo
time $TE$.

\subsection{Gradient Pulse $\delta$ without Relaxation or Diffusion}

In general there will be $T_{1}$and $T_{2}^{*}$ relaxation occurring
after excitation, but we will neglect this for the moment. Also, we
will neglect diffusion for the moment and look at the solution to
the Bloch equation in the presence of the gradient pulse. In the rotating
frame the Bloch equation is

\[
\frac{d\vec{M}}{dt}=\gamma\,\vec{M}\times[G_{s}(t)\, s\,\hat{z}],\]
which is equation \ref{eq:Bloch_eq_tran} with only the gradient term.
The gradient is along the arbitrary direction $\hat{s}$ and $s$
is the distance along $\hat{s}$ from the origin. 

By neglecting relaxation, we need only consider the Bloch equation
for the transverse component (in complex form)\begin{equation}
\frac{\partial M_{\bot}}{\partial t}=-i\,\gamma\, M_{\bot}G_{s}(t)\, s.\label{eq:Tran_gradient}\end{equation}
 We can divide both sides by $M_{\bot}$\begin{equation}
\frac{\partial M_{\bot}}{M_{\bot}}=-i\,\gamma\, G_{s}(t)\, s\,\partial t,\label{eq:Tran_gradient}\end{equation}
and integrate to get\begin{equation}
ln(\frac{M_{\bot}}{M_{\perp}^{initial}})=-i\,\gamma\, s\int_{0}^{t}G_{s}(t')\, dt'.\label{eq:Mtran_log}\end{equation}
 The solution to \ref{eq:Tran_gradient} becomes\begin{equation}
M_{\bot}(s,\, t)=M_{\perp}^{initial}e^{-i\,2\,\pi\, q(t)\, s}.\label{eq:Mtran_staircase}\end{equation}
Equation \ref{eq:Mtran_staircase} is {}``staircase'' twisted transverse
magnetization as shown in figure \ref{cap:Staircase}. We have defined\begin{equation}
q(t)\equiv\frac{\gamma}{2\,\pi}\int_{0}^{t}G_{s}(t')\, dt'.\label{eq:q_again}\end{equation}
This says that the instantaneous pitch of the magnetization twist
along $\hat{s}$ is equal to the integral over time of the gradient.

\subsection{Gradient Pulse $\delta$ with Diffusion}

The effect of diffusion will be to introduce a time-dependent term
to the solution in equation \ref{eq:Mtran_staircase}, which for now
we can assume to be complex valued (it could alter the phase of the
magnetization)\begin{equation}
M_{\bot}(s,\, t)=A(t)\, M_{\perp initial}e^{-i\,2\,\pi\, q(t)\, s}.\label{eq:Mtran_attenuated}\end{equation}
We insert this into the transverse Bloch equation in the rotating
frame (this time with the diffusion term) \begin{equation}
\frac{\partial M_{\bot}}{\partial t}=-i\,\gamma\, M_{\bot}G_{s}(t)\, s+\nabla\cdot\mathbf{D}\nabla M_{\perp}.\label{eq:Bloch_grad_diff}\end{equation}
Since the only spatial variation in $M_{\bot}$ is along $\hat{s}$
we can replace $\nabla\cdot\mathbf{D}\nabla M_{\perp}$ with $D_{s}\frac{\partial^{2}M_{\bot}}{\partial s^{2}}$
to get\begin{equation}
\frac{\partial M_{\bot}}{\partial t}=-i\,\gamma\, M_{\bot}G_{s}(t)\, s+D_{s}\frac{\partial^{2}M_{\bot}}{\partial s^{2}}.\label{eq:Bloch_grad_diff_s}\end{equation}
Substituting \ref{eq:Mtran_attenuated} into the spatial and temporal
derivatives in \ref{eq:Bloch_grad_diff_s} we have\begin{equation}
\frac{\partial M_{\bot}}{\partial t}=-i\,2\,\pi\frac{\partial q(t)}{\partial t}A(t)\, M_{\perp initial}e^{-i\,2\,\pi\, q(t)\, s}+\frac{\partial A(t)}{\partial t}\, M_{\perp initial}e^{-i\,2\,\pi\, q(t)\, s}\label{eq:dMtran_dt}\end{equation}
and\begin{equation}
D_{s}\frac{\partial^{2}M_{\bot}}{\partial s^{2}}=-4\, D_{s}\pi^{2}q^{2}(t)\, A(t)\, M_{\perp initial}e^{-i\,2\,\pi\, q(t)\, s}.\label{eq:Dsd2Mtrands2}\end{equation}
 These lead to the following equation for $A(t)$\begin{equation}
\frac{\partial A(t)}{\partial t}=-4\, D_{s}\pi^{2}q^{2}(t)\, A(t).\label{eq:dAdt_st}\end{equation}
We can divide both sides by $A(t)$ and integrate to get\begin{equation}
ln[\frac{A(t)}{A_{0}}]=-4\, D_{s}\pi^{2}\int_{0}^{t}q^{2}(t')\, dt.\label{eq:A_log}\end{equation}
We can set $A_{0}=1$ and put this into the form\begin{equation}
A(t)=e^{-b(t)\, D_{s}},\label{eq:AebD}\end{equation}
with the $b-value$\index{b-value} defined as\begin{equation}
b(t)\equiv4\,\pi^{2}\int_{0}^{t}q^{2}(t')\, dt.\label{eq:b_defined}\end{equation}
Our general solution for the transverse magnetization in the presence
of diffusion and an applied gradient becomes\begin{equation}
M_{\bot}(s,\, t)=M_{\perp initial}e^{-b(t)\, D_{s}}e^{-i\,2\,\pi\, q(t)\, s}.\label{eq:Mtran_se_diffusion}\end{equation}

After the gradient of duration $\delta$ in the superscript notation
we have\begin{equation}
M_{\parallel}^{\delta}=M_{\parallel}^{90}=0,\label{eq:Ml_delta_st}\end{equation}
\begin{equation}
M_{\bot}^{\delta}=M_{\perp}^{90}e^{-b(\delta)\, D_{s}}e^{-i\,2\,\pi\, q(\delta)\, s}=i\, M_{0}e^{-b(\delta)\, D_{s}}e^{-i\,2\,\pi\, q(\delta)\, s}.\label{eq:Mt_delta_st}\end{equation}

\subsection{$\frac{TE}{2}$ Delay}

The situation during the rest of $\frac{TE}{2}$ is much the same
as during $\delta$ except there is no gradient so $q$ is constant.
The $b-value$, however, will continue to evolve during this period.
We have\begin{equation}
M_{\parallel}^{\frac{TE}{2}}=0\label{eq:Ml_TE2_st}\end{equation}
and\begin{equation}
M_{\bot}^{\frac{TE}{2}}=i\, M_{0}e^{-b(\frac{TE}{2})\, D_{s}}e^{-i\,2\,\pi\, q(\delta)\, s}\label{eq:Mt_TE2_st}\end{equation}
at the time $\frac{TE}{2}$ just before the $180^{\circ}$ pulse.

\subsection{$180^{\circ}$ Pulse}

The effect of the $180^{\circ}$ pulse is to invert the $\hat{y}$
component of the transverse magnetization. It would also invert the
longitudinal magnetization if present. Note that the sign of the imaginary
argument of the exponential (the gradient twist) is reversed. We can
think of this as a change of the sign of $q$, giving\begin{equation}
M_{\parallel}^{180}=0\label{eq:Ml_180_st}\end{equation}
and\begin{equation}
M_{\bot}^{180}=-i\, M_{0}e^{-b(\frac{TE}{2})\, D_{s}}e^{i\,2\,\pi\, q(\frac{TE}{2})\, s}.\label{eq:Mt_180_st}\end{equation}

\subsection{Second $\frac{TE}{2}$ Delay}

At the end of the second $\frac{TE}{2}$ delay, the phase acquired
during the first $\frac{TE}{2}$ delay due to inhomogeneity (or chemical
shift) will cancel. This is due to the change in the sense of the
helix due to any field (we have only included the Gradient explicitly)
by the $180^{\circ}$ Pulse. Since attenuation due to diffusion depends
on $q^{2}$, the attenuation continues to accumulate as during the
first $\frac{TE}{2}$ delay. Just before the second gradient pulse
$\delta2$ we have\begin{equation}
M_{\parallel}^{TE}=0\label{eq:Ml_TE_st}\end{equation}
and\begin{equation}
M_{\bot}^{TE-\delta}=-i\, M_{0}e^{-b(TE-\delta)\, D_{s}}e^{i\,2\,\pi\, q(\delta)\, s}.\label{eq:Mt_TE_st}\end{equation}

\subsection{Second Gradient Pulse $\delta_{2}$}

First we will make a few observations. We will want $q(\Delta+\delta_{2})=0$
when we acquire the FID, otherwise the transverse magnetization is
still twisted, and the signal is spoiled. This means that the area
of the first gradient should be equal and opposite to the area of
the second gradient or $\delta=\delta_{2}$. However, in figure \ref{cap:st_ps}
the two gradients have the same positive area. What we must remember
is the effect of the $180^{\circ}$ pulse. The effect of the pulse
is to reverse the imaginary ($\hat{y}$) component of $M_{\bot}$
which in our complex notation meant changing the sign of the $q$
accumulated before the pulse. Now the two positive gradients will
cancel since they are on opposite sides of the $180^{\circ}$ pulse.
We see this mathematically as

\begin{equation}
M_{\parallel}^{\delta_{2}}=0\label{eq:Ml_TE_st}\end{equation}
and\begin{equation}
M_{\bot}^{\delta_{2}}=-i\, M_{0}e^{-b(\Delta+\delta_{2})\, D_{s}}e^{i\,2\,\pi\, q(\delta)\, s}e^{-i\,2\,\pi\, q(\Delta+\delta_{2})\, s}=-i\, M_{0}e^{-b(\Delta+\delta_{2})\, D_{s}}.\label{eq:Mt_TE_st}\end{equation}

\subsection{Acquisition of FID}

\begin{table}
\begin{tabular}{c|c|c|c|}
&
$G_{s}(t)$&
$q(t)$&
$b(t)$\tabularnewline
\hline
$t<0$&
$0$&
$0$&
$0$\tabularnewline
$0\leq t<\delta$&
$G_{s}$&
$\frac{\gamma}{2\pi}G_{s}t$&
$\gamma^{2}G_{s}^{2}\frac{t^{3}}{3}$\tabularnewline
$\delta\leq t<\frac{TE}{2}$&
$0$&
$\frac{\gamma}{2\pi}G_{s}\delta$&
$\gamma^{2}G_{s}^{2}[\frac{\delta^{3}}{3}+\delta^{2}(t-\delta)]$\tabularnewline
$\frac{TE}{2}\leq t<\Delta$&
$0$&
-$\frac{\gamma}{2\pi}G_{s}\delta$&
$\gamma^{2}G_{s}^{2}[\frac{\delta^{3}}{3}+\delta^{2}(t-\delta)]$\tabularnewline
$\Delta\leq t<\Delta+\delta_{2}$&
$G_{s}$&
-$\frac{\gamma}{2\pi}G_{s}[\delta-(t-\Delta)]$&
$\gamma^{2}G_{s}^{2}[\frac{\delta^{3}}{3}-\delta^{3}+\delta^{2}\Delta+\frac{(t-\Delta-\delta)^{3}}{3}]$\tabularnewline
$t\geq\Delta+\delta_{2}$&
$0$&
$0$&
$\gamma^{2}G_{s}^{2}\delta^{2}(\Delta-\frac{\delta}{3})$\tabularnewline
\end{tabular}

\caption{$G_{s}(t)$, $q(t)$ and $b(t)$. $\delta_{1}=\delta_{2}=\delta$\label{cap:Gqb}}
\end{table}
Now we evaluate $q(t)$ and $b(t)$ given $G_{s}(t)$ at different
time points along the sequence. Table \ref{cap:Gqb} shows the results.
The $b-value$ at the spin echo time $TE=\Delta+\delta_{2}$ is then\begin{equation}
b(TE)=\gamma^{2}G_{s}^{2}\delta^{2}(\Delta-\frac{\delta}{3}),\label{eq:b_st}\end{equation}
the well known result for a ST sequence. The $b-value$ determines
the sensitivity of the sequence to diffusion. If the $b-value$ is
small or zero then the sequence is not sensitive to diffusion. If
the sequence has a significant $b-value$ it is called {}``diffusion
weighted\index{diffusion weighted}.''

\subsection{$T_{2}$ Relaxation}

We will show in the next section \ref{sec:Stimulated Echo} equation
\ref{eq:Mt_delta_ste} that $T_{2}$ relaxation and diffusion effects
are separable so that we can write the final FID signal for our ST
spin-echo as\begin{equation}
M_{\bot}^{\delta2}=-i\, M_{0}e^{-b(TE)\, D_{s}}e^{-\frac{TE}{T_{2}}}.\label{eq:Mt_T2_st}\end{equation}

\section{Stimulated Echo\label{sec:Stimulated Echo}}

\begin{figure}
\includegraphics[%
  width=1.0\columnwidth,
  keepaspectratio]{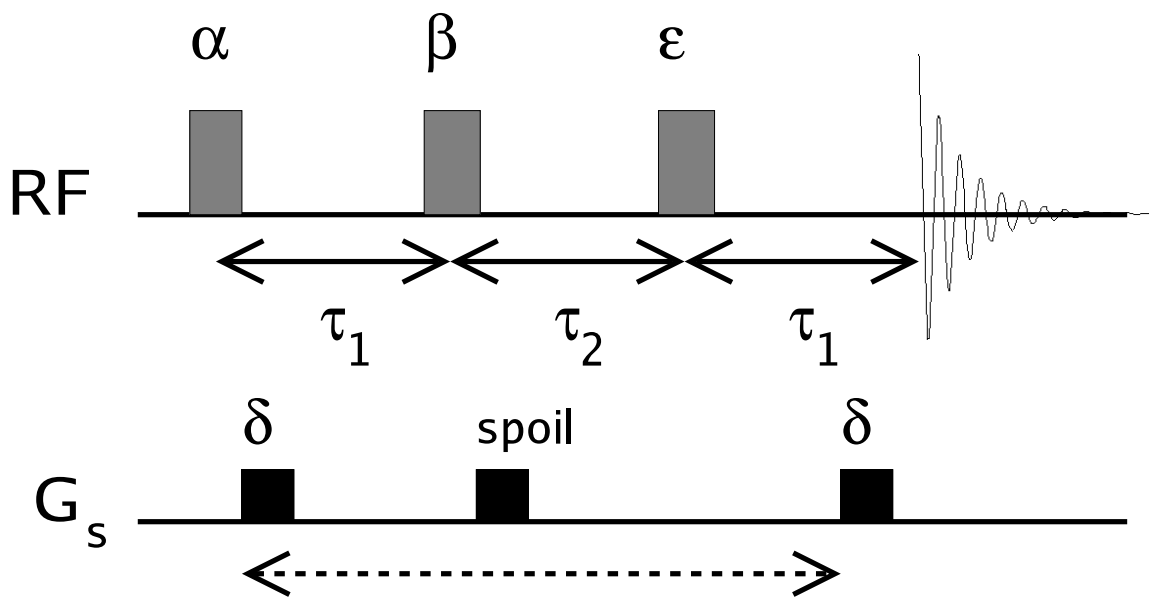}

\caption[Stimulated-Echo sequence.]{\label{cap:ste_ps}Stimulated-Echo sequence. Again, the large time
interval $\Delta$ includes one of the small gradient pulse duration
intervals $\delta$.}
\end{figure}
The stimulated echo was first reported by E. L. Hahn \cite[fig. 6g]{Hah50b}.
The sequence (see figure \ref{cap:ste_ps}) is similar to a spin echo
sequence. The major difference is that the $180^{\circ}$ pulse is
split into two pulses with a delay in between. The path that the magnetization
leading to the final FID takes is longitudinal between the last two
pulses. Also there is a 50\% loss of signal in all but the ideal perfectly
homogeneous no-gradient case, due to the process of rotating the twisted
transverse magnetization helix into the longitudinal direction. The
major advantage of the stimulated echo sequence is that during the
$\tau_{2}$ time period there is no $T_{2}$ relaxation\cite{Tan70,Tan72}.
This potentially allows a long $\Delta$ without losing as much signal
as in a spin echo if $T_{2}<T_{1}$.

We can analyze the stimulated echo in a similar manner to the spin
echo, using some of our previous results for the attenuation of transverse
magnetization due to diffusion. In addition we will see that we need
a similar expression for the attenuation due to diffusion of longitudinal
magnetization.

\subsection{Initial Magnetization}

We will introduce relaxation and arbitrary RF pulse angles and phases
into the analysis. Before the $\alpha$ pulse we have\begin{equation}
M_{\Vert}^{0}=M_{0}\label{eq:Ml_alpha_init_ste}\end{equation}
and\begin{equation}
M_{\bot}^{0}=0.\label{eq:Mt_alpha_init_ste}\end{equation}

\subsection{$\alpha$ Pulse}

After the $\alpha$ pulse we have\begin{equation}
M_{\Vert}^{\alpha}=cos(\alpha)\, M_{\Vert}^{0}-sin(\alpha)\, Im(e^{-i\,\phi_{\alpha}}M_{\bot}^{0})=cos(\alpha)\, M_{0}\label{eq:Ml_alpha_ste}\end{equation}
and\begin{equation}
M_{\bot}^{\alpha}=[Re(e^{i\,\phi_{\alpha}}M_{\bot}^{0})+i\, cos(\alpha)\, Im(e^{i\,\phi_{\alpha}}M_{\bot}^{0})]\, e^{-i\,\phi_{\alpha}}+i\, sin(\alpha)\, e^{i\,\phi_{\alpha}}M_{\Vert}^{0}=+i\, sin(\alpha)\, e^{i\,\phi_{\alpha}}M_{0}.\label{eq:Mt_alpha_ste}\end{equation}

This assumes that during the RF pulse all other terms (relaxation,
diffusion) in the Bloch equations are negligible and that the pulse
is on-resonance. $\phi_{\alpha}$ is the phase of the RF pulse, corresponding
to the orientation of the $B_{1}$ field in the rotating frame. $\alpha$
is the flip angle. $Re$ and $Im$ correspond to the real and imaginary
parts of their argument. The above are solutions to the Bloch equations
for the condition\begin{equation}
\frac{\partial\alpha}{\partial t}=\gamma B_{1},\label{eq:dalpha_dt_ste}\end{equation}
leading to the equation for $\alpha$\begin{equation}
\alpha=\gamma\int_{0}^{t}B_{1}dt.\label{eq:alpha_ste}\end{equation}

\subsection{1st Gradient $\delta$}

The effect of the gradient is to twist the transverse magnetization
along the direction $\hat{s}$. $q$ is $q(\delta)$ as defined in
equation \ref{eq:q_again}. We already know that we need to consider
diffusion during the period $\delta$ from section \ref{sec:Stejskal-Tanner-Sequence}.
We will also consider relaxation, but neglect off-resonance and inhomogeneity
effects. We will assume a solution of the form\begin{equation}
M_{\perp}=e^{-i\,2\pi\, q\, s}A_{\perp}(\tau_{1})\, R_{\perp}(\tau_{1})\, M_{\perp}^{\alpha},\label{eq:Mt_assumed_ste}\end{equation}
and substitute into the rotating frame Bloch equation\begin{equation}
\frac{dM_{\perp}}{dt}=\gamma\,[\vec{M}_{\perp}\times\vec{G}_{s}s]_{\perp}-\frac{M_{\perp}}{T_{2}}+\nabla\cdot\mathbf{D}\nabla M_{\perp},\label{eq:Mt_eq_ste}\end{equation}
where $A$ is the attenuation due to diffusion and $R$ due to relaxation.
We can expand out the cross product (using equation \ref{eq:Bloch_eq_expand_tran})
and since the spatial variation of $M_{\perp}$ is only along $\hat{s}$
we get\begin{equation}
\frac{dM_{\perp}}{dt}=-i\,\gamma\, M_{\perp}G_{s}s-\frac{M_{\perp}}{T_{2}}+D_{s}\frac{\partial^{2}M_{\perp}}{\partial s^{2}}.\label{eq:Mt_eq_expand_ste}\end{equation}
Substituting \ref{eq:Mt_assumed_ste} into \ref{eq:Mt_eq_expand_ste}
we get constraint equations for $q$, $A_{\perp}$and $R_{\perp}$\begin{equation}
\frac{\partial q}{\partial t}=\frac{\gamma}{2\pi}G_{s},\label{eq:dqdt_ste}\end{equation}
 \begin{equation}
\frac{\partial R_{\perp}}{\partial t}=-\frac{R_{\perp}}{T_{2}},\label{eq:dRdt_ste}\end{equation}
and

\begin{equation}
\frac{\partial A_{\perp}}{\partial t}=-4\,\pi^{2}q^{2}A_{\perp},\label{eq:dAdt_ste}\end{equation}

These have the corresponding solutions (with the additional constraint
that all go to $1$ at $t=0$)

\begin{equation}
q=\frac{\gamma}{2\pi}\int_{0}^{t}G_{s}dt',\label{eq:q_ste}\end{equation}

\begin{equation}
R_{\perp}=e^{-\frac{t}{T_{2}}},\label{eq:R_ste}\end{equation}
and

\begin{equation}
A_{\perp}=e^{-b(t)\, D_{s}}\label{eq:A_ste}\end{equation}

with\begin{equation}
b(t)\equiv4\,\pi^{2}\int_{0}^{t}q^{2}dt'.\label{eq:b_ste}\end{equation}

All of this is consistent with the results in section \ref{sec:Stejskal-Tanner-Sequence}
except that we have allowed the excitation pulse to have arbitrary
rotation angle and phase and added $T_{2}$ relaxation. The solution
is

\begin{equation}
M_{\bot}^{\delta}=e^{-i\,2\pi\, q\, s}A_{\perp}(\delta)\, R_{\perp}(\delta)\, M_{\bot}^{\alpha}=i\, sin(\alpha)\, e^{i\,\phi_{\alpha}}e^{-i\,2\pi\, q\, s}e^{-b(\delta)\, D_{s}}e^{-\frac{\delta}{T_{2}}}M_{0}.\label{eq:Mt_delta_ste}\end{equation}
The longitudinal magnetization is not affected by the gradient, and
will only experience $T_{1}$ relaxation. The result is

\begin{equation}
M_{\Vert}^{\delta}=M_{0}-(M_{0}-M_{\Vert}^{\alpha})\, e^{-\frac{\delta}{T_{1}}}=M_{0}-[M_{0}-cos(\alpha)\, M_{0}]\, e^{-\frac{\delta}{T_{1}}},\label{eq:Ml_delta_ste}\end{equation}
which is a solution to the equation

\begin{equation}
\frac{dM_{\parallel}}{dt}=\frac{(M_{0}-M_{\parallel})}{T_{1}}.\label{eq:Ml_eq_ste}\end{equation}
The longitudinal magnetization has no spatial variation so there will
be no diffusional effects at this point ($\nabla\cdot\mathbf{D}\nabla M_{\parallel}=0$).

\subsection{$\tau_{1}$ Delay}

We now consider the delay $\tau_{1}$, which we will make inclusive
of the delay $\delta$. We will again neglect off-resonance effects,
and since there are no gradient or RF pulses we will be left with
relaxation and diffusion. This is the same situation as during the
$\delta$ delay, except that $q$ is constant after $\delta$. We
can re-use the results from above to get

\begin{equation}
M_{\bot}^{\tau_{1}}=e^{-i\,2\pi\, q\, s}A_{\perp}(\tau_{1})\, R_{\perp}(\tau_{1})\, M_{\bot}^{\alpha}=i\, sin(\alpha)\, e^{i\,\phi_{\alpha}}e^{-i\,2\pi\, q\, s}e^{-b(\tau_{1})\, D_{s}}e^{-\frac{\tau_{1}}{T_{2}}}M_{0}\label{eq:Mt_tau1_ste}\end{equation}
and

\begin{equation}
M_{\Vert}^{\tau_{1}}=M_{0}-(M_{0}-M_{\Vert}^{\alpha})\, e^{-\frac{\tau_{1}}{T_{1}}}=M_{0}-[M_{0}-cos(\alpha)\, M_{0}]\, e^{-\frac{\tau_{1}}{T_{1}}}.\label{eq:Ml_tau1_ste}\end{equation}

\subsection{$\beta$ Pulse}

We handle the $\beta$ RF pulse similarly to $\alpha$, assuming that
the pulse is short enough so that there is no relaxation or diffusional
attenuation. The difference is that now we have non-zero transverse
magnetization which will be rotated into the longitudinal direction.
We will see that this longitudinal magnetization has spatially varying
amplitude (and will be subject to diffusional attenuation during subsequent
delays).

We will assume that the transverse magnetization after the $\beta$
pulse is immediately spoiled by the gradient, we now have\[
M_{\Vert}^{\beta}=cos(\beta)\, M_{\Vert}^{\tau_{1}}-sin(\beta)\, Im(e^{-i\,\phi_{\beta}}M_{\bot}^{\tau_{1}})\]
\begin{equation}
=cos(\beta)\,\{ M_{0}-[M_{0}-cos(\alpha)\, M_{0}]\, e^{-\frac{\tau_{1}}{T_{1}}}\}-sin(\beta)\, Im(e^{-i\,\phi_{\beta}}i\, sin(\alpha)\, e^{i\,\phi_{\alpha}}e^{-i\,2\pi\, q\, s}e^{-b(\tau_{1})\, D_{s}}e^{-\frac{\tau_{1}}{T_{2}}}M_{0})\label{eq:Ml_beta_ste}\end{equation}
and\begin{equation}
M_{\bot}^{\beta}=0.\label{eq:Mt_beta_ste}\end{equation}

\subsection{$\tau_{2}$ Delay\label{sub:ste_tau_2_Delay}}

During the $\tau_{2}$ delay the longitudinal component obeys the
Bloch equation\begin{equation}
\frac{dM_{\parallel}}{dt}=\frac{(M_{0}-M_{\parallel})}{T_{1}}+D_{s}\frac{\partial^{2}M_{\parallel}}{\partial s^{2}},\label{eq:Ml_tau2_eq_ste}\end{equation}
where we have made the substitution $\nabla\cdot\mathbf{D}\nabla M_{\parallel}=D_{s}\frac{\partial^{2}M_{\parallel}}{\partial s^{2}}$
since all spatial variation of $M_{\parallel}$ is along $\hat{s}$.
We can further break $M_{\parallel}$ into a spatially constant $M_{\parallel cnst}$
part and a spatially varying $M_{\parallel}(s)$ part. After substituting
$M_{\parallel}=M_{\parallel cnst}+M_{\parallel}(s)$ into equation
\ref{eq:Ml_tau2_eq_ste} we have\begin{equation}
\frac{dM_{\parallel cnst}}{dt}=\frac{(M_{0}-M_{\parallel cnst})}{T_{1}}\label{eq:Mlcnst_tau2_eq_ste}\end{equation}
and

\begin{equation}
\frac{dM_{\parallel}(s)}{dt}=-\frac{M_{\parallel}(s)}{T_{1}}+D_{s}\frac{\partial^{2}M_{\parallel}(s)}{\partial s^{2}}.\label{eq:Mls_tau2_eq_ste}\end{equation}
 We already know the solution to equation \ref{eq:Mlcnst_tau2_eq_ste}:
it is just $T_{1}$ relaxation. As it turns out we also know the solution
to \ref{eq:Mls_tau2_eq_ste}. It has exactly the same form as equation
\ref{eq:Mt_eq_expand_ste} but with no gradient (which means no change
in $q$) and with $T_{2}$ replaced by $T_{1}$. The solutions are
then\begin{equation}
M_{\parallel cnst}=M_{0}-(M_{0}-M_{\parallel cnst,initial})e^{-\frac{\tau_{2}}{T_{1}}}\label{eq:Mlcnst_tau2_ste}\end{equation}
and\begin{equation}
M_{\parallel}(s)=e^{-b_{\parallel}(\tau_{2})\, D_{s}}e^{-\frac{\tau_{2}}{T_{1}}}M_{\parallel initial}(s),\label{eq:Mls_tau2_ste}\end{equation}
where $b_{\parallel}(t)$ has the same form as equation \ref{eq:b_ste}
but refers to the spatial variation $q$ of the longitudinal magnetization
and the time interval $\tau_{2}$ only. Making the substitutions\begin{equation}
M_{\parallel cnst,initial}=cos(\beta)\,\{ M_{0}-[M_{0}-cos(\alpha)\, M_{0}]\, e^{-\frac{\tau_{1}}{T_{1}}}\}\label{eq:Ml_cnst_initial_ste}\end{equation}
and\begin{equation}
M_{\parallel initial}(s)=-sin(\alpha)\, sin(\beta)\, cos[\phi_{\alpha}-\phi_{\beta}-2\pi q(\delta)\, s]\, e^{-b(\tau_{1})\, D_{s}}e^{-\frac{\tau_{1}}{T_{2}}}M_{0}\label{eq:Mls_initital_ste}\end{equation}

we get the rather complicated expression\begin{multline}
M_{\parallel}^{\tau_{2}}=M_{0}-\{ M_{0}-cos(\beta)\,\{ M_{0}-[M_{0}-cos(\alpha)\, M_{0}]\, e^{-\frac{\tau_{1}}{T_{1}}}\}\} e^{-\frac{\tau_{2}}{T_{1}}}\\
-sin(\alpha)\, sin(\beta)\, cos[\phi_{\alpha}-\phi_{\beta}-2\pi q(\delta)\, s]\, e^{-[b(\tau_{1})+b_{\parallel}(\tau_{2})]\, D_{s}}e^{-\frac{\tau_{1}}{T_{2}}-\frac{\tau_{2}}{T_{1}}}M_{0}\label{eq:Ml_tau2_solution_ste}\end{multline}

and the relatively simple\begin{equation}
M_{\perp}^{\tau_{2}}=0.\label{eq:Mt_tau2_solution_ste}\end{equation}

\subsection{$\epsilon$ Pulse}

The $\epsilon$ pulse is the final RF pulse. Before the $\epsilon$
pulse we have no transverse magnetization. Our observable signal must
then come from magnetization that was longitudinal at the end of the
$\tau_{2}$ delay. The effect of the $\epsilon$ pulse is to give\begin{equation}
M_{\Vert}^{\epsilon}=cos(\epsilon)\, M_{\Vert}^{\tau_{2}}\label{eq:Ml_epsilon_ste}\end{equation}
and\begin{equation}
M_{\bot}^{\epsilon}=i\, sin(\epsilon)\, e^{i\,\phi_{\epsilon}}M_{\Vert}^{\tau_{2}}.\label{eq:Mt_epsilon_ste}\end{equation}

After substitution of $M_{\Vert}^{\tau_{2}}$ we have

\begin{multline}
M_{\perp}^{\epsilon}=i\, sin(\epsilon)\, e^{i\,\phi_{\epsilon}}\{ M_{0}-\{ M_{0}-cos(\beta)\,\{ M_{0}-[M_{0}-cos(\alpha)\, M_{0}]\, e^{-\frac{\tau_{1}}{T_{1}}}\}\}\, e^{-\frac{\tau_{2}}{T_{1}}}\\
-sin(\alpha)\, sin(\beta)\, cos[\phi_{\alpha}-\phi_{\beta}-2\pi q(\delta)\, s]\, e^{-[b(\tau_{1})+b_{\parallel}(\tau_{2})]\, D_{s}}e^{-\frac{\tau_{1}}{T_{2}}-\frac{\tau_{2}}{T_{1}}}M_{0}\}.\label{eq:Mt_epsilon_expand1_ste}\end{multline}
We can ignore the longitudinal component at this point as it will
not contribute to the final signal. We would need to consider it if
we were interested in the steady state magnetization for partial recovery.

\subsection{Final Gradient and Delay}

We now have enough information to know how the transverse component
will behave without further derivation. We note that the final delay
period is the same as the first period, $\tau_{1}$. Any phase acquired
due to inhomogeneity during the first delay will be re-phased during
the second delay. There is no phase acquired during the center delay,
since the magnetization leading to the final observable signal interest
was longitudinal.

During the final delay, which we take to include the last gradient
$\delta$ at the end, we have\begin{equation}
M_{\perp{}}^{\tau_{1}last}=e^{-i\,2\pi\, q(\delta)\, s}e^{-b(\tau_{1}last)\, D_{s}}e^{-\frac{\tau_{1}}{T_{2}}}M_{\perp}^{\epsilon}.\label{eq:Mt_tau1_last_ste}\end{equation}

Substituting $cos(a)=\frac{e^{i\, a}+e^{-i\, a}}{2}$ for the $cos[\phi_{\alpha}-\phi_{\beta}-2\pi q(\delta)\, s]$
term in $M_{\perp}^{\epsilon}$ we get{\small \begin{multline}
M_{\perp}^{\tau_{1}last}=\\
i\, c_{\epsilon}\{ M_{0}-\{ M_{0}-cos(\beta)\,\{ M_{0}-[M_{0}-cos(\alpha)\, M_{0}]\, e^{-\frac{\tau_{1}}{T_{1}}}\}\}\} e^{-\frac{\tau_{2}}{T_{1}}}e^{-i\,2\pi\, q(\delta)\, s}e^{-b(\tau_{1}last)\, D_{s}}e^{-\frac{\tau_{1}}{T_{2}}}\\
-i\, c_{\epsilon}sin(\alpha)\, sin(\beta)\,\frac{e^{i\,[\phi_{\alpha}-\phi_{\beta}-2\pi q(\delta)\, s]}}{2}\, e^{-[b(\tau_{1})+b_{\parallel}(\tau_{2})]\, D_{s}}e^{-\frac{\tau_{1}}{T_{2}}-\frac{\tau_{2}}{T_{1}}}M_{0}\, e^{-i\,2\pi\, q(\delta)\, s}e^{-b(\tau_{1}last)\, D_{s}}e^{-\frac{\tau_{1}}{T_{2}}}\label{eq:Mt_epsilon_expand_ste}\\
-i\, c_{\epsilon}sin(\alpha)\, sin(\beta)\,\frac{e^{-i\,[\phi_{\alpha}-\phi_{\beta}-2\pi q(\delta)\, s]}}{2}\, e^{-[b(\tau_{1})+b_{\parallel}(\tau_{2})]\, D_{s}}e^{-\frac{\tau_{1}}{T_{2}}-\frac{\tau_{2}}{T_{1}}}M_{0}\, e^{-i\,2\pi\, q(\delta)\, s}e^{-b(\tau_{1}last)\, D_{s}}e^{-\frac{\tau_{1}}{T_{2}}},\end{multline}
}{\small \par}

where\begin{equation}
c_{\epsilon}\equiv sin(\epsilon)\, e^{i\,\phi_{\epsilon}}.\label{eq:c_eps}\end{equation}
We notice that only in the last term do the gradient twists cancel.
We now assume that $q$ is large enough so that any signal that is
twisted is spoiled and we end up with\[
M_{\perp}^{ste}=-\frac{i}{2}\, sin(\alpha)\, sin(\beta)\, sin(\epsilon)\, e^{-i\,(\phi_{\alpha}-\phi_{\beta}-\phi_{\epsilon})}e^{-b\, D_{s}}e^{-2\frac{\tau_{1}}{T_{2}}-\frac{\tau_{2}}{T_{1}}}M_{0},\]
where we have combined all the $b-values$ into\[
b=\gamma^{2}G^{2}\delta^{2}(\Delta-\frac{\delta}{3}).\]
The accumulated $b-value$ is identical to the ST sequence since during
the $\tau_{2}$ period the longitudinal magnetization undergoes the
same attenuation due to diffusion.

The pre-factor of $\frac{1}{2}$, corresponding to a \%50 loss of
signal not attributable to relaxation or diffusion, is due to the
gradient not being able to simultaneously re-phase the counter-twisted
components embodied in the 2nd term of equation \ref{eq:Mt_epsilon_expand_ste}.

\part{Distant Dipolar Field Effects\label{par:DDF}}

\chapter{DISTANT DIPOLAR FIELD\index{distant dipolar field}\index{DDF}}

\section{Introduction}

In Part \ref{par:The-Basics} we introduced the various physical effects
considered relevant in liquid-state NMR, culminating in the Bloch
equations (see chapter \ref{cha:Bloch-Equations}) for describing
the classical macroscopic behavior of an ensemble of spins. Up to
this point we have neglected the explicit contribution of the field
from nuclear magnetization originating in the sample on other parts
of the sample. This gives rise to two new effects.

One is called radiation damping, and is not directly felt by the sample,
but requires a receiver coil to {}``feed back'' an RF field into
the sample. For the most part we will discuss radiation dampening
as a nuisance to be avoided (Chapter \ref{cha:RADIATION-DAMPING}).

Another is the distant dipolar field\index{distant dipolar field}
or DDF\index{DDF}. The term DDF has actually been modified from {}``dipolar
demagnetization field\index{dipolar demagnetizing field}'' \cite[p. 49-61]{Cul72}
and both are used in the literature, the former being an innovation
of NMR researchers investigating dipolar field-induced echoes in liquids,
or biological samples. It had been thought in liquids that static%
\footnote{as opposed to dynamic dipolar fields which contribute to relaxation%
} dipolar field effects, which are the source of many useful and confounding
effects in spectroscopy and imaging of solids \cite{WA98}, could
largely be neglected due to the averaging effects of diffusion.

The first sign that this was not true came from low-temperature physics
experiments using solid $^{3}He$ in the late 1970's and early '80's.
Deville et al. observed unexpected {}``multiple spin echoes'' in
low temperature solid%
\footnote{although solid, $^{3}He$ has significant {}``exchange narrowing'',
analogous to motional narrowing in liquids%
} $^{3}He$, \cite{DBD79,DE84,DE85}. One reason for this is that even
at low field the magnetization is very large due to the extremely
low $\mu K$ temperatures.

Observation of multiple spin echoes in water at room temperature came
in the early 1990s when Bowtell\index{Bowtell, R.}, Korber\index{Korber, H.},
and Warren\index{Warren, W.}, \cite{BBG90,KDE91,WHM+92,QH93} all
reported echoes or effects they attributed to sample nuclear magnetization,
coupled by the dipolar field. At first these claims were sometimes
disputed and attributed to other sources, especially in the case of
the Warren and collaborators 2d spectroscopy experiments \cite{ADL92,WHM+92}.

There has also been a lively discussion of the necessity to treat
the DDF classically or quantum mechanically \cite{Jee00} as intermolecular
multiple quantum coherence\index{intermolecular multiple quantum coherence}
(iMQC\index{iMQC}). In general it has been shown that the classical
description is adequate under most conditions, and in fact has lead
to the quantification of many effects, such as diffusion weighting
\cite{AKK01,CG04b}, that have so far been intractable in the quantum
picture.

Interest has grown steadily over the intervening years due to novel
application possibilities. One of the first was the realization that
signal weighting (contrast) was sensitive to so-called {}``meso-scale''\index{mesoscale}
structure \cite{RB96,SCMA+01,BRW02}. Meso-scale is the term used
to distinguish the scale intermediate between micro-scale\index{microscale}
processes, such as diffusion, $T_{1}$, and $T_{2}$, and macro-scale\index{macroscale},
such as a resolvable imaging voxel. In other words, DDF based sequences
could probe sub-voxel structure with scale larger than the diffusion
distance. This novel imaging contrast mechanism has continued to be
pursued \cite{WAM+98,RAA+00,SC02,CTB+03b,MB04,BW05}.

The Holy Grail of in-vivo magnetic resonance spectroscopy\index{in-vivo magnetic resonance spectroscopy}
(MRS\index{MRS}) is the ability to localize and quantify metabolite
peaks at high resolution and high signal-to-noise ratio. Several DDF
sequences offer the possibility of obtaining higher resolution spectra
than obtainable with conventional NMR\index{NMR} sequences. The first
implemented was HOMOGENIZED\index{HOMOGENIZED} \cite{VLW96,YYL00}
which stands for {}``homogeneity enhancement by intermolecular zero-quantum
detection.'' Its usefulness has been demonstrated already for non-localized
spectroscopy of live animals and excised tissue \cite{FPH03}. HOMOGENIZED
continues to be an active research area with improved understanding
of relaxation and diffusion effects and water suppression being recently
reported \cite{CG04,CG04b,BF04b,CHC+04}. Other variations of HOMOGENIZED
have been proposed as well \cite{ZCC+03,BF04b}.

\section{Field of a Dipole }

\begin{figure}
\begin{center}

\includegraphics[%
  width=5in,
  height=5in,
  angle=270]{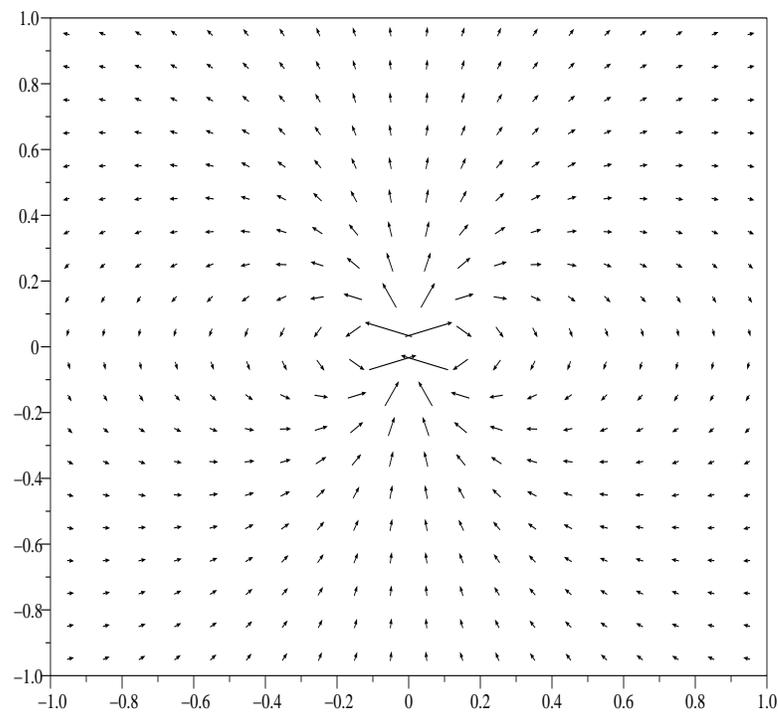}\end{center}

\caption[Field due to single point dipole $\vec{\mu}=\mu\,\hat{z}$.]{\label{cap:Field-due-to-Dipole}Field due to single point dipole
$\vec{\mu}=\mu\,\hat{z}$. The $\hat{x}$ axis is along the horizontal,
$\hat{z}$ along the vertical. The field is symmetric about the $\hat{z}$
axis. The plotted $r$ dependence has been changed from $\frac{1}{r^{3}}$
to $\frac{1}{r}$ in order to aid visualization. }
\end{figure}

The field due to a single magnetic dipole\index{magnetic dipole}
is\begin{equation}
\vec{B}_{dip}=\frac{\mu_{0}}{4\pi}\frac{3\,(\vec{\mu}\cdot\hat{r})\,\hat{r}-\vec{\mu}}{r^{3}}.\label{eq:B_dipole_SI}\end{equation}
and is plotted in figure \ref{cap:Field-due-to-Dipole}.

\begin{figure}
\includegraphics[%
  width=0.45\columnwidth,
  keepaspectratio]{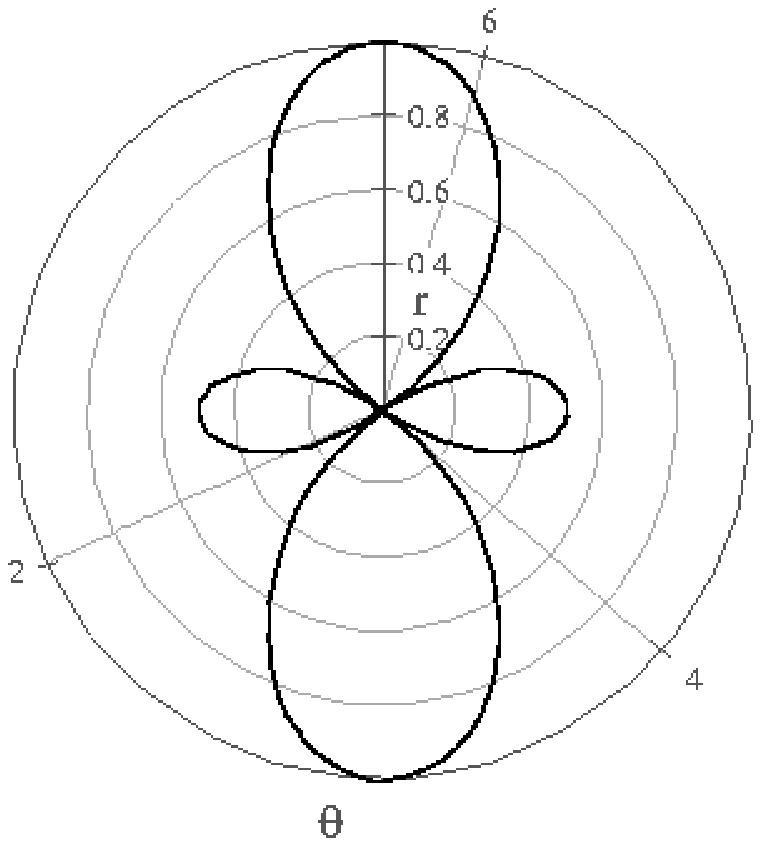}\hfill{}\includegraphics[%
  width=0.50\columnwidth,
  keepaspectratio]{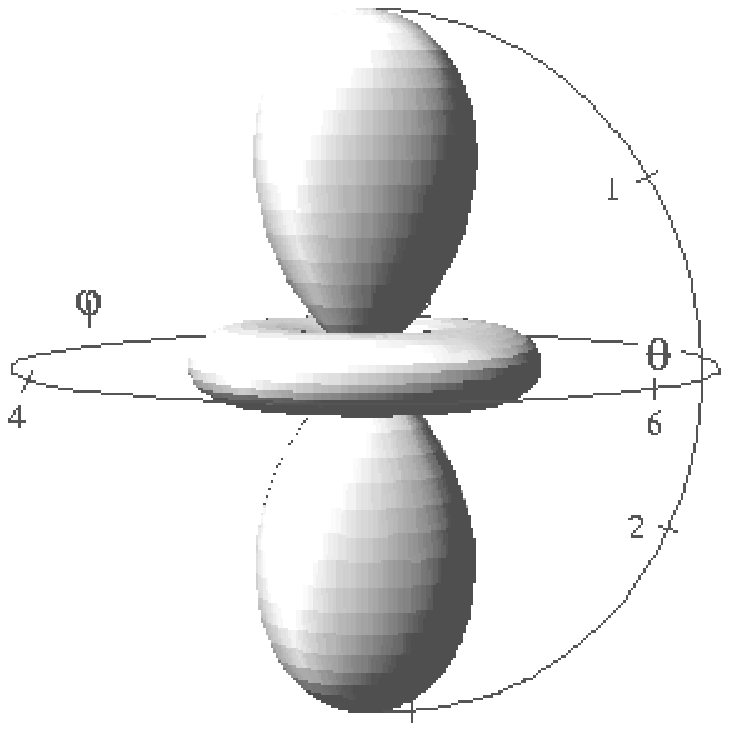}

\caption[Angular dependence, $\Lambda(\vec{r})$, of the secular field of a
dipole.]{\label{cap:Angular-dependence-Lambda}Angular dependence, $\Lambda(\vec{r})$,
of the secular field of a dipole. Note that the polarity of $\Lambda$
is positive in the upper and lower lobes and negative in the side
lobes (toroidal lobe in 3d).}
\end{figure}

In most circumstances of interest, the secular component\index{secular component}
(see \ref{sec:Secular-Component-Derived}) of $\vec{B}_{dip}$ is
the only component that will contribute in the presence of a much
stronger externally applied field $\vec{B}_{0}=B_{0}\hat{z}$. The
secular component of the field is\begin{equation}
\vec{B}_{secular}=\frac{\mu_{0}}{4\pi}\frac{1}{r^{3}}[\frac{3\, cos^{2}(\theta)-1}{2}]\,(3\,\mu_{z}\hat{z}-\vec{\mu}).\label{eq:B_dipole_secular_defined}\end{equation}
The range of validity of the approximation (that the non-secular components
are negligible) can be estimated from the condition,\begin{equation}
B_{0}\gg\frac{\mu_{0}}{4\pi}\frac{1}{r^{3}}|\vec{\mu}|.\label{eq:Secular_validity}\end{equation}
In a liquid, diffusion will determine the minimum $r$ that need be
considered. In a solid it is lattice parameters and exchange. The
angular dependence of the secular field deserves some attention. First
of all we notice that it is the Legendre polynomial\begin{equation}
P_{2}[cos(\theta)]=\frac{3\, cos^{2}(\theta)-1}{2}.\label{eq:P2_cos_theta}\end{equation}
In the DDF/iMQC literature the angular dependence is often defined
as%
\footnote{The origin of the definition is unknown to this author, but it may
be that the $\Lambda$ refers to Legendre.%
}\begin{equation}
\Lambda(\vec{r})\equiv\frac{3\,(\hat{r}\cdot\hat{z})^{2}-1}{2}=\frac{3\, cos^{2}(\theta)-1}{2}.\label{eq:Lambda_defined_first}\end{equation}
The zeros of $\Lambda$ are at the so called {}``magic angle''\index{magic angle}\begin{equation}
cos(\theta_{magic})=_{-}^{+}\sqrt{\frac{1}{3}}\label{eq:cos_theta_magic}\end{equation}
or\begin{equation}
\theta_{magic}=_{-}^{+}54.73561^{\circ}.\label{eq:theta_magic}\end{equation}
At this angle the secular field of a dipole disappears, regardless
of the orientation or magnitude of $\vec{\mu}$. We plot $\Lambda$
in figure \ref{cap:Angular-dependence-Lambda}.

\section{Secular Dipolar Demagnetizing Field\index{dipolar demagnetizing field}\label{sec:Secular-Dipolar-Field}}

The secular dipolar demagnetizing field from a distribution of magnetization
takes the form \cite{DBD79,RB97b,LSB+04}\begin{equation}
\vec{B}_{d}(\vec{r})=-\frac{\mu_{0}}{4\pi}\int_{\infty}d^{3}r'\,\frac{\Lambda(\vec{r}-\vec{r}')}{|\vec{r}-\vec{r}'|^{3}}\,[3\, M_{z}(\vec{r}')\,\hat{z}-\vec{M}(\vec{r}')],\label{eq:Bd_integral}\end{equation}
with\begin{equation}
\Lambda(\vec{r})\equiv\frac{3\,(\hat{r}\cdot\hat{z})^{2}-1}{2}.\label{eq:Lambda_defined_again}\end{equation}
This is the field that a spin or small ensemble of spins {}``feels''
due to all other spins (magnetization) in the sample.

\ref{eq:Bd_integral} is in fact the convolution\begin{equation}
\vec{B}_{d}(\vec{r})=-\frac{\mu_{0}}{4\pi}\frac{\Lambda(\vec{r})}{r^{3}}*[3\, M_{z}(\vec{r})\,\hat{z}-\vec{M}(\vec{r})].\label{eq:Bd_convolution}\end{equation}

We then take the three-dimensional Fourier transform of $\vec{B}_{d}(\vec{r})$\begin{equation}
\mathcal{F}_{3}\{\vec{B}_{d}(\vec{r})\}\equiv\vec{B}_{d}(\vec{\rho})=\int_{\infty}d^{3}r\, e^{-i\,2\pi\,\vec{\rho}\cdot\vec{r}}\vec{B}_{d}(\vec{r}),\label{eq:Bd_transform}\end{equation}
which by the convolution theorem \cite[section 3.3.6, p. 124-28]{BM03}
is\begin{equation}
\vec{B}_{d}(\vec{\rho})=-\frac{\mu_{0}}{4\pi}\mathcal{F}_{3}\{\frac{\Lambda(\vec{r})}{r^{3}}\}\,\mathcal{F}_{3}\{3\, M_{z}(\vec{r})\,\hat{z}-\vec{M}(\vec{r})\}.\label{eq:Bd_transform_product}\end{equation}
For now we will not worry about the explicit form of $\vec{M}(\vec{r})$
and use the general form\begin{equation}
\mathcal{F}_{3}\{3\, M_{z}(\vec{r})\,\hat{z}-\vec{M}(\vec{r})]\}=3\, M_{z}(\vec{\rho})\,\hat{z}-\vec{M}(\vec{\rho}).\label{eq:Bd_rho}\end{equation}
The transform of the convolution kernel $\frac{\Lambda(\vec{r})}{r^{3}}$
from reference \cite{DBD79} and Appendix \ref{sec:Fourier-Transform-of-kernel}
is\begin{equation}
\mathcal{F}_{3}\{\frac{\Lambda(\vec{r})}{r^{3}}\}=-\frac{4\pi}{3}\Lambda(\vec{\rho}).\label{eq:F3_lambda_over_r3}\end{equation}
and the result for the transform of\[
\vec{B}_{d}(\vec{\rho})=\frac{1}{3}\Lambda(\vec{\rho})\,[3\, M_{z}(\vec{\rho})\,\hat{z}-\vec{M}(\vec{\rho})].\]

\section{{}``local'' form\index{local form}\label{sec:local-form}}

Deville\index{Deville} et al. \cite[section B]{DBD79} noted that
if the sample magnetization is periodic the contribution of the sample
magnetization to the dipolar field becomes localized. 

\begin{figure}
\includegraphics[%
  width=0.45\columnwidth,
  keepaspectratio]{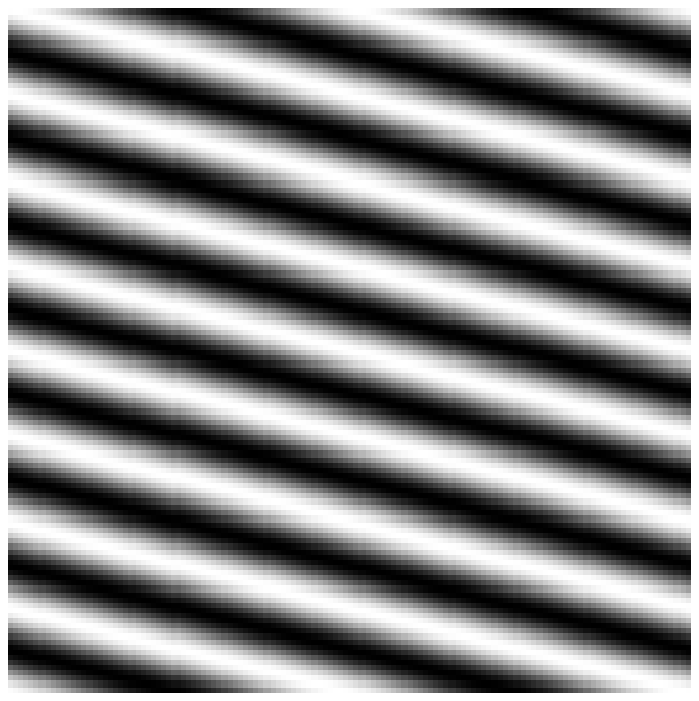}\hfill{}\includegraphics[%
  width=0.45\columnwidth,
  keepaspectratio]{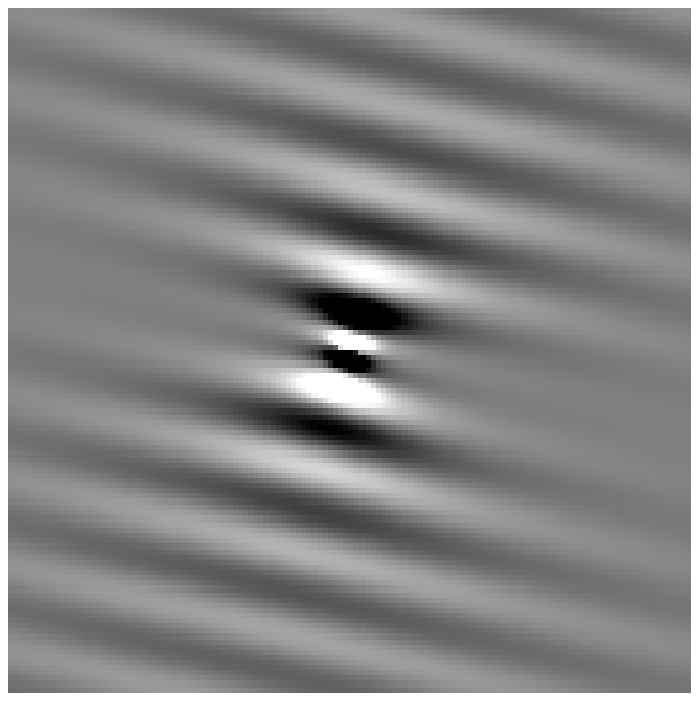}

\caption[Modulation and Localization.]{\label{cap:Localization_of_Field}Modulated Magnetization. \hfill{}Weighted
Contribution to DDF.}

Magnetization is shown on the left. The weighted contribution of the
magnetization to the dipolar field at the center is shown on the right.
Far from the center there is a weight of zero. While the figure is
shown in 2d, the localization applies in 3d as well, the weighting
being symmetric about the gradient axis through the center. The $\vec{B}_{d}$
experienced at the center of the plotted region will be the integral
over the 3d volume.
\end{figure}

One can visualize this as follows%
\footnote{The visualization and first order derivation is not how Deville et
al. justified the localization, but in the author's opinion expands
on and clarifies the phenomenon.%
} (see figure \ref{cap:Localization_of_Field}). When one looks far
from the point of interest where one is computing the field, there
are regions of positive and negative magnetization, at approximately
the same distance and angle. This leads to an {}``effective magnetization''
which is the spatial average. The effective magnetization is zero,
leading to a contribution to the dipolar field of zero. Close to the
point of interest the differing regions of magnetization have significantly
different distance or angle, and do not cancel. This is a {}``Sphere
of Lorentz''\index{sphere of Lorentz} argument, similar to the line
of reasoning presented in \cite{CXB+90,DHK03}. This line of reasoning
applies to the dipolar field from both longitudinal and transverse
magnetization. In the transverse case, the magnetization is complex-valued
and we can visualize the real and imaginary component separately.

Mathematically we can state the localization as follows. Consider
two regions of the sample, separated by half the modulation period.
The two location vectors are $\vec{r}_{1}$and $\vec{r}_{2}$. First,
assuming they have the same equilibrium magnetization and relaxation
properties, we can write their contribution to the dipolar field (at
$r$=0 for convenience) as\begin{equation}
\vec{b}_{d,\,1}=\frac{\mu_{0}}{4\pi}\delta v\frac{\Lambda(\vec{r}_{1})}{r_{1}^{3}}\,\vec{M}'(\vec{r}_{1}),\label{eq:bd1_def}\end{equation}
\begin{equation}
\vec{b}_{d,\,2}=\frac{\mu_{0}}{4\pi}\delta v\frac{\Lambda(\vec{r}_{2})}{r_{2}^{3}}\,\vec{M'}(\vec{r}_{2}),\label{eq:bd2_def}\end{equation}
with\begin{equation}
\vec{M'}(\vec{r})\equiv3\, M_{z}(\vec{r})\,\hat{z}-\vec{M}(\vec{r})\label{eq:M'_def}\end{equation}
and\begin{equation}
\vec{r}_{2}=\vec{r}_{1}+\frac{\vec{\delta r}}{2}\label{eq:r2_def}\end{equation}
where $\vec{\delta r}$ is the period of the modulation. If the magnetization
is smoothly varying compared to the scale of modulation we have\begin{equation}
\vec{M}(\vec{r}_{1})\approx-\vec{M}(\vec{r}_{2}).\label{eq:Mr1_approx}\end{equation}
The volume of the region under consideration, $\delta v$, is such
that $\delta v\leq(\frac{\delta r}{2})^{3}$.

We write \begin{equation}
\vec{\delta b}=\frac{\vec{b}_{d,\,2}+\vec{b}_{d,\,1}}{2}\label{eq:delta_b}\end{equation}
keeping in mind that we can consider all such pairs of magnetized
regions in the sample once (i.e. don't double count).

Substituting into \ref{eq:delta_b} and keeping all first-order terms
in $\vec{\delta r}$ gives us

\[
\vec{\delta b}\approx\frac{\mu_{0}}{4\pi}\delta v\,\{\frac{1}{r_{1}^{3}}[1-\frac{3}{2}(\hat{r}_{1}\cdot\hat{\delta r})\,\frac{\delta r}{r_{1}}]-\frac{1}{r_{1}^{3}}\}\,\Lambda(\vec{r}_{1})\,\vec{M}'(\vec{r}_{1}),\]
 which after simplification gives\[
\vec{\delta b}\approx\frac{\mu_{0}}{4\pi}\delta v\,\frac{1}{r_{1}^{3}}[-\frac{3}{2}(\hat{r}_{1}\cdot\hat{\delta r})\,\frac{\delta r}{r_{1}}]\,\Lambda(\vec{r}_{1})\,\vec{M}'(\vec{r}_{1}).\]
When considering integration over the entire sample we see that modulation
has introduced a weighting factor of \[
W=-\frac{3}{2}(\hat{r}\cdot\hat{\delta r})\,\frac{\delta r}{r}\]
when we consider magnetized regions of the sample in the pairwise
manner above.

At this point we can say that the dipolar field originates primarily
from magnetization within a radius of $r\sim\delta r$ . Magnetization
from outside that radius contributes less significantly. The weighting
further favors magnetization along the direction of the modulation,
and penalizes magnetization orthogonal to the modulation (see right
figure \ref{cap:Localization_of_Field}). 

Although not arrived at by this argument, $\delta r$ is also the
so-called {}``correlation distance''\index{correlation distance}
used in the DDF literature. The correlation distance is the distance
over which the DDF\index{DDF} $\vec{B}_{d}$ is assumed to act in
a structured sample. Contributions from farther than $\delta r$ are
assumed to be negligible.

\section{When does this break down?}

There are several conditions under which the localization effect of
modulating the magnetization will break down. The underlying cause
of the breakdown is that nearby regions of magnetization far from
the point of interest for the $B_{d}$ calculation no longer cancel.

At the edges of a sample of finite extent (in other words, all real
samples) there will be a volume whose paired volume needed for cancellation
lies outside the sample boundary, which is assumed to have zero magnetization.
This can result in magnetization far from the point of interest contributing
to $B_{d}$. One good assumption is that if the sample boundary is
far enough away (so that $r\gg\delta r$) from the point of interest
the contribution of this magnetization will be small.

Another case is when the underlying structure of the sample has variation
near the scale of the modulation. This is a violation of the {}``slowly
varying'' condition for $\vec{M}(\vec{r})$. This results in a failure
of the cancellation condition, potentially over large volumes of the
sample and not necessarily far from the point of interest. This effect
had been predicted \cite{RB97b}, and more recently observed \cite{LdS+04,TCB+04}
and dubbed {}``NMR Diffraction\index{nmr diffraction}.''

It is ironic that this sensitivity to underlying magnetization modulation
or structure relative to the applied modulation has also been proposed
as a contrast method for DDF weighted MRI \cite{BTC+03}, or for potential
quantization of bone density \cite{BCT+03}. The irony comes from
the fact that there is a desire to localize a contrast that has an
inherent non-locality associated with it. There has been some reporting
of the difficulties due to this \cite{CPL+02}, but it is still an
active topic of investigation.

\section{{}``point'' form\index{point form}}

We can gain further insight for calculation of $\vec{B}_{d}(\vec{r})$.
This was again first suggested by Deville\index{Deville} et al. \cite[eq. (9)]{DBD79}%
\footnote{Deville et al. developed this relationship for homogeneous magnetization,
and noted that the relationship is approximate for a sample of finite
extent, but it still holds for the less stringent condition of slowly
varying magnetization.%
}. The idea is that when $\vec{M}(\vec{r})$ is constant (or stretched
to the less stringent condition of changing slowly or is {}``slowly
varying\index{slowly varying}'' compared to $\hat{\delta r}$) we
can approximate the dipolar field $\vec{B}_{d}(\vec{r})$ as proportional
to $\vec{M}'(\vec{r})$. In other words, the spatial integration of
equation \ref{eq:Bd_integral} (or convolution of \ref{eq:Bd_convolution})
disappears, and we have (using equation \ref{eq:M'_def}) the proportional
relationship \begin{equation}
\vec{B}_{d}(\vec{r})\propto\Lambda(\vec{r})\,\vec{M}'(\vec{r}).\label{eq:Bd_proportional}\end{equation}

Deville et al. justify this as follows (in our notation). 

First we note that the Fourier transform of $\vec{B}_{d}(\vec{r})$
has the form\begin{equation}
\mathcal{F}_{3}\{\vec{B}_{d}(\vec{r})\}\equiv\vec{B}_{d}(\vec{\rho})=\frac{\mu_{0}}{3}\Lambda(\vec{\rho})\,[3\, M_{z}(\vec{\rho})\,\hat{z}-\vec{M}(\vec{\rho})],\label{eq:Bd_transform_2}\end{equation}
and note that the convolution operation leads to a product in the
transform space. We can write as\begin{equation}
\vec{B}_{d}(\vec{\rho})=\mathbf{C_{\hat{\rho}}}\vec{M}(\vec{\rho}),\label{eq:Deville_tensor}\end{equation}
where $\mathbf{C}_{\hat{\rho}}$ is a 3x3 matrix or tensor that depends
only on the direction $\hat{\rho}$, not on the radius.

When the dominant variation in $\vec{M}(\vec{r})$ is one-dimensional,
with constant (or slowly varying) value orthogonal to direction $\hat{s}$,
we can define

\begin{equation}
\vec{M}(\vec{r})\approx\vec{M}(\hat{s}\cdot\vec{r})\equiv\vec{M}(s).\label{eq:M_of_s}\end{equation}

We perform the 3d Fourier transform of $\vec{M}(s)$,\begin{equation}
\mathcal{F}_{3}\{\vec{M}(s)\}=\vec{M}(\rho_{s})\,\delta_{2}(\rho_{s}),\label{eq:Bd_sub_M_of_S_transform}\end{equation}
where $\delta_{2}(\rho_{s})$ represents a plane delta function orthogonal
to the direction $\hat{s}$, and $\rho_{s}\equiv\vec{\rho}\cdot\hat{s}$.
The $\rho_{s}$ dependence is not altered by $\mathbf{C}$. We now
have \begin{equation}
\vec{B}_{d}(\vec{\rho})=\mathbf{C_{\hat{\rho}}}\vec{M}(\rho_{s})\,\delta_{2}(\rho_{s}),\label{eq:Bd_rho}\end{equation}
and note that $|\vec{B}_{d}(\vec{\rho})|=0$ when $\rho_{s}\neq0$,
due to $\delta_{2}(\rho_{s})$. 

We can perform the inverse 3d Fourier transform to get, \begin{equation}
\vec{B}_{d}(\vec{r})=\mathbf{C_{\hat{s}}}\vec{M}(s)\equiv\vec{B}_{d}(s),\label{eq:Bd_rho}\end{equation}
noting that $\mathbf{C_{\hat{s}}}$ and $\mathbf{C_{\hat{\rho}}}$
are identical except for the naming of the associated polar angle,
which is identical in both spaces.

$\vec{B}_{d}(s)$ is no longer a convolution, and is a function only
of the parameters $s$ and $\hat{s}$,\begin{equation}
\vec{B}_{d}(\vec{r})=\frac{\mu_{0}}{3}\Lambda(\hat{s})\,[3\, M_{z}(s)\,\hat{z}\vec{-M}(s)].\label{eq:Bd_s}\end{equation}

This relationship has been stretched by some investigators%
\footnote{including the author of this dissertation%
} to the relation\begin{equation}
\vec{B}_{d}(\vec{r})=\frac{\mu_{0}}{3}\Lambda(\hat{s})\,[3\, M_{z}(\vec{r})\,\hat{z}-\vec{M}(\vec{r})],\label{eq:Bd_point_form}\end{equation}
 with $\hat{s}$ being the direction of the applied modulation, and
$\vec{M}(\vec{r})$ including the applied modulation. The subtle difference
here is that variation other than the induced one-dimensional modulation
on an otherwise homogeneous magnetization profile is now allowed.
In other words there is an underlying magnetization profile. There
is not yet rigorous theoretical justification for equation \ref{eq:Bd_point_form},
section \ref{sec:local-form} being the beginning of such justification.

\chapter{{}``NON-LINEAR'' BLOCH EQUATIONS\index{non-linear bloch equations}}

\section{Adding the \index{DDF}}

We now include $\vec{B}_{d}(\vec{r})$ into the vector Bloch equation
introduced in chapter \ref{cha:Bloch-Equations},

\begin{equation}
\frac{d\vec{M}}{dt}=\gamma\,\vec{M}\times\vec{B}+\frac{(\vec{M}_{0}-\vec{M}_{\Vert})}{T_{1}}-\frac{\vec{M}_{\perp}}{T_{2}}+\nabla\cdot\mathbf{D}\nabla\vec{M},\label{eq:Bloch_eq_nonlinear}\end{equation}
where we have added $\vec{B}_{d}(r)$ to the magnetic field term,

\begin{equation}
\vec{B}=\vec{B}_{0}+\vec{B}_{1}+\vec{B}_{d}(\vec{r})+\vec{G}_{s}s+\bigtriangleup\vec{B}.\label{eq:B_total_again}\end{equation}
We have explicitly left the $\vec{r}$ dependence on $\vec{B}_{d}(\vec{r})$
as a reminder of the possibility for spatially-dependent modulation\index{spatially-dependent modulation}.

As a start we will look in the rotating frame with $\omega=\omega_{0}$,
with no gradient, RF field, or field inhomogeneity terms, and no relaxation
or diffusion; in other words the only term being $\vec{B}_{d}(\vec{r})$.
We write out the vector Bloch equation, \begin{equation}
\frac{d\vec{M}(\vec{r})}{dt}=\gamma\,\vec{M}\times\vec{B}_{d}\label{eq:Bloch_Bd}\end{equation}

and substitute in equation \ref{eq:Bd_point_form},\begin{equation}
\frac{d\vec{M}(\vec{r})}{dt}=\frac{\mu_{0}}{3}\Lambda(\hat{r})\,\gamma\,\vec{M}(\vec{r})\times[3\, M_{z}(\vec{r})\,\hat{z}-\vec{M}(\vec{r})].\label{eq:Bloch_eq_nonlinear}\end{equation}
We can immediately simplify, since the cross product of a vector with
itself is zero ($\vec{M}\times\vec{M}=0$), to\begin{equation}
\frac{d\vec{M}(\vec{r})}{dt}=\gamma\,\mu_{0}\Lambda(\hat{r})\, M_{z}(\vec{r})\,\vec{M}(\vec{r})\times\hat{z}.\label{eq:Bloch_eq_nonlinear}\end{equation}

At first this appears to be a non-linear differential equation for
$\vec{M}$. We can look at the longitudinal and transverse components
(as introduced in section \ref{sec:Longitudinal-and-Transverse},
and using the substitutions from appendix \ref{sec:Cross-Product-with-Ml-and-Mt}
)\begin{equation}
\frac{dM_{\perp}(\vec{r})}{dt}=-i\,\gamma\,\mu_{0}\Lambda(\hat{r})\, M_{\parallel}(\vec{r})\, M_{\perp}(\vec{r})\label{eq:dMperp_dt}\end{equation}
 and\begin{equation}
\frac{dM_{\parallel}(\vec{r})}{dt}=0.\label{eq:dMpar_dt}\end{equation}

For a single-component spin system the DDF has no effect on the longitudinal
magnetization state. The DDF acts like a longitudinal field term in
the transverse Bloch equation\begin{equation}
\frac{dM_{\perp}(\vec{r})}{dt}=-i\,\gamma\, B_{\parallel d\, eff}(\vec{r})\, M_{\perp}(\vec{r}),\label{eq:dMperp_dt}\end{equation}
 with an effective field of 

\begin{equation}
B_{\parallel d\, eff}(\vec{r})=\mu_{0}\Lambda(\hat{r})\, M_{\parallel}(\vec{r}).\label{eq:dMperp_dt}\end{equation}

The addition of the DDF to the Bloch equations has been said to lead
to {}``non-linear'' Bloch equations. But after the above discussion
this is seen not always to be so, at least when relaxation and diffusion
are neglected. The DDF merely acts as an additional spatially dependent
field term on the transverse magnetization.

\section{The Z magnetization {}``Gradient''\index{DDF as an effective gradient}\label{sec:The-Z-magnetization-Gradient}}

Before our discussion of the DDF\index{DDF} we introduced another
spatially dependent field term, the gradient field (see chapter \ref{cha:Gradients}).
If we could somehow control $M_{\parallel}(\vec{r})$, we could use
it as if it were a gradient. We will see in chapter \ref{cha:HOMOGENIZED}
how the HOMOGENIZED pulse sequence accomplishes this.

It is helpful to have an idea of the potential strength of the $B_{\parallel d\, eff}(\vec{r})$
field. It will be dependent on the concentration of spins, the $B_{0}$
field, and temperature. It is also dependent on the direction of applied
modulation $\hat{s}$. For pure water at room temperature we have\begin{equation}
\frac{\mu_{0}M_{0}}{B_{0}}=2.35\times10^{-7}.\label{eq:Tesla_per_Tesla}\end{equation}
We calculate $B_{\parallel d\, eff}(\vec{r})$ for a $400MHz$ system
$(9.4T)$ and $\hat{s}=\hat{z}$ this gives\[
B_{\parallel d\, eff}(\vec{r})=2.21\,\mu T.\]
This is a very small field. We have\[
\gamma\, B_{\parallel d\, eff}(\vec{r})=10\, Hz\]

which is the rate at which transverse magnetization will precess in
the rotating frame under the influence of $B_{\parallel d\, eff}(\vec{r})$.
The reciprocal of this value is defined as the {}``dipolar demagnetization
time\index{dipolar demagnetization time}'' and is\[
\tau_{d}\equiv\frac{1}{\gamma\, B_{\parallel d\, eff}}=100\, ms\]
for our $400\, MHz$ example.

Looking back to equation \ref{eq:B_total_again} we can understand
the reason the distant dipolar field had been thought insignificant
and ignored until recently. If there is no modulation on $\vec{B}_{d}(\vec{r})$,
the field looks for all purposes like a $\bigtriangleup\vec{B}$ term,
causing potentially very small (much smaller than electronic susceptibility
induced) inhomogeneous broadening or a small homogeneous field shift.
The field shift especially is not usually noticed, since it is compensated
by referencing to a known spectral line.

\section{Two Component System\label{sec:Two-Component-System}}

When there are two types of spins the situation gets more complicated.
We start with\begin{equation}
\vec{M}=\vec{M}^{I}+\vec{M}^{S},\label{eq:M_I_and_S}\end{equation}
the two spin types being labeled $I$ and $S$. The DDF\index{DDF}
has two contributions\begin{equation}
\vec{B}_{d}(\vec{r})=\vec{B}_{d}^{I}(\vec{r})+\vec{B}_{d}^{S}(\vec{r}).\label{eq:Bd_I_and_S}\end{equation}

We substitute into the longitudinal and transverse Bloch equations
and carry out the cross product operation. This gives components, 

\begin{equation}
\frac{dM_{\perp}^{I}(\vec{r})}{dt}=-i\,\gamma_{I}\,\mu_{0}\Lambda(\hat{s})\,[M_{\parallel}^{I}(\vec{r})\, M_{\perp}^{I}(\vec{r})+\frac{1}{3}M_{\perp}^{S}(\vec{r})\, M_{\parallel}^{I}(\vec{r})+\frac{2}{3}M_{\parallel}^{S}(\vec{r})\, M_{\perp}^{I}(\vec{r})]+i\,\Delta\omega_{0}^{I}\, M_{\perp}^{I}(\vec{r})\label{eq:dMtran_dt_two}\end{equation}
and\begin{equation}
\frac{dM_{\parallel}^{I}(\vec{r})}{dt}=i\,\gamma_{I}\,\mu_{0}\Lambda(\hat{s})\,[\frac{1}{6}M_{\perp}^{S}(\vec{r})\,\{ M_{\perp}^{I}(\vec{r})\}^{\star}-\frac{1}{6}\{ M_{\perp}^{S}(\vec{r})\}^{\star}\, M_{\perp}^{I}(\vec{r})].\label{eq:dMlong_dt_two}\end{equation}
Exchanging the labels $I$ and $S$ gives the components for $\frac{d\vec{M}^{S}(\vec{r})}{dt}$.
Note that $\gamma$ is also labeled, and that we have explicitly included
the resonance offset, since in general one of the spins must have
an offset in the rotating frame.

A few things to notice:

The DDF field terms are first in each product of magnetizations, so
that $M_{\parallel}^{I}(\vec{r})\, M_{\perp}^{I}(\vec{r})$ denotes
the DDF due to $M_{\parallel}^{I}(\vec{r})$ causing $M_{\perp}^{I}(\vec{r})$
to rotate. Looking again at the equations we see that the DDF does
not {}``transfer'' magnetization from one spin to the other. This
is as expected due to the cross-product in the Bloch equations. The
DDF from one spin only causes rotation of the other spin's magnetization.

If the spins are significantly different in resonance frequency (either
heteronuclear or homonuclear chemical shift), only the longitudinal
magnetizations will cause a net time average DDF%
\footnote{It is also possible to use a {}``mixing'' sequence to {}``spin-lock''
the transverse magnetization to allow a DDF interaction due to transverse
magnetization. This was recently demonstrated in reference \cite{ZHC+05}.%
}. This eliminates all the terms having prefactor $\frac{1}{3}$ (and
$\frac{1}{6}$ for the longitudinal form), greatly simplifying the
situation.

Also note that the cross terms due to longitudinal magnetization between
$I$ and $S$ have prefactor $\frac{2}{3}$. The heteronuclear (or
chemically shifted) interaction is intrinsically weaker (for the same
magnetization magnitude) than the homonuclear interaction which has
prefactor $1$.

\chapter{RADIATION DAMPING\index{radiation damping}\label{cha:RADIATION-DAMPING}}

\section{What is it?}

The phenomenon of radiation damping was recognized early on in NMR
research \cite{BP54,Blo57,SM59}. It is caused by the field created
by the receiver coil, and can be considered a type of (undesired)
feedback from the receiver back into the sample. The term {}``radiation
damping'' originates from the days of steady state induction experiments
where the signal was smaller or {}``damped'' when the sample to
coil coupling was high.

We can write \cite{VJB95} \begin{equation}
V_{S}(t)=-(\mu_{0}\eta\, L\, V)^{\frac{1}{2}}\frac{dM_{x}}{dt}\label{eq:Rdamp_VS}\end{equation}
for the EMF voltage in the coil due to precessing magnetization in
the sample. We have made several simplifying assumptions. First, the
coil response is uniform. Second, the magnetization in the sample
is uniform. $M_{x}$ is the $\hat{x}$ component of the magnetization
and thirdly we have assumed that the coil is only sensitive to this
component. $\eta$ is the filling factor, representing the fraction
of the sensitive volume of the coil filled by the sample. $L$ is
the inductance of the receiver coil, and $V$ is its sensitive volume.

The voltage induced in the coil will now induce a current in the receiver
coil. This will in turn produce a field in the sample\begin{equation}
\vec{B}_{rd}=\hat{x}\,(\frac{\mu_{0}\eta\, L}{V})^{\frac{1}{2}}\frac{V_{S}(t)}{Z}=-\hat{x}\frac{\mu_{0}\eta\, L}{Z}\frac{dM_{x}}{dt}.\label{eq:Rdamp_B}\end{equation}

We have introduced the impedance of the coil%
\footnote{Strictly speaking we need to consider the parameters $L,\, C,\, R,\, Q,\, Z,\, and\,\omega_{LC}$
for the system consisting of the coil and sample together.%
}\begin{equation}
Z=(\omega_{LC}\frac{L}{Q})(1+\Delta^{2})^{\frac{1}{2}}.\label{eq:Rdamp_Z}\end{equation}
The quality factor is\begin{equation}
Q=\omega_{LC}\frac{L}{R},\label{eq:Rdamp_Q}\end{equation}
and the resonant frequency of the coil is\begin{equation}
\omega_{LC}=\sqrt{\frac{1}{L\, C}},\label{eq:Rdamp_omega}\end{equation}
where $C$ is the capacitance. Finally, we define the off-resonance
or {}``detuning'' parameter\begin{equation}
\Delta=Q\frac{\omega_{LC}}{\omega_{0}}(\frac{\omega_{0}^{2}-\omega_{LC}^{2}}{\omega_{LC}^{2}}).\label{eq:Rdamp_Delta}\end{equation}

When we add $\vec{B}_{rd}$ as a source term to the Bloch equations
(e.g. add $\vec{B}_{rd}$ as a term in equation \ref{eq:B_total})
the equations become nonlinear, and in general much more difficult
to solve.

\section{What does it do?}

We can decompose the $\frac{dM_{x}}{dt}$ term in equation \ref{eq:Rdamp_B}
to get a glimpse of the effects of radiation damping. Assuming we
have just issued a $90^{\circ}\hat{x}$ pulse giving\begin{equation}
M_{\perp}=i\, M_{0}e^{i\,\omega_{0}t}\label{eq:Rdamp_Mt}\end{equation}
in the laboratory frame. We can decompose this into the two linear
oscillating components (the $Re$ and $Im$ components of equation
\ref{eq:Rdamp_Mt})\begin{equation}
M_{x}=-M_{0}sin(\omega_{0}t)\label{eq:Rdamp_Mx}\end{equation}
and\begin{equation}
M_{y}=M_{0}cos(\omega_{0}t).\label{eq:Rdamp_My}\end{equation}
$M_{y}$ is orthogonal to the sensitive axis of the receiver coil
and does not contribute to the signal. We compute the derivative\begin{equation}
\frac{dM_{x}}{dt}=\omega_{0}M_{0}cos(\omega_{0}t),\label{eq:Rdamp_dMdx}\end{equation}
which gives us\begin{equation}
\vec{B}_{rd}=-\hat{x}\frac{\mu_{0}\eta\, L\,\omega_{0}M_{0}}{Z}cos(\omega_{0}t).\label{eq:Rdamp_B_cos}\end{equation}

We have seen a similar expression before, in our discussion of RF
fields in section \ref{sec:RF-Field}. It is a linearly oscillating
RF field, which we can now decompose into its rotating components\begin{equation}
B''_{\perp rd}=-\frac{\mu_{0}\eta\, L\,\omega_{0}M_{0}}{2\, Z}e^{i\,\omega_{0}t}\label{eq:Rdamp_B''}\end{equation}
and\begin{equation}
B'{}_{\perp rd}=-\frac{\mu_{0}\eta\, L\,\omega_{0}M_{0}}{2\, Z}e^{-i\,\omega_{0}t}.\label{eq:Rdamp_B'}\end{equation}

The counter-rotating component $B'{}_{\perp rd}$ has negligible effect
\cite{BS40} so we will drop it from $B{}_{\perp rd}$ and use \begin{equation}
B{}_{\perp rd}=-\frac{\mu_{0}\eta\, L\,\omega_{0}M_{0}}{2\, Z}e^{i\,\omega_{0}t}.\label{eq:Rdamp_B_rot}\end{equation}
Finally we can substitute equation \ref{eq:Rdamp_Z} for $Z$ into
\ref{eq:Rdamp_B_rot} to get\begin{equation}
B{}_{\perp rd}=-\frac{\mu_{0}\eta\, Q\,\omega_{0}M_{0}}{2\,\omega_{LC}(1+\Delta^{2})^{\frac{1}{2}}}e^{i\,\omega_{0}t}.\label{eq:Rdamp_B_Q}\end{equation}

Unlike our transmitter-induced RF field, $B_{1}$, $B{}_{\perp rd}$
is not under direct control of the pulse sequence. It tends to oppose
the effects of the applied $B_{1}$ field (note that when our initial
pulse was along the positive $\hat{x}$ axis $B{}_{\perp rd}$ is
along the negative $\hat{x}$ axis). This is not surprising, as it
is a manifestation of Lenz's Law.

We can define a parameter called the radiation damping time\[
\frac{1}{\tau_{rd}}\equiv\frac{\gamma}{2\pi}\,|B{}_{\perp rd}|\]
which gives us a measure of the strength of radiation damping in relation
to other processes such as applied RF fields and relaxation.

When $\tau_{rd}\sim\tau_{90}$ (strong radiation damping), where $\tau_{90}$
is the duration of a $90^{\circ}$ pulse, there can be significant
problems in obtaining consistent excitation. Changes in recovery time
could lead to over or under excitation of the sample. One can potentially
use special pulses to compensate \cite{WHB89}.

When $\tau_{rd}\sim\tau_{mix}$ in various sequences (such as 2d spectroscopy,
or magnetization transfer) erroneous (or at least unexpected) cross-peaks\cite{MW90,BBH+96}
or signal behavior can be observed.

When $\tau_{rd}\sim T_{1}\, or\, T_{2}$ in inversion/saturation recovery
measurements of $T_{1}$, or spin echo measurements of $T_{2}$, there
can be significant errors \cite{aug02}.

When $\tau_{rd}\sim\tau_{build}$ in a DDF experiment there can be
significant signal attenuation.

Also there has been theoretical as well as experimental studies of
chaotic dynamics that can result when both radiation damping and distant
dipolar field effects are present \cite{LLAW00,Abe02,HL03}.

\section{How to avoid it?}

In most cases it is desirable to avoid radiation damping effects.
First one must estimate the magnitude or test for the presence of
radiation damping \cite{SD99} to see if one need take special precautions.

A quick test is to detune the receiver coil (usually called the probe
in high resolution systems) while watching the shape of the FID. If
the decay seems to lessen as the probe is detuned then there is likely
radiation damping%
\footnote{Thanks to Norbert Lutz, Ph.D. for suggesting this.%
}.

Examination of equation \ref{eq:Rdamp_B_Q} suggests several remedies.

High resolution spectroscopists have avoided the problem of radiation
damping even as $B_{0}$ fields and $Q$ factors have increased in
NMR spectrometers. This is due to the fact that it it desirable (usually
for reasons of eliminating the water signal from the spectrum) to
use deuterated solvents. This corresponds to reducing $M_{0}$ in
equation \ref{eq:Rdamp_B_Q}.

Use of a deuterated solvent is usually not possible for biological
samples. In this case one can increase $\Delta$ by detuning the receiver
coil . This also has the effect of reducing the signal to noise ratio,
which is not desirable.

There has been some investigation of active feedback systems that
cancel the induced current in the receiver coil \cite{BJ95}. This
is still an active (no pun intended) research topic.

A very effective means of radiation damping suppression is to use
balanced diffusion weighting gradients (see section \ref{sec:Stejskal-Tanner-Sequence}).
In the time period between the balanced gradients the transverse magnetization
in in a helical state (spoiled). The average magnetization $M_{0}$
that the receiver coil experiences is near zero, as long as there
are many twists of the helix across the sample. Note however that
this does not suppress radiation damping after the second gradient
pulse, such as during acquisition. In distant dipolar field based
sequences this is usually not an issue as the distant dipolar field
re-phased transverse magnetization during the acquisition period is
usually much smaller than the directly excited transverse magnetization
after a $90^{\circ}$ pulse.

\chapter{HOMOGENIZED\label{cha:HOMOGENIZED}}

\section{Acronyms}

There are many acronyms used to describe NMR and MRI pulse sequences.
The two most common in DDF-based activities are CRAZED and HOMOGENIZED\index{HOMOGENIZED}.
CRAZED stands for {}``COSY Revamped by Asymmetric Z Gradient Echo
Detection'' \cite{RLW+95}. HOMOGENIZED is the variation of CRAZED
we are interested in. It stands for {}``HOMOGeneity ENhancement by
Intermolecular ZEro quantum Detection'' \cite{VLW96}.

\section{Sequence}

\begin{figure}
\begin{center}\includegraphics[%
  width=1.0\columnwidth,
  keepaspectratio]{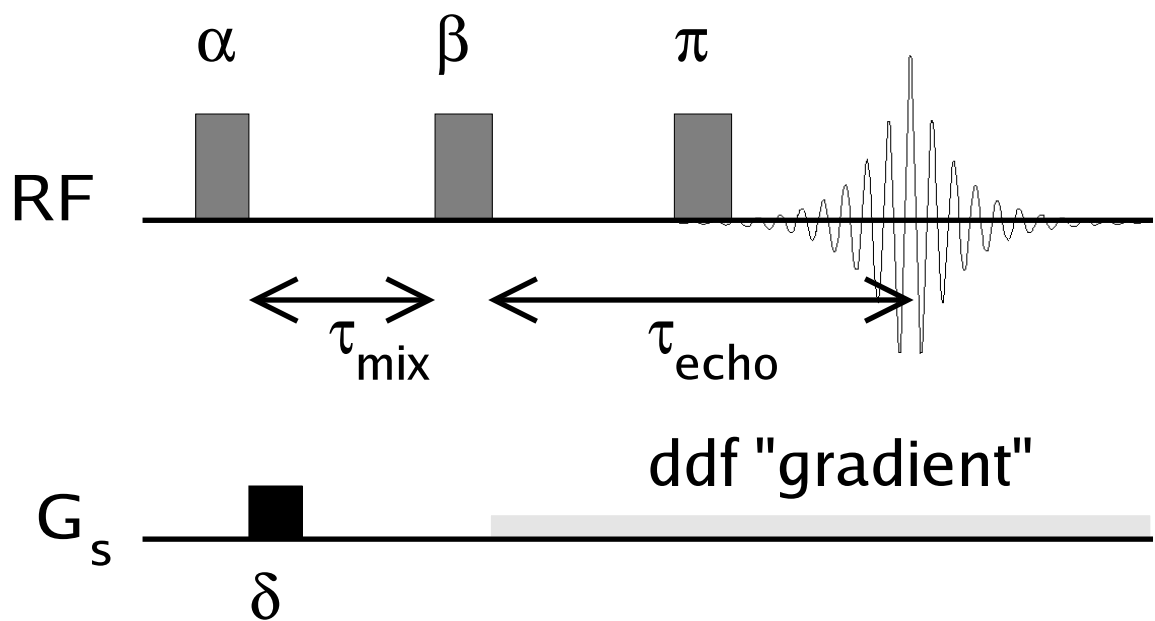}\end{center}

\caption{\label{cap:HOMOGENIZED_ps}The HOMOGENIZED pulse sequence.}
\end{figure}
The sequence shown in Figure \ref{cap:HOMOGENIZED_ps} consists of
three RF pulses, $\alpha$ for excitation, \textbf{$\beta$} to convert
helical transverse magnetization to $M_{z}$ modulation, and $\pi$
to form a spin-echo. The $G_{s}$ gradient in combination with $\beta$
creates spatially-modulated longitudinal magnetization whose magnetic
field causes unwinding (and eventually rewinding) of helically twisted
transverse magnetization \cite{CG04c}.

The goal of the HOMOGENIZED sequence is to obtain a high resolution
1d spectrum by performing a 2d acquisition and projecting along the
$F2$ dimension.

\section{Step by Step HOMOGENIZED}

It is helpful to visualize the HOMOGENIZED sequence in a step by step
manner, noting analogies to more commonly understood spin behavior.
The process is twofold, first, noting when the sequence does not utilize
the DDF and so behaves as conventionally expected, and second, noting
when the sequence utilizes the DDF\index{DDF} and how the behavior
departs from commonly understood behavior. We will also note simplifying
assumptions that have been made along the way. The following description
is done in the rotating frame of the solvent $S$.

\subsection{Excitation by the $\alpha$ pulse\label{sub:alpha}}

First, the HOMOGENIZED sequence excites the system with the $\alpha$
pulse. This will usually be a $90^{\circ}$ pulse, unless utilizing
a short $TR$ in which case one might use the Ernst angle\index{Ernst angle}
(see section \ref{sub:Repetition-and-Recovery}). We will use a non-selective
$90^{\circ}$, phase $\phi_{\alpha}$, {}``hard'' pulse, although
variations of HOMOGENIZED have used selective pulses \cite{CHC+04}.
We start with fully relaxed magnetization, with two spin types $I$
and $S$. After the $\alpha$ pulse we have, using the longitudinal/transverse
notation,

\begin{equation}
M_{\perp}^{\alpha}=i\,[M_{0}^{I}+M_{0}^{S}]\, e^{i\,\phi_{\alpha}},\label{eq:homo_alpha_tran}\end{equation}
and\begin{equation}
M_{\parallel}^{\alpha}=0.\label{eq:homo_alpha_long}\end{equation}
This is for a homonuclear system where the hard pulse has bandwidth
to cover both $I$ and $S$ magnetization.

\subsection{$G_{s}$gradient}

The $G_{s}$ gradient of duration $\delta$ causes the transverse
magnetization to {}``twist'' along the $\hat{s}$ spatial direction.
The pitch of the twist is

\begin{equation}
q=\frac{\gamma}{2\pi}G_{s}\delta.\label{eq:homo_Gs_q}\end{equation}

After the applied gradient we have\begin{equation}
M_{\perp}^{G}=i\,[M_{0}^{I}+M_{0}^{S}]\, e^{i\,\phi_{\alpha}-i\,2\pi\, q\, s},\label{eq:homo_Gs_tran}\end{equation}
and

\begin{equation}
M_{\parallel}^{G}=0.\label{eq:homo_Gs_long}\end{equation}

\subsection{$\tau_{mix}$ delay\label{sub:homo_tau_mix_delay}}

During the $\tau_{mix}$delay (which includes $\delta$) we will have
phase {}``evolution'' of transverse magnetization due to resonance
offsets, in this case the chemical shift. We will assume that the
$S$ magnetization is on-resonance and that there is a chemical shift
of $I$ relative to $S$ of $\sigma$. We will neglect other effects
such as field inhomogeneity, relaxation, and diffusion for now. This
gives us

\begin{equation}
M_{\perp}^{\tau_{mix}}=i\,[M_{0}^{I}e^{i\,\sigma\,\omega_{0}\tau_{mix}}+M_{0}^{S}]\, e^{i\,\phi_{\alpha}-i\,2\pi\, q\, s},\label{eq:homo_tau_mix_tran}\end{equation}
and

\begin{equation}
M_{\parallel}^{\tau_{mix}}=0.\label{eq:homo_tau_mix_long}\end{equation}

We will assume that $\sigma$ is large enough to meet the condition\begin{equation}
\frac{\sigma\,\omega_{0}}{2\pi}\gg\frac{1}{\tau_{d}}\label{eq:homo_tau_sigma}\end{equation}
where $\tau_{d}$, defined in section \ref{sec:The-Z-magnetization-Gradient},
is the timescale it takes the DDF to cause significant precession
of magnetization. $\tau_{d}$ needs to be computed from the sum of
both $S$ and $I$ magnetizations when both are significant. The condition
states that the chemical shift will dephase or average to zero any
DDF component originating from transverse magnetization. As a practical
guide, the $I$ and $S$ peaks in the spectrum should not significantly
overlap due to inhomogeneous broadening ($T_{2}$relaxation) for this
condition to be met.

\subsection{$\beta$ pulse}

The purpose of the $\beta$ pulse is to form spatially-modulated longitudinal
magnetization from the {}``twisted'' transverse magnetization. $\beta$
is again assumed to be a {}``hard'' pulse with sufficient bandwidth
to cover both $I$ and $S$ resonance offsets.

The flip angle of $\beta$ controls the {}``depth'' or magnitude
of the modulation. We will see that different flip angles for $\beta$
can lead to different classes of signals being optimized at the end
of the sequence.

We will look at the case where $\beta$ has phase $-\hat{y}$ or $\phi_{\beta}=-\frac{\pi}{2}$.
We now have\begin{multline}
M_{\perp}^{\beta}=i\, M_{0}^{I}cos(\phi_{\alpha}+\tau_{mix}\sigma\,\omega_{0}-2\pi\, q\, s)-M_{0}^{I}sin(\phi_{\alpha}+\tau_{mix}\sigma\,\omega_{0}-2\pi\, q\, s)\, cos(\beta)\label{eq:homo_beta_tran}\\
+i\, M_{0}^{S}cos(\phi_{\alpha}-2\pi\, q\, s)-M_{0}^{S}sin(\phi_{\alpha}-2\pi\, q\, s)\, cos(\beta)\end{multline}
and

\begin{equation}
M_{\parallel}^{\beta}=-M_{0}^{I}sin(\phi_{\alpha}+\tau_{mix}\sigma\,\omega_{0}-2\pi\, q\, s)\, sin(\beta)-M_{0}^{S}sin(\phi_{\alpha}-2\pi\, q\, s)\, sin(\beta).\label{eq:homo_beta_long}\end{equation}

We notice immediately that the longitudinal component amplitude is
maximized when $\beta=\frac{\pi}{2}$ or $\frac{3\pi}{2}$. The real
($\hat{x}$) transverse component is concomitantly minimized. Both
transverse and longitudinal components have spatial variation (modulation)
due to the gradient, and chemical shift variation is present for the
$I$ magnetization.

\subsection{$\frac{\tau_{echo}}{2}$ Delay}

The $\tau_{echo}$ delay is split in half by the $\pi$ RF pulse,
forming a spin-echo\index{spin-echo}. During the first half of the
delay period $\frac{\tau_{echo}}{2}$ we will again have chemical
shift evolution, but this will be refocused at the spin echo point.
Since we now have longitudinal magnetization, we will have a DDF\index{DDF}.
This DDF is not averaged away by off-resonance effects as described
in section \ref{sub:homo_tau_mix_delay}.

We restate (from section \ref{sec:Two-Component-System} equations
\ref{eq:dMtran_dt_two} and \ref{eq:dMlong_dt_two}) the Bloch equations
for the DDF in transverse and longitudinal form 

\begin{equation}
\frac{dM_{\perp}^{I}(\vec{r})}{dt}=-i\,\gamma_{I}\,\mu_{0}\Lambda(\hat{s})\,[M_{\parallel}^{I}(\vec{r})\, M_{\perp}^{I}(\vec{r})+\frac{1}{3}M_{\perp}^{S}(\vec{r})\, M_{\parallel}^{I}(\vec{r})+\frac{2}{3}M_{\parallel}^{S}(\vec{r})\, M_{\perp}^{I}(\vec{r})]+i\,\Delta\omega_{0}^{I}\, M_{\perp}^{I}(\vec{r}),\label{eq:dMtran_dt_two_again}\end{equation}

and

\begin{equation}
\frac{dM_{\parallel}^{I}(\vec{r})}{dt}=i\,\gamma_{I}\,\mu_{0}\Lambda(\hat{s})\,[\frac{1}{6}M_{\perp}^{S}(\vec{r})\,\{ M_{\perp}^{I}(\vec{r})\}^{\star}-\frac{1}{6}\{ M_{\perp}^{S}(\vec{r})\}^{\star}\, M_{\perp}^{I}(\vec{r})].\label{eq:dMlong_dt_two_again}\end{equation}

We assume that the condition \ref{sub:homo_tau_mix_delay} \begin{equation}
\frac{\sigma\,\omega_{0}}{2\pi}\gg\frac{1}{\tau_{d}}\label{eq:homo_tau_sigma_again}\end{equation}
is met, which leads to any transverse {}``field'' term (the left
term in each pair of magnetization products) that is off-resonance
giving a time average DDF of zero. This simplifies the equations to\begin{equation}
\frac{dM_{\perp}^{I}(\vec{r})}{dt}=-i\,\gamma_{I}\,\mu_{0}\Lambda(\hat{s})\,[M_{\parallel}^{I}(\vec{r})\, M_{\perp}^{I}(\vec{r})+\frac{2}{3}M_{\parallel}^{S}(\vec{r})\, M_{\perp}^{I}(\vec{r})]-i\,\Delta\omega_{0}^{I}\, M_{\perp}^{I}(\vec{r}),\label{eq:dMtran_dt_two_simplified}\end{equation}
and\begin{equation}
\frac{dM_{\parallel}^{I}(\vec{r})}{dt}=0\label{eq:dMlong_dt_two_simplified}\end{equation}
We now have again no change in the longitudinal component due to the
DDF, similar to the single component system (equation \ref{eq:dMpar_dt}).

We use the following notation for the DDF, which is subtly different
than that introduced in equation \ref{eq:Bd_I_and_S}. This is the
total DDF {}``felt'' by the $I$ magnetization, not the DDF due
to $I$ magnetization alone,

\begin{equation}
B_{d,\, I}(\vec{r})=\mu_{0}\Lambda(\hat{s})\,[M_{\parallel}^{I}(\vec{r})+\frac{2}{3}M_{\parallel}^{S}(\vec{r})].\label{eq:BdI_simplified}\end{equation}
 It is a longitudinal field only, and does not change due to DDF or
off-resonance effects %
\footnote{We will see later in chapter \ref{cha:ISMRMposter} that it can change
due to relaxation and diffusion.%
}. Substituting \ref{eq:BdI_simplified} into \ref{eq:dMtran_dt_two_simplified}
gives\begin{equation}
\frac{dM_{\perp}^{I}(\vec{r})}{dt}=-i\,[\gamma_{I}B_{d,\, I}(\vec{r})-\Delta\omega_{0}^{I}]\, M_{\perp}^{I}(\vec{r})\label{eq:dMtran_dt_two_Bd}\end{equation}
This represents transverse $I$ magnetization precessing due to resonance
offset and a spatially dependent longitudinal field (like a gradient).
It has the solution\[
M_{\perp}^{I}(\vec{r})=M_{\perp,\,0}^{I}(\vec{r})\, e^{-i\,[\gamma_{I}B_{d,\, I}(\vec{r})-\Delta\omega_{0}^{I}]\, t}\]

We can get similar expressions for $S$ by exchanging $I$ and $S$.

We now have, after the $\frac{\tau_{echo}}{2}$ portion of the HOMOGENIZED
sequence, the magnetization states

\begin{multline}
M_{\perp}^{\frac{\tau_{echo}}{2}}=M_{0}^{I}[i\, cos(\phi_{\alpha}+\tau_{mix}\sigma\,\omega_{0}-2\pi\, q\, s)\\
-sin(\phi_{\alpha}+\tau_{mix}\sigma\,\omega_{0}-2\pi\, q\, s)\, cos(\beta)]\, e^{-i\,[\gamma_{I}B_{d,\, I}(\vec{r})-\sigma\,\omega_{0}]\,\frac{\tau_{echo}}{2}}\\
+M_{0}^{S}[i\, cos(\phi_{\alpha}-2\pi\, q\, s)-sin(\phi_{\alpha}-2\pi\, q\, s)\, cos(\beta)]\, e^{-i\,\gamma_{S}B_{d,\, S}(\vec{r})\,\frac{\tau_{echo}}{2}}\label{eq:homo_tau_echo_ov2_tran}\end{multline}
and

\begin{equation}
M_{\parallel}^{\frac{\tau_{echo}}{2}}=-M_{0}^{I}sin(\phi_{\alpha}+\tau_{mix}\sigma\,\omega_{0}-2\pi\, q\, s)\, sin(\beta)-M_{0}^{S}sin(\phi_{\alpha}-2\pi\, q\, s)\, sin(\beta).\label{eq:homo_tau_echo_ov2_long}\end{equation}
We have made the substitution $\sigma\,\omega_{0}=\Delta\omega_{0}$
in equation \ref{eq:homo_tau_echo_ov2_tran}.

\subsection{$\pi$ pulse and spin echo}

We will simplify the effects of the $\pi$ pulse by fixing its flip
angle ( to $180^{\circ}$) and phase to $-\hat{y}$. This inverts
the longitudinal magnetization%
\footnote{The actual value of the DDF $B_{d,\, I}(\vec{r})$ as defined in equation
\ref{eq:BdI_simplified} changes, but in our notation there is no
sign change in the exponents of equation \ref{eq:homo_pi_tran}.%
} and inverts the $\hat{x}$ component of the transverse magnetization.
We have

\begin{multline}
M_{\perp}^{\pi}=M_{0}^{I}[i\, cos(\phi_{\alpha}+\tau_{mix}\sigma\,\omega_{0}-2\pi\, q\, s)\\
+sin(\phi_{\alpha}+\tau_{mix}\sigma\,\omega_{0}-2\pi\, q\, s)\, cos(\beta)]\, e^{-i\,[\gamma_{I}B_{d,\, I}(\vec{r})-\sigma\,\omega_{0}]\,\frac{\tau_{echo}}{2}}\\
+M_{0}^{S}[i\, cos(\phi_{\alpha}-2\pi\, q\, s)+sin(\phi_{\alpha}-2\pi\, q\, s)\, cos(\beta)]\, e^{-i\,\gamma_{S}B_{d,\, S}(\vec{r})\,\frac{\tau_{echo}}{2}}\label{eq:homo_pi_tran}\end{multline}
and

\begin{equation}
M_{\parallel}^{\pi}=M_{0}^{I}sin(\phi_{\alpha}+\tau_{mix}\sigma\,\omega_{0}-2\pi\, q\, s)\, sin(\beta)+M_{0}^{S}sin(\phi_{\alpha}-2\pi\, q\, s)\, sin(\beta).\label{eq:homo_pi_long}\end{equation}

\subsection{Second $\frac{\tau_{echo}}{2}$ Delay}

The second $\frac{\tau_{echo}}{2}$ is similar to the first. At the
echo point the chemical shift is refocused. The DDF-induced phase
is not canceled since the longitudinal magnetization causing the DDF
has been inverted as well. We end up with\begin{multline}
M_{\perp}^{\tau_{echo}}=M_{0}^{I}[i\, cos(\phi_{\alpha}+\tau_{mix}\sigma\,\omega_{0}-2\pi\, q\, s)\\
+sin(\phi_{\alpha}+\tau_{mix}\sigma\,\omega_{0}-2\pi\, q\, s)\, cos(\beta)]\, e^{-i\,\gamma_{I}B_{d,\, I}(\vec{r})\,\tau_{echo}}\\
+M_{0}^{S}[i\, cos(\phi_{\alpha}-2\pi\, q\, s)+sin(\phi_{\alpha}-2\pi\, q\, s)\, cos(\beta)]\, e^{-i\,\gamma_{S}B_{d,\, S}(\vec{r})\,\tau_{echo}}\label{eq:homo_tau_echo_tran}\end{multline}
and\begin{equation}
M_{\parallel}^{\tau_{echo}}=M_{0}^{I}sin(\phi_{\alpha}+\tau_{mix}\sigma\,\omega_{0}-2\pi\, q\, s)\, sin(\beta)+M_{0}^{S}sin(\phi_{\alpha}-2\pi\, q\, s)\, sin(\beta).\label{eq:homo_tau_echo_long}\end{equation}

\section{Signal growth due to the DDF%
\footnote{follows the classical calculation of Ahn et al. \cite{alw98b}%
}}

The spatially-varying longitudinal DDF\index{DDF} will alter the
phase of the transverse magnetization, which also has spatial variation.
At this point we will make the simplifying assumption that all spatial
variation is due to the applied gradient. We will also conduct the
analysis first for transverse $I$ magnetization.

The $I$ transverse magnetization is\begin{multline}
M_{\perp}^{I}=M_{0}^{I}[i\, cos(\phi_{\alpha}+\tau_{mix}\sigma\,\omega_{0}-2\pi\, q\, s)\\
+sin(\phi_{\alpha}+\tau_{mix}\sigma\,\omega_{0}-2\pi\, q\, s)\, cos(\beta)]\, e^{-i\,\gamma_{I}B_{d,\, I}(\vec{r})\, t+i\,\sigma\,\omega_{0}(t-\tau_{echo})},\label{eq:MtranI_after_beta}\end{multline}
for the time period $t$ after the $\beta$ pulse. We have taken the
final form of the magnetization after the $\pi$ pulse. Substituting
\ref{eq:homo_tau_echo_long} into \ref{eq:BdI_simplified} gives\begin{equation}
B_{d,\, I}(\vec{r})=\mu_{0}\Lambda(\hat{s})\,[M_{0}^{I}sin(\phi_{\alpha}+\tau_{mix}\sigma\,\omega_{0}-2\pi\, q\, s)\, sin(\beta)+\frac{2}{3}M_{0}^{S}sin(\phi_{\alpha}-2\pi\, q\, s)\, sin(\beta)].\label{eq:BdI_after_beta}\end{equation}

We make the following substitutions\begin{equation}
x_{I}=\phi_{\alpha}+\sigma\,\omega_{0}\tau_{mix}-2\pi\, q\, s,\label{eq:DDF_build_xI}\end{equation}
\begin{equation}
x_{S}=\phi_{\alpha}-2\pi\, q\, s,\label{eq:DDF_build_xS}\end{equation}

\begin{equation}
z_{I}=\gamma_{I}\mu_{0}\Lambda(\hat{s})\, M_{0}^{I}sin(\beta)\, t,\label{eq:DDF_build_zI}\end{equation}
and\begin{equation}
z_{S}=\frac{2}{3}\gamma_{I}\mu_{0}\Lambda(\hat{s})\, M_{0}^{S}sin(\beta)\, t\label{eq:DDF_Build_zS}\end{equation}
where $x$ and $z$ are variables of convenience only, and do not
designate coordinates or directions. Substitution into \ref{eq:BdI_after_beta}
and \ref{eq:MtranI_after_beta} gives us\begin{equation}
M_{\perp}^{I}=M_{0}^{I}[i\, cos(x_{I})+sin(x_{I})\, cos(\beta)]\, e^{i\,\sigma\,\omega_{0}(t-\tau_{echo})}e^{i\, z_{I}sin(x_{I})}e^{i\, z_{S}sin(x_{S})}.\label{eq:MtranI_substituted}\end{equation}
Now we use a form of the generating function for Bessel functions\index{Bessel function}%
\footnote{Note that this is a Fourier series expansion.%
}

\begin{equation}
e^{i\, z\, sin(x)}=\sum_{m=-\infty}^{\infty}e^{i\, m\, x}J_{m}(z),\label{eq:DDF_build_Bessel_sin}\end{equation}
which is obtained by substituting $cos(x-\frac{\pi}{2})=sin(x)$ into
\cite[8.511 4. p973]{GRG+80}\begin{equation}
e^{i\, z\, cos(x)}=\sum_{m=-\infty}^{\infty}i^{m}e^{i\, m\, x}J_{m}(z).\label{eq:DDF_build_bessel_cos}\end{equation}
Substitution of \ref{eq:DDF_build_Bessel_sin} into \ref{eq:MtranI_substituted}
gives\begin{equation}
M_{\perp}^{I}=M_{0}^{I}[i\, cos(x_{I})-sin(x_{I})\, cos(\beta)]\, e^{i\,\sigma\,\omega_{0}(t-\tau_{echo})}\sum_{m=-\infty}^{\infty}\sum_{n=-\infty}^{\infty}e^{i\, m\, x_{I}}e^{i\, n\, x_{S}}J_{m}(z_{I})J_{n}(z_{S}).\label{eq:MtranI_double_series}\end{equation}

The detected signal amplitude in magnetic resonance is proportional
to the spatial integral of the transverse magnetization over the sample

\begin{equation}
A\propto\int_{sample}M_{\perp}(\vec{r})\, d^{3}r.\label{eq:nmr_signal}\end{equation}
The proportionality relation takes into account the coil sensitivity
and amplifier gain.

The only terms in the double sum of equation \ref{eq:MtranI_double_series}
that will lead to significant signal are those that have no spatial
variation (are not spoiled)%
\footnote{It is possible to refocus these other terms by an additional gradient
after the $\beta$ RF pulse, corresponding to the selection of different
orders of coherence.%
}. The spatial variation is found in the $-2\pi\, q\, s$ terms of
$x_{I}$ and $x_{s}$. We can therefore search for terms where $x_{I}$
and $x_{S}$ cancel in the exponent. We must take into account the
$cos(x_{I})$ and $sin(x_{I})$ terms in front as well. We make the
substitutions\begin{equation}
cos(x)=\frac{e^{i\, x}+e^{-i\, x}}{2}\label{eq:euler_cos}\end{equation}
and\begin{equation}
sin(x)=-i\frac{e^{i\, x}-e^{-i\, x}}{2}\label{eq:euler_sin}\end{equation}
into \ref{eq:MtranI_double_series}, which yields\begin{multline}
M_{\perp}^{I}=i\, M_{0}^{I}[\frac{e^{i\, x_{I}}+e^{-i\, x_{I}}}{2}-\frac{e^{i\, x_{I}}-e^{-i\, x_{I}}}{2}cos(\beta)]\\
e^{i\,\sigma\,\omega_{0}(t-\tau_{echo})}\sum_{m=-\infty}^{\infty}\sum_{n=-\infty}^{\infty}e^{i\, m\, x_{I}}e^{i\, n\, x_{S}}J_{m}(z_{I})\, J_{n}(z_{S}).\label{eq:MtranI_double_series_exp}\end{multline}

There are two cases where the net $x$ power is zero (note that $x_{S}$
is equivalent to $x_{I}$ for spatial dependence $-2\, q\, s$),\begin{equation}
n=-(m-1)\label{eq:n_p}\end{equation}
and

\begin{equation}
n=-(m+1).\label{eq:n_n}\end{equation}

We separate these two classes of terms to get

\begin{equation}
M_{\perp p}^{I}=\frac{1}{2}i\, M_{0}^{I}[1+cos(\beta)]e^{i\,\sigma\,\omega_{0}(t-\tau_{echo})}\sum_{m=-\infty}^{\infty}e^{i\,(m+1)\, x_{I}}e^{-i\,(m+1)\, x_{S}}J_{m}(z_{I})\, J_{-(m+1)}(z_{S})\label{eq:MtranI_p}\end{equation}
and\begin{equation}
M_{\perp n}^{I}=\frac{1}{2}i\, M_{0}^{I}[1-cos(\beta)]e^{i\,\sigma\,\omega_{0}(t-\tau_{echo})}\sum_{m=-\infty}^{\infty}e^{i\,(m-1)\, x_{I}}e^{-i\,(m-1)\, x_{S}}J_{m}(z_{I})\, J_{-(m-1)}(z_{S}).\label{eq:MtranI_n}\end{equation}
finally we can substitute back our values of $x$ and $z$ to get\begin{multline}
M_{\perp p}^{I}=\frac{1}{2}i\, M_{0}^{I}[1+cos(\beta)]\, e^{i\,\sigma\,\omega_{0}(t-\tau_{echo})}\\
\sum_{m=-\infty}^{\infty}e^{i\,(m+1)\,\sigma\,\omega_{0}\tau_{mix}}J_{m}[\gamma_{I}\mu_{0}\Lambda(\hat{s})\, M_{0}^{I}sin(\beta)\, t]\, J_{-(m+1)}[\frac{2}{3}\gamma_{I}\mu_{0}\Lambda(\hat{s})\, M_{0}^{S}sin(\beta)\, t]\label{eq:Mtranl_p_sub}\end{multline}

\begin{multline}
M_{\perp n}^{I}=\frac{1}{2}i\, M_{0}^{I}[1-cos(\beta)]\, e^{i\,\sigma\,\omega_{0}(t-\tau_{echo})}\\
\sum_{m=-\infty}^{\infty}e^{i\,(m-1)\,\sigma\,\omega_{0}\tau_{mix}}J_{m}[\gamma_{I}\mu_{0}\Lambda(\hat{s})\, M_{0}^{I}sin(\beta)\, t]\, J_{-(m-1)}[\frac{2}{3}\gamma_{I}\mu_{0}\Lambda(\hat{s})\, M_{0}^{S}sin(\beta)\, t]\label{eq:Mtranl_n_sub}\end{multline}
We can make two more substitutions\begin{equation}
\tau_{dII}=\frac{1}{\gamma_{I}\,\mu_{0}M_{0}^{I}}\label{eq:tau_dII}\end{equation}
and\begin{equation}
\tau_{dIS}=\frac{1}{\gamma_{I}\,\mu_{0}M_{0}^{S}}\label{eq:tau_dIS}\end{equation}
to get\begin{multline}
M_{\perp p}^{I}=\frac{1}{2}i\, M_{0}^{I}[1+cos(\beta)]\, e^{i\,\sigma\,\omega_{0}(t-\tau_{echo})}\\
\sum_{m=-\infty}^{\infty}e^{i\,(m+1)\,\sigma\,\omega_{0}\tau_{mix}}J_{m}[\Lambda(\hat{s})\, sin(\beta)\frac{t}{\tau_{dII}}]\, J_{-(m+1)}[\frac{2}{3}\Lambda(\hat{s})\, sin(\beta)\frac{t}{\tau_{dIS}}]\label{eq:MtranI_t_d_p}\end{multline}
and\begin{multline}
M_{\perp n}^{I}=\frac{1}{2}i\, M_{0}^{I}[1-cos(\beta)]\, e^{i\,\sigma\,\omega_{0}(t-\tau_{echo})}\\
\sum_{m=-\infty}^{\infty}e^{i\,(m-1)\,\sigma\,\omega_{0}\tau_{mix}}J_{m}[\Lambda(\hat{s})\, sin(\beta)\frac{t}{\tau_{dII}}]\, J_{-(m-1)}[\frac{2}{3}\Lambda(\hat{s})\, sin(\beta)\frac{t}{\tau_{dIS}}].\label{eq:MtranI_t_d_n}\end{multline}

The results for $S$ ( obtained by interchanging $I$ and $S$ in
equations \ref{eq:MtranI_p} and \ref{eq:MtranI_n}) are

\begin{multline}
M_{\perp p}^{S}=\frac{1}{2}i\, M_{0}^{S}[1+cos(\beta)]\\
\sum_{m=-\infty}^{\infty}e^{-i\,(m+1)\,\sigma\,\omega_{0}\tau_{mix}}J_{m}[\Lambda(\hat{s})\, sin(\beta)\frac{t}{\tau_{dSS}}]\, J_{-(m+1)}[\frac{2}{3}\Lambda(\hat{s})\, sin(\beta)\frac{t}{\tau_{dSI}}]\label{eq:MtranS_t_d_p}\end{multline}

and\begin{multline}
M_{\perp n}^{S}=\frac{1}{2}i\, M_{0}^{S}[1-cos(\beta)]\\
\sum_{m=-\infty}^{\infty}e^{-i\,(m-1)\,\sigma\,\omega_{0}\tau_{mix}}J_{m}[\Lambda(\hat{s})\, sin(\beta)\frac{t}{\tau_{dSS}}]\, J_{-(m-1)}[\frac{2}{3}\Lambda(\hat{s})\, sin(\beta)\frac{t}{\tau_{dSI}}].\label{eq:MtranS_t_d_n}\end{multline}
Note that the phase of the $\alpha$ RF pulse $\phi_{\alpha}$ has
dropped out of the equations through cancellation.

\section{Interpreting the results\label{sec:Interpreting-the-results}}

The equations \ref{eq:MtranI_t_d_p} and \ref{eq:MtranI_t_d_n} lead
to a series of peaks in a two dimensional spectrum. The $e^{i\,\sigma\,\omega_{0}(t-\tau_{echo})}$
term causes the shift in the directly detected $F_{2}$ dimension.
This term is missing from the $S$ magnetization which for simplicity
was made on resonance. Performing multiple acquisitions while incrementing
$\tau_{mix}$ provides the indirect, or $F_{1}$ dimension. The shift
in the $F_{1}$ dimension is determined by the $e^{i\,(m-1)\,\sigma\,\omega_{0}\tau_{mix}}$and
similar terms.

\begin{figure}
\begin{center}\includegraphics[%
  width=1.0\columnwidth,
  keepaspectratio]{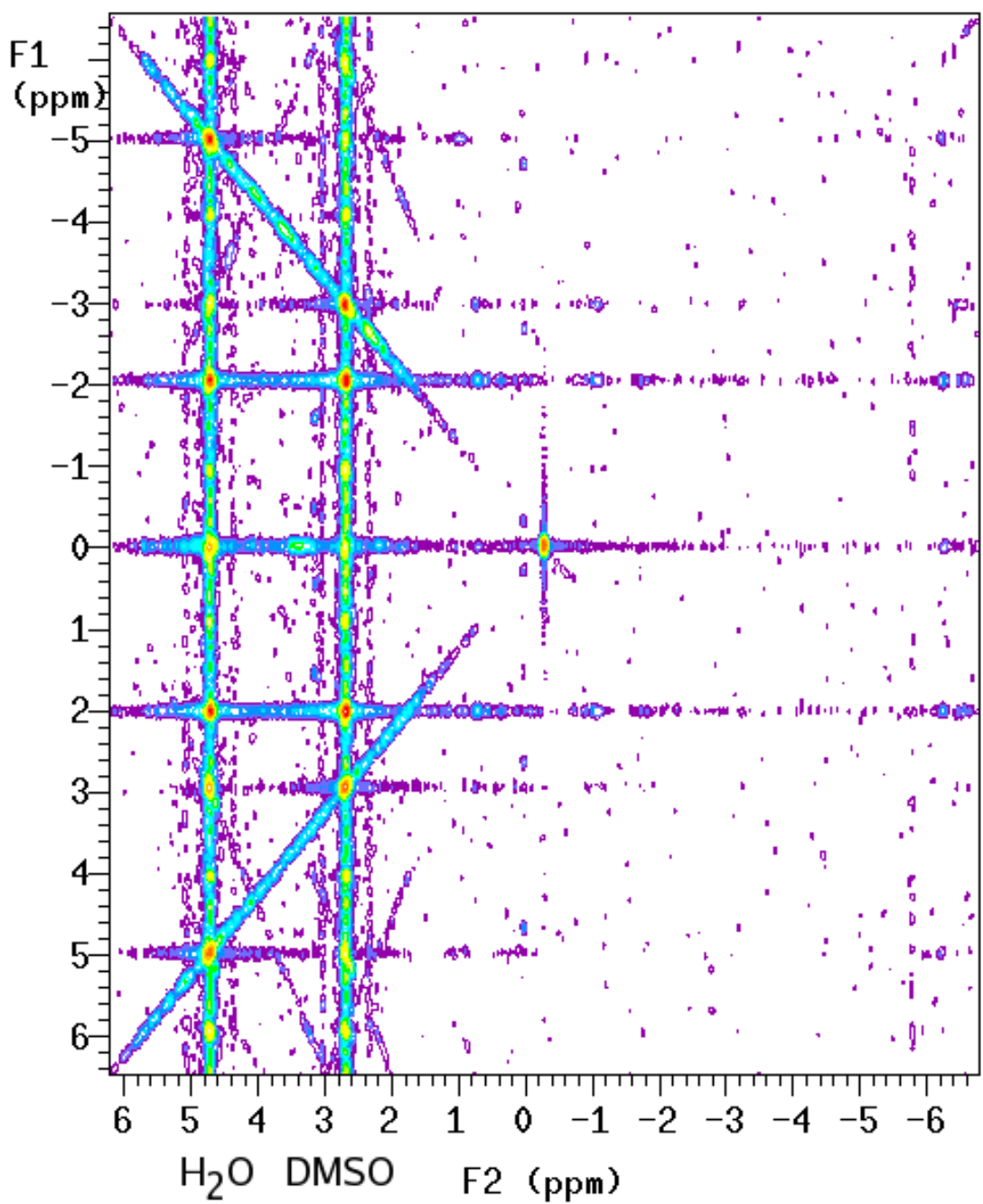}\end{center}

\caption[Example HOMOGENIZED spectrum of 50\% $H_{2}O$ (4.8 ppm) and 50\%
$DMSO$ (2.8 ppm) at 4 Tesla.]{\label{cap:HOMOGENIZED_h2o_dmso}Example HOMOGENIZED spectrum of
50\% $H_{2}O$ (4.8 ppm) and 50\% $DMSO$ (2.8 ppm) at 4 Tesla. $F_{2}$
resolution is 1024 points, $F_{1}$ resolution 256 points. $\beta=90^{\circ}$
yields symmetry along $F_{1}$. The diagonal lines and peaks that
lie on them are incompletely spoiled magnetization {}``artifacts''.
Vertical lines are the magnitude tails as well as{}``T1'' noise
which results from slight phase errors of pulses and incomplete crushing.
Horizontal lines are magnitude tails along $F_{2}$ and {}``zero
quantum'' noise.}
\end{figure}

\begin{figure}
\begin{center}\includegraphics[%
  width=1.0\columnwidth,
  keepaspectratio]{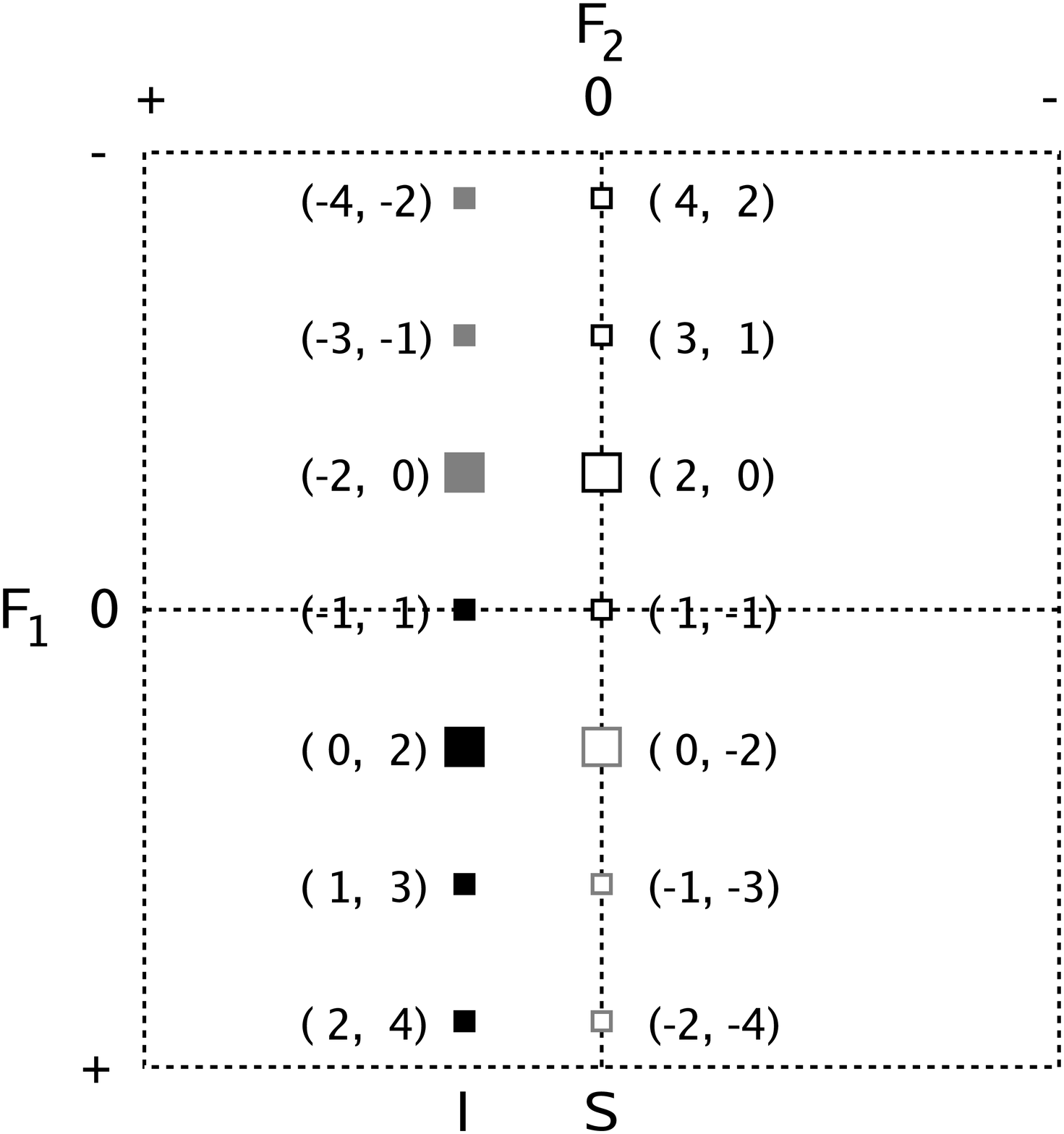}\end{center}

\caption[Schematic 2d HOMOGENIZED spectrum.]{\label{cap:schematic_generic} Schematic 2d HOMOGENIZED spectrum
- Cross peaks are labeled as $(m_{p},\, m_{n})$. The dominant peaks
(with $m_{p}$or $m_{n}=0$) are shown as large squares. Peaks with
major contribution of $p$ type for spin $I$ are filled in black.
$n$ type peaks are shown as solid gray. Peaks corresponding to $m\neq0$
are small squares. For spin $S$ the peaks are not filled.}
\end{figure}

There are theoretically an infinite number of peaks (which could alias
along the $F_{1}$ dimension) of each type $p$ or $n$ but in practice
relaxation will limit the number of peaks observed. Also the relative
concentration of $S$ and $I$ will limit the number, the largest
number of peaks observed when $S$ and $I$ magnetizations are in
the ratio 1 to 1 \cite{alw98b}. Those peaks corresponding to the
lowest order Bessel functions are most easily observed, as they build
the fastest, before relaxation and diffusion effects can attenuate
the signal. For example, if the $I$ spin is present in low concentration,
only the term $J_{0}[\Lambda(\hat{s})\, sin(\beta)\frac{t}{\tau_{dII}}]$
corresponding to $m=0$ will have significant amplitude for the $M^{I}$
cross peaks. We summarize in figure \ref{cap:schematic_generic} with
a corresponding experimental example in figure \ref{cap:HOMOGENIZED_h2o_dmso}.

\begin{figure}
\begin{center}\includegraphics[%
  width=1.0\columnwidth,
  keepaspectratio]{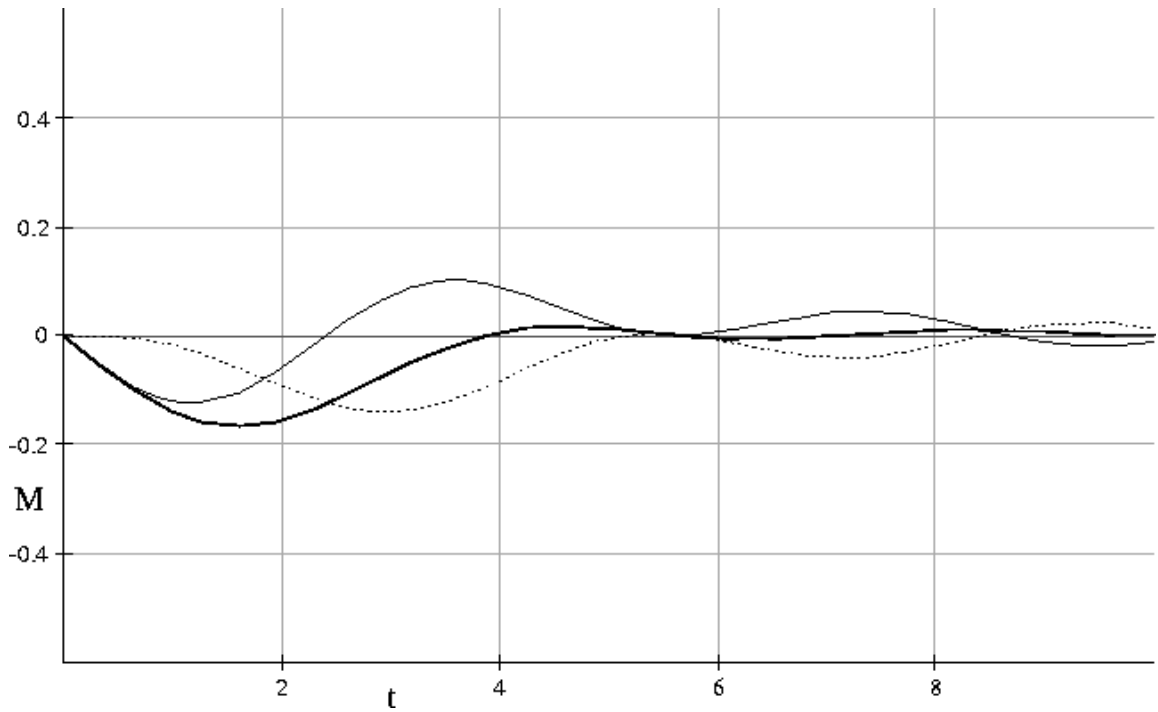}\end{center}

\caption[$I$ peak amplitude for $(m_{p},\, m_{n})=(0,\,2)$, $\beta=\frac{\pi}{2}$,
$M_{0}^{S}=1.0$, $M_{0}^{I}=1.0$.]{\label{cap:HOMOGENIZED_cross_0_2_90}$I$ peak amplitude for $(m_{p},\, m_{n})=(0,\,2)$,
$\beta=\frac{\pi}{2}$, $M_{0}^{S}=1.0$, $M_{0}^{I}=1.0$. The time
scale is arbitrary. The \textbf{heavy} curve is the sum of the $p$
and $n$ type contributions (net peak amplitude). The normal curve
is the $p$ type contribution. The dotted curve is the $n$ type contribution.
At short times this peak is dominated by the $p$ type signal.}
\end{figure}

\begin{figure}
\begin{center}\includegraphics[%
  width=1.0\columnwidth,
  keepaspectratio]{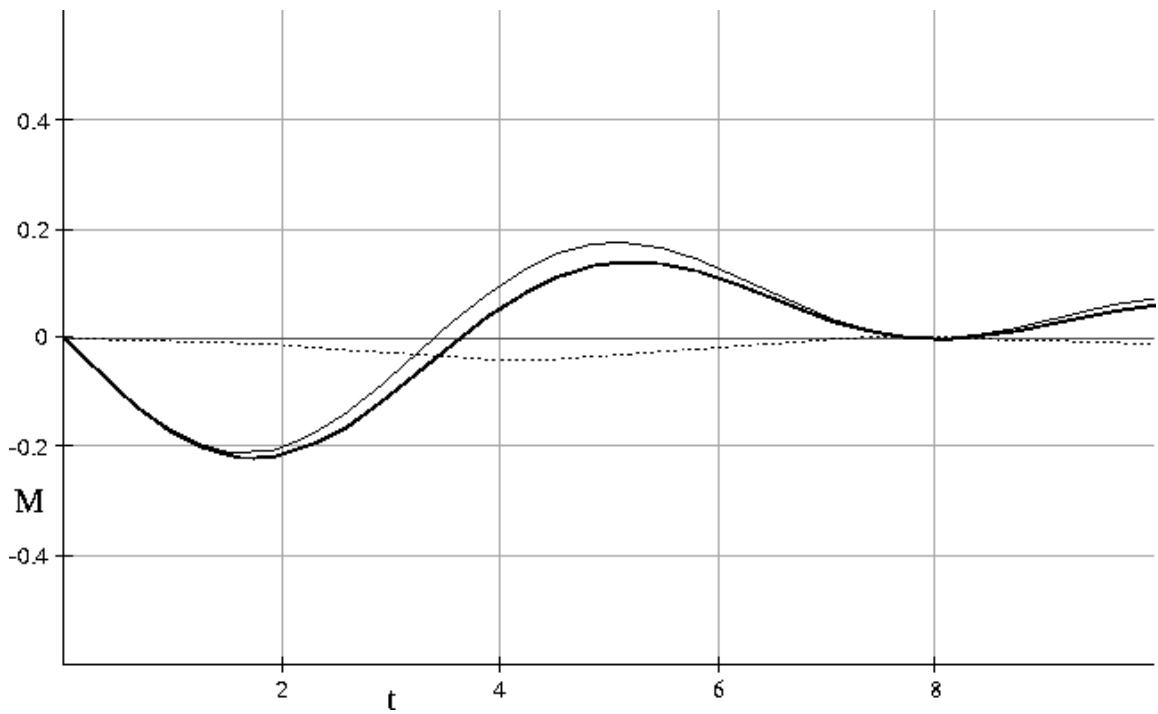}\end{center}

\caption[$I$ peak amplitude for $(m_{p},\, m_{n})=(0,\,2)$, $\beta=\frac{\pi}{4}$,
$M_{0}^{S}=1.0$, $M_{0}^{I}=1.0$.]{\label{cap:HOMOGENIZED_cross_0_2_45}$I$ peak amplitude for $(m_{p},\, m_{n})=(0,\,2)$,
$\beta=\frac{\pi}{4}$, $M_{0}^{S}=1.0$, $M_{0}^{I}=1.0$. Same labeling
as in figure \ref{cap:HOMOGENIZED_cross_0_2_90}. Changing $\beta$
has increased the $p$ type contribution in this peak and decreased
the $n$ type, raising the overall maximum amplitude.}
\end{figure}

\begin{figure}
\begin{center}\includegraphics[%
  width=1.0\columnwidth,
  keepaspectratio]{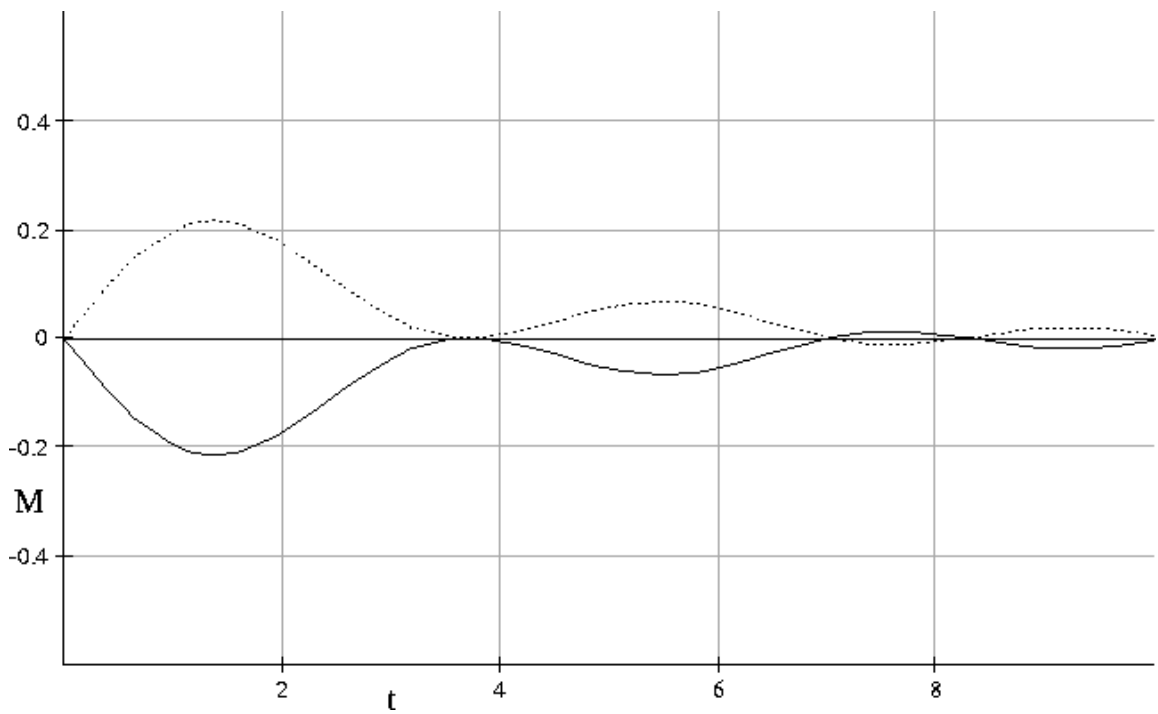}\end{center}

\caption[$I$ peak amplitude for $(m_{p},\, m_{n})=(-1,\,1)$, $\beta=\frac{\pi}{2}$,
$M_{0}^{S}=1.0$, $M_{0}^{I}=1.0$.]{\label{cap:HOMOGENIZED_cross_m1_1_90}$I$ peak amplitude for $(m_{p},\, m_{n})=(-1,\,1)$,
$\beta=\frac{\pi}{2}$, $M_{0}^{S}=1.0$, $M_{0}^{I}=1.0$. Same labeling
as in figure \ref{cap:HOMOGENIZED_cross_0_2_90}. This is a so called
{}``axial'' peak. The $p$ and $n$ type contributions cancel for
$\beta=\frac{\pi}{2}$.}
\end{figure}

Figures \ref{cap:HOMOGENIZED_cross_0_2_90}, \ref{cap:HOMOGENIZED_cross_0_2_45}
and \ref{cap:HOMOGENIZED_cross_m1_1_90} show the temporal behavior
of relative peak amplitudes for specific cases. For the case of $\beta=\pm45^{\circ}$the
$p$-type crosspeak magnitude is maximized, and for $\beta=\pm135^{\circ}$
the $n$ type crosspeak magnitude is maximized.

We can think of the terms of the type $\Lambda(\hat{s})\, sin(\beta)\frac{\tau_{echo}}{\tau_{dII}}$
inside the Bessel function\index{Bessel function} in equations \ref{eq:MtranI_t_d_p},
\ref{eq:MtranI_t_d_n}, etc. as a linear time proportional {}``unwinding''
parameter, which depends on field, concentration, and $\gamma$ (through
$M_{0})$, the flip angles (only $\beta$ in this case) and on the
applied gradient angle (through $\Lambda(\hat{s})$).


\begin{figure}
\begin{center}\includegraphics[%
  width=1.0\columnwidth,
  keepaspectratio]{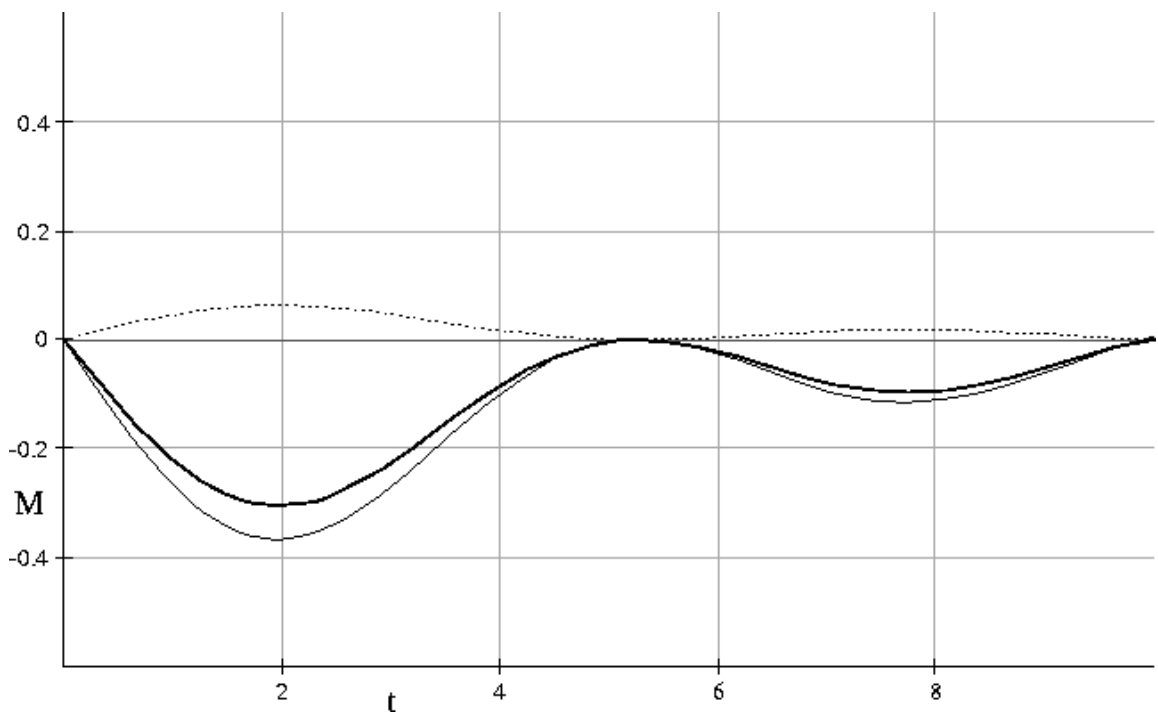}\end{center}

\caption[$I$ peak amplitude for $(m_{p},\, m_{n})=(-1,\,1)$, $\beta=\frac{\pi}{4}$,
$M_{0}^{S}=1.0$, $M_{0}^{I}=1.0$.]{\label{cap:HOMOGENIZED_cross_m1_1_45}$I$ peak amplitude for $(m_{p},\, m_{n})=(-1,\,1)$,
$\beta=\frac{\pi}{4}$, $M_{0}^{S}=1.0$, $M_{0}^{I}=1.0$. Same labeling
as in figure \ref{cap:HOMOGENIZED_cross_0_2_90}. This axial peak
can have non-zero amplitude even when the concentration of $S$ spins
is zero, due to $I,\, I$ spin interaction. }
\end{figure}

\section{Why HOMOGENIZED homogenizes...}

\begin{figure}
\begin{center}\includegraphics[%
  width=1.0\columnwidth,
  keepaspectratio]{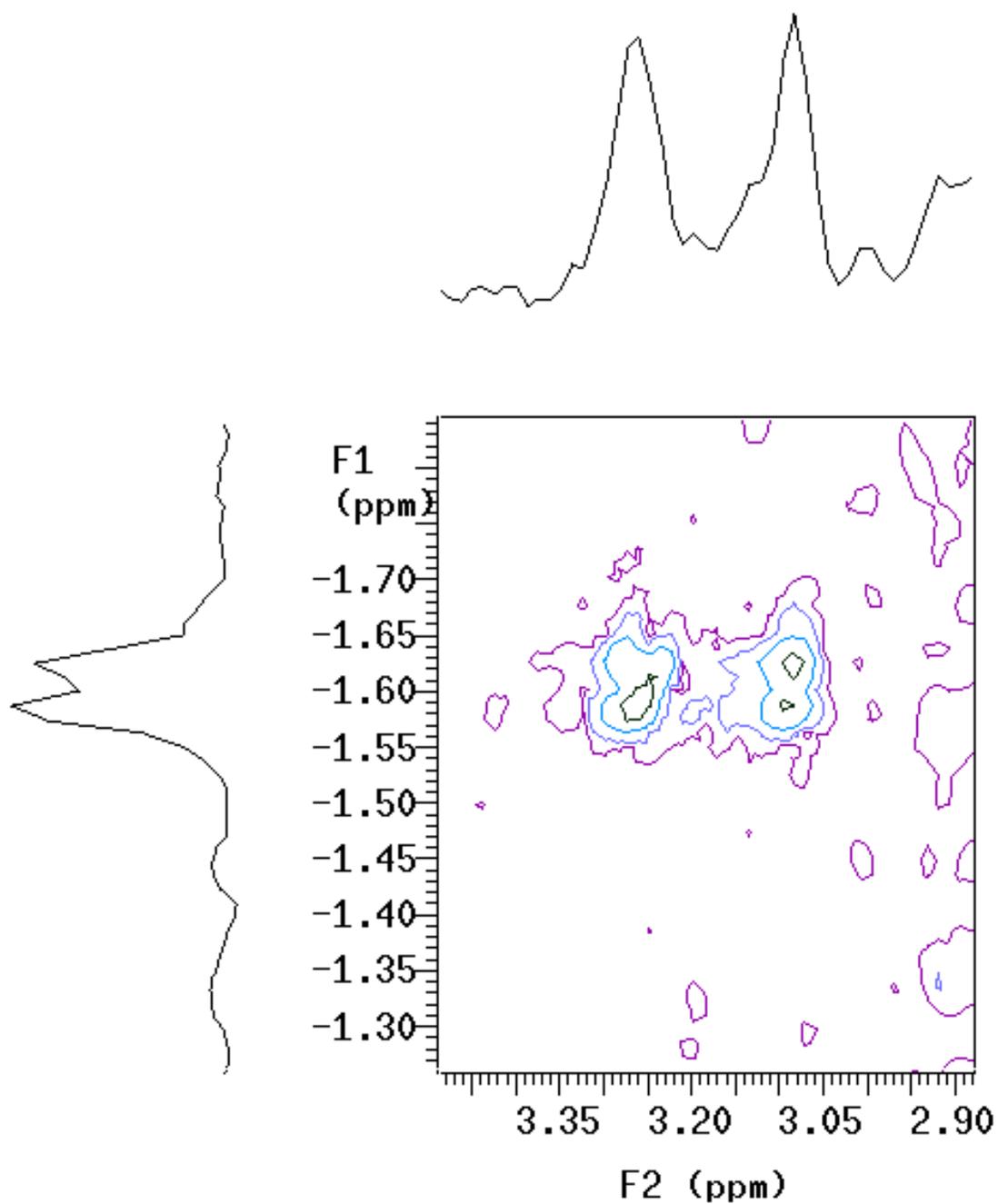}\end{center}

\caption[HOMOGENIZED spectrum of 99.9\% $H_{2}O$ (2.8 ppm) and \textasciitilde{}20
mM Choline Chloride (3.2 ppm) at 4 Tesla.]{\label{cap:HOMOGENIZEing}HOMOGENIZED spectrum of 99.9\% $H_{2}O$
(2.8 ppm) and \textasciitilde{}20 mM Choline Chloride (3.2 ppm) at
4 Tesla, showing the region around the (0, 2) Choline peak. $F_{2}$
resolution is 1024 points, $F_{1}$ resolution 512 points. The $X$
shim has been deliberately offset to give a very broad line in the
$F_{2}$ projection. It has actually split the peak into two peaks
(for unknown reasons, possibly spoiling during acquisition). The $F_{1}$projection
peak is much narrower.}
\end{figure}

The HOMOGENIZED signal results from refocusing due to the DDF. As
long as the DDF has not been perturbed significantly by any susceptibility
or inhomogeneity fields (during the $\tau_{mix}$ period) there will
not be significant broadening of the chemical shifts in the $F1$
dimension. This is not true of the $F2$ dimension, and we still have
$T_{2}^{*}$ effects determining the SNR of the acquired FID.

The condition can be stated as follows\[
\Delta B\,\tau_{mix}\ll G\,\delta\,\Delta s,\]
where $\Delta B\,\tau_{mix}$ is the total magnitude of field inhomogeneity
over the sample acting over the mixing time, and $G\,\delta\,\Delta s$
is the HOMOGENIZED gradient strength duration product multiplied by
the dimension of the sample. This condition means that the modulation
pattern is undisturbed by inhomogeneity.

Another way of looking at this effect is to say that only the inhomogeneity
on the scale of $q$ (see equation \ref{eq:homo_Gs_q}) matters, and
that HOMOGENIZED {}``shrinks'' the sample down to size $q$.

We show an example in figure \ref{cap:HOMOGENIZEing}.

\chapter{HOMOGENIZED WITH $T_{2}$ RELAXATION AND DIFFUSION\label{cha:ISMRMposter}%
\footnote{This chapter is expanded from ISMRM 2004, Poster 2323 \cite{CG04b}%
}}

\section{Introduction}

An analytical expression, equation (\ref{eq:M_tran_I_p_diffusion}),
for the HOMOGENIZED cross peak amplitude in the presence of diffusion
and $T_{2}$ relaxation has been developed%
\footnote{While this work was conceived of and executed independently, the author
is now aware of the work of I. Ardelean and collaborators in references
\cite{AK00b,AK00e}. Their analysis is similar, but covers the single
component case for double quantum DDF sequences.%
}. 

HOMOGENIZED \cite{VLW96} and its variants \cite{CHC+04,FB04} and
the recently proposed IDEAL \cite{ZCC+03} sequences have great potential
for in-vivo magnetic resonance spectroscopy \cite{FPH03} (MRS\index{MRS}).
Diffusion weighting in HOMOGENIZED is present both to give intentional
diffusion weighting and as a side effect of the various gradients
present. Stejskal-Tanner (ST) diffusion weighting \cite{ST65} during
the $\tau_{mix}$ and $\tau_{echo}$ periods of the sequence can also
be used to suppress radiation dampening. {}``Enhanced'' diffusion
weighting \cite{JZ01,ZCJZ01,ZCGL01} is obtained by reducing the DDF\index{DDF}
during $\tau_{echo}$. There is an additional $\tau_{echo}$ dependent
diffusion weighting possible, due to the iZQC\index{iZQC} (intermolecular
zero quantum coherence\index{intermolecular zero quantum coherencecoherence})
gradient $G_{zq}$ and $\beta$ pulse combination. The weighting results
from diffusing modulated longitudinal magnetization. Kennedy et al.
\cite{KRC+03} have shown recently that this diffusion weighting has
the novel property of being insensitive to object motion. $T_{2}$
relaxation also attenuates the signal.

It is desirable to have an analytical signal equation describing HOMOGENIZED
cross peaks. This is a necessary first step to using HOMOGENIZED for
quantitative in-vitro and in-vivo spectroscopy\index{in-vivo spectroscopy}.

\section{Step by Step HOMOGENIZED with $T_{2}$ Relaxation and Diffusion}

\begin{figure}
\includegraphics[%
  width=1.0\columnwidth,
  keepaspectratio]{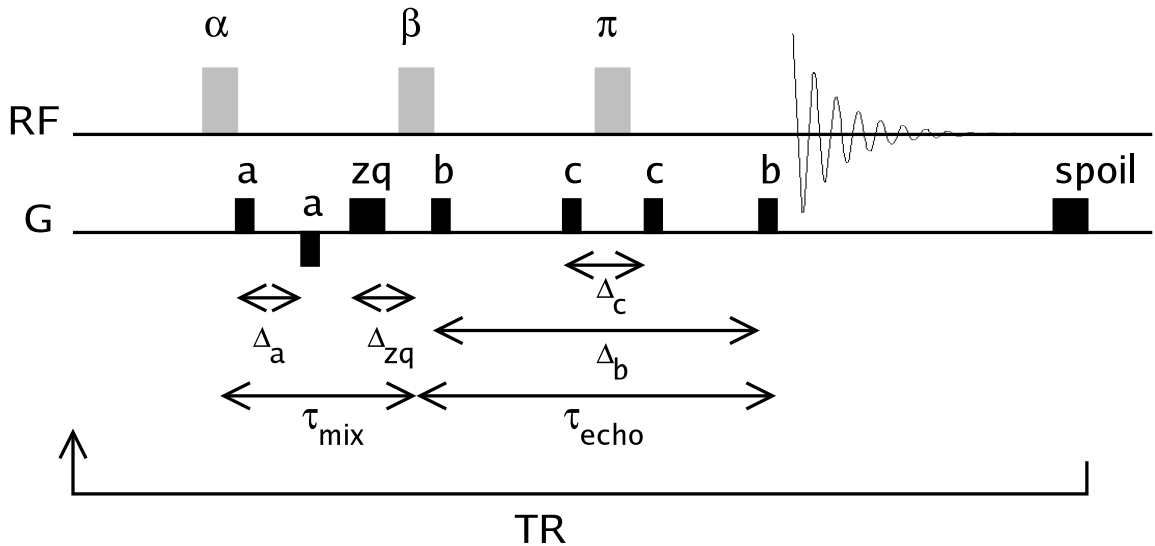}

\caption[HOMOGENIZED pulse sequence with crushing and diffusion weighting.]{\label{cap:Pulse-Sequence} HOMOGENIZED pulse sequence with Stejskal-Tanner
diffusion weighting during $\tau_{mix}$ and $\tau_{echo}$ and crusher
gradients on the $\pi$ RF pulse.}
\end{figure}

We will concentrate our discussion on the 2d HOMOGENIZED sequence
shown in figure \ref{cap:Pulse-Sequence}. This sequence is very similar
to the HOMOGENIZED sequence discussed in chapter \ref{cha:HOMOGENIZED}
and shown in figure \ref{cap:HOMOGENIZED_ps}. The difference is that
there are several additional gradient pairs to allow control of Stejskal
Tanner diffusion weighting during the $\tau_{mix}$ and $\tau_{echo}$
periods and to allow separate control of the diffusion weighting due
to $G_{zq}$. We have also added crusher gradients around the $\pi$
RF pulse.

\subsection{Excitation by the $\alpha$ pulse}

First, the HOMOGENIZED sequence excites the system with the $\alpha$
pulse. This is unchanged from section \ref{sub:alpha}. For simplicity
will assume that the system starts fully relaxed, and that we are
using $\alpha=90^{\circ}$with phase $\phi_{\alpha}$. One could substitute
the steady state values for $M^{I}$ and $M^{S}$ , and consider $\alpha\neq90^{\circ}$
which will lead to additional effects discussed in chapter \ref{cha:JMRpaper}.
The transverse and longitudinal magnetization after the $\alpha$
pulse are

\begin{equation}
M_{\perp}^{\alpha}=i\,[M_{0}^{I}+M_{0}^{S}]\, e^{i\,\phi_{\alpha}},\label{eq:homo_alpha_tran_again}\end{equation}
and\begin{equation}
M_{\parallel}^{\alpha}=0.\label{eq:homo_alpha_long_again}\end{equation}
This is for a homonuclear system where the hard pulse has bandwidth
to cover both $I$ and $S$ magnetization.

\subsection{$G_{a}$ gradients and delay $\Delta_{a}$}

Here is the first departure from chapter \ref{cha:HOMOGENIZED}. The
first $G_{a}$ is half of a gradient pair designed to give Stejskal-Tanner\index{Stejskal-Tanner}
diffusion weighting\index{diffusion weighting} to transverse magnetization
during a significant portion of the $\tau_{mix}$ time period $\Delta_{a}$.
It also serves to keep the transverse magnetization in a spoiled state
to reduce radiation damping\index{radiation damping} effects%
\footnote{as discussed in chapter \ref{cha:RADIATION-DAMPING}%
}.

The first $G_{a}$ gradient of duration $\delta_{a}$ and direction
$\hat{s}_{a}$results in

\begin{equation}
q=\frac{\gamma}{2\pi}G_{a}\delta_{a}\label{eq:homo_Gs_q}\end{equation}
for both $I$ and $S$ magnetization. The second $G_{a}$ gradient
has the same duration $\delta_{a}$ and opposite magnitude, refocusing
the magnetization giving $q=0$. The combined effects of the $G_{a}$
gradients and $\Delta_{a}$ delay is Stejskal-Tanner diffusion weighting
as discussed in section \ref{sec:Stejskal-Tanner-Sequence}. The $b-value$
for this diffusion weighting is\begin{equation}
b_{a}=\gamma^{2}G_{a}^{2}\delta_{a}^{2}(\Delta_{a}-\frac{\delta_{a}}{3}).\label{eq:homo_b_a}\end{equation}

\subsection{$G_{zq}$ gradient}

The $G_{zq}$ gradient (applied along direction $\hat{s}_{zq}$) has
multiple effects. It selects the desired zero-quantum coherence pathway
during $\tau_{mix}$ by twisting transverse magnetization, and ensuring
that only untwisting by the DDF (Distant Dipolar Field) leads to observable
signal at the end of the sequence. The $G_{zq}$ gradient also introduces
diffusion weighting. Finally it determines the spatial scale (correlation
distance\index{correlation distance}) of the DDF as discussed in
section \ref{sec:local-form}.

The diffusion weighting $b-value$ from $G_{zq}$ up until the $\beta$
pulse (duration $\Delta_{zq}$) is of the Stejskal-Tanner type, but
we omit the final gradient contribution (see table \ref{cap:Gqb})

\begin{equation}
b_{zq}=\gamma^{2}G_{zq}^{2}\delta_{zq}^{2}(\Delta_{zq}-\frac{2\,\delta_{zq}}{3}).\label{eq:homo_b_zq}\end{equation}

The spatial frequency $q_{zq}$ of transverse magnetization is now\begin{equation}
q_{zq}=\frac{\gamma}{2\pi}G_{zq}\delta_{zq}.\label{eq:homo_q_zq}\end{equation}

\subsection{$\tau_{mix}$ time period}

The $\tau_{mix}$ time period is inclusive of the $G_{a}$ gradient
pair, $G_{zq}$ gradient and their associated delays. We assume that
the duration of the $\alpha$ and $\beta$ RF pulses is short compared
to the other delays, and include half of each pulse duration in $\tau_{mix}$.
During these pulses and delays there is also $T_{2}$ relaxation (and
$T_{1}$ relaxation which we will neglect). We can now write the transverse
magnetization state immediately before the $\beta$ RF pulse\begin{equation}
M_{\perp}^{\tau_{mix}}=i\, e^{i\,\phi_{\alpha}}e^{-i\,2\,\pi\, q_{zq}\, s_{zq}}[M_{0}^{I}e^{i\,\omega_{0}\sigma\,\tau_{mix}}e^{-(b_{a}+b_{zq})\, D^{I}}e^{-\frac{\tau_{mix}}{T_{2}^{I}}}+M_{0}^{S}e^{-(b_{a}+b_{zq})\, D^{S}}e^{-\frac{\tau_{mix}}{T_{2}^{S}}}].\label{eq:homo_tau_mix_tran}\end{equation}
 $T_{2}$ is labeled similarly for each spin type. $\sigma$ is the
chemical shift difference. 

We have\begin{equation}
M_{\parallel}^{\tau_{mix}}=0\label{eq:homo_tau_mix_long}\end{equation}
for the longitudinal magnetization state, which is a valid approximation
when $\tau_{mix}\ll T_{1}.$

$D$ is the (isotropic%
\footnote{The following results can be generalized to anisotropic diffusion
by calculating and substituting the tensor product of the type $\mathbf{b:D}$
for the scalar terms $b\, D$.\cite[equation 5]{BML94} \cite[equation 2]{MBL94}%
}) diffusion coefficient, labeled with a superscript for each spin
type.

\subsection{$\beta$ Pulse}

The $\beta$ pulse forms modulated longitudinal magnetization, creating
a net DDF\index{DDF} which will refocus twisted magnetization during
$\tau_{echo}$. Immediately after the $\beta$ pulse of phase $\phi_{\beta}=-90^{\circ}$
or $-\hat{y}$ (which is considered to have $~0$ duration) we have

\begin{multline}
M_{\perp}^{\beta}=i\, M_{0}^{I}cos(\phi_{\alpha}+\tau_{mix}\sigma\,\omega_{0}-2\pi\, q_{zq}\, s_{zq})\, e^{-(b_{a}+b_{zq})\, D^{I}}e^{-\frac{\tau_{mix}}{T_{2}^{I}}}\\
-M_{0}^{I}sin(\phi_{\alpha}+\tau_{mix}\sigma\,\omega_{0}-2\pi\, q_{zq}\, s_{zq})\, cos(\beta)\, e^{-(b_{a}+b_{zq})\, D^{I}}e^{-\frac{\tau_{mix}}{T_{2}^{I}}}\\
+i\, M_{0}^{S}cos(\phi_{\alpha}-2\pi\, q_{zq}\, s_{zq})\, e^{-(b_{a}+b_{zq})\, D^{S}}e^{-\frac{\tau_{mix}}{T_{2}^{S}}}\\
-M_{0}^{S}sin(\phi_{\alpha}-2\pi\, q_{zq}\, s_{zq})\, cos(\beta)\, e^{-(b_{a}+b_{zq})\, D^{S}}e^{-\frac{\tau_{mix}}{T_{2}^{S}}}\label{eq:homo_beta_tran_again}\end{multline}
and

\begin{multline}
M_{\parallel}^{\beta}=-M_{0}^{I}sin(\phi_{\alpha}+\tau_{mix}\sigma\,\omega_{0}-2\pi\, q_{zq}s_{zq})\, sin(\beta)\, e^{-(b_{a}+b_{zq})\, D^{I}}e^{-\frac{\tau_{mix}}{T_{2}^{I}}}\label{eq:homo_beta_long_again}\\
-M_{0}^{S}sin(\phi_{\alpha}-2\pi\, q_{zq}s_{zq})\, sin(\beta)\, e^{-(b_{a}+b_{zq})\, D^{S}}e^{-\frac{\tau_{mix}}{T_{2}^{S}}}\end{multline}

\subsection{$G_{b}$ gradients, $\Delta_{b}$ delay, $G_{c}$ gradients and $\Delta_{c}$delay\label{sub:Delta_b}}

A lot is going on during the $\tau_{echo}$ period, the DDF is beginning
to refocus our desired signal. We have $T_{2}$ relaxation of transverse
magnetization, $T_{1}$ relaxation of longitudinal magnetization (which
we will again neglect), and attenuation due to diffusion for longitudinal
and transverse magnetization. For simplicity we will consider only
diffusional attenuation, and discuss DDF refocusing and $T_{2}$ relaxation,
which are separable, during our discussion of signal build during
$\tau_{echo}$ in section \ref{sub:tau_echo}.

We make the assumption that the diffusion attenuation due to the presence
of the spatially varying DDF field $\vec{B}_{d}$ (which has a spatially
varying gradient) is negligible, a point discussed in reference \cite[section I]{AKK01}.

The diffusion weighting becomes more complicated, as we are now concerned
with the longitudinal and transverse components, and we have applied
gradients in differing directions. This leads to increasingly complicated
expressions for $q$ and $b$.

The longitudinal magnetization is not affected by the $G_{b}$ or
$G_{c}$ gradients, and its $q$ stays as $q_{zq}$. We have attenuation
due to diffusion, like in the stimulated echo sequence (section \ref{sub:ste_tau_2_Delay}),
with $b-value$\begin{equation}
b_{\parallel}=\gamma^{2}G_{zq}^{2}\delta_{zq}^{2}\Delta_{b}.\label{eq:b_long}\end{equation}

The transverse magnetization is affected by the $G_{b}$and $G_{c}$gradient
pairs, and for simplicity we will assume that $\hat{s}_{zq}\bot\hat{s}_{b}\bot\hat{s}_{c}$.
In this case the attenuation can be described by independent $b-values$
from each gradient pair\begin{equation}
b_{\bot}=b_{\bot zq}+b_{\bot b}+b_{\bot c}\label{eq:b_tran}\end{equation}
with\begin{equation}
b_{\bot zq}=\gamma^{2}G_{zq}^{2}\delta_{zq}^{2}\Delta_{b},\label{eq:b_tran_b_zq}\end{equation}

\begin{equation}
b_{\bot b}=\gamma^{2}G_{b}^{2}\delta_{b}^{2}(\Delta_{b}-\frac{\delta_{b}}{3}),\label{eq:b_tran_b_b}\end{equation}
and\begin{equation}
b_{\bot c}=\gamma^{2}G_{c}^{2}\delta_{c}^{2}(\Delta_{c}-\frac{\delta_{c}}{3}).\label{eq:b_tran_c}\end{equation}
Note that we have included the last $\delta_{b}$ time period in $b_{\bot b,\, b}$.

\subsection{$\tau_{echo}$\label{sub:tau_echo} and final magnetization components}

During the $\tau_{echo}$ time period we have $T_{1}$ relaxation
of the longitudinal magnetization (which we will neglect for now),
$T_{2}$ relaxation of the transverse magnetization, and the DDF re-phasing
of our desired signal. The $\pi$ pulse also has the effect of inverting
longitudinal magnetization, and the $x$ component of the transverse
magnetization. From chapter \ref{cha:HOMOGENIZED} we know that the
$\pi$ RF pulse does not reverse the signal re-phasing due to the
DDF. In addition we have the diffusion weighting discussed in section
\ref{sub:Delta_b}.

We end up with the following transverse magnetization%
\footnote{Chemical shift is refocused at $\tau_{echo}$ but will reappear during
the acquisition period $t2$.%
}

\begin{multline}
M_{\perp}^{\tau_{echo}}=i\, M_{0}^{I}cos(\phi_{\alpha}+\tau_{mix}\sigma\,\omega_{0}-2\pi\, q_{zq}\, s_{zq})\, e^{-(b_{a}+b_{zq}+b_{\bot})\, D^{I}}e^{-\frac{\tau_{mix}+\tau_{echo}}{T_{2}^{I}}}\\
+M_{0}^{I}sin(\phi_{\alpha}+\tau_{mix}\sigma\,\omega_{0}-2\pi\, q_{zq}\, s_{zq})\, cos(\beta)\, e^{-(b_{a}+b_{zq}+b_{\bot})\, D^{I}}e^{-\frac{\tau_{mix}+\tau_{echo}}{T_{2}^{I}}}\\
+i\, M_{0}^{S}cos(\phi_{\alpha}-2\pi\, q_{zq}\, s_{zq})\, e^{-(b_{a}+b_{zq}+b_{\bot})\, D^{S}}e^{-\frac{\tau_{mix}+\tau_{echo}}{T_{2}^{S}}}\\
+M_{0}^{S}sin(\phi_{\alpha}-2\pi\, q_{zq}\, s_{zq})\, cos(\beta)\, e^{-(b_{a}+b_{zq}+b_{\bot})\, D^{S}}e^{-\frac{\tau_{mix}+\tau_{echo}}{T_{2}^{S}}}\label{eq:homo_tau_echo_tran}\end{multline}

The longitudinal component is\begin{multline}
M_{\parallel}^{\tau_{echo}}=M_{0}^{I}sin(\phi_{\alpha}+\tau_{mix}\sigma\,\omega_{0}-2\pi\, q_{zq}s_{zq})\, sin(\beta)\, e^{-(b_{a}+b_{zq})\, D^{I}}e^{-\frac{\tau_{mix}}{T_{2}^{I}}}\, e^{-b_{\parallel}D^{I}}\\
+M_{0}^{S}sin(\phi_{\alpha}-2\pi\, q_{zq}s_{zq})\, sin(\beta)\, e^{-(b_{a}+b_{zq})\, D^{S}}e^{-\frac{\tau_{mix}}{T_{2}^{S}}}\, e^{-b_{\parallel}D^{S}}\label{eq:homo_tau_echo_long}\end{multline}
We have explicitly placed $e^{-b_{\parallel}D}$ separately in each
term as we will need to consider its value (which attenuates the DDF)
throughout the $\tau_{echo}$ period (and subsequent acquisition period
$t2$) rather than just its final value.

\section{Signal}

First, in order to obtain an analytical solution for a system of biological
interest, we assume that $M_{0}^{I}\ll M_{0}^{S}$. Making this assumption
implies that only the DDF\index{DDF} due to $M^{S}$ leads to significant
refocused signal and we can neglect the DDF due to $M^{I}$. This
collapses the sums in equation \ref{eq:MtranI_t_d_p} (and similarly
for the others) and leads to only the terms with $J_{0}(\Lambda(\hat{s})\, sin(\beta)\frac{t}{\tau_{dII}})\approx1$
surviving, since $\Lambda(\hat{s})\, sin(\beta)\frac{t}{\tau_{dII}}\approx0$
when $\tau_{dII}\rightarrow\infty$. Terms of the type $J_{n}(\Lambda(\hat{s})\, sin(\beta)\frac{t}{\tau_{dII}})\approx0$,
for $n\neq0$.

We can define some more terms that will help us see the effects of
diffusion and $T_{2}$ relaxation.

\begin{equation}
\tau_{dIS,\, eff}\equiv\tau_{dIS}e^{(b_{a}+b_{zq})\, D_{S}}e^{\frac{\tau_{mix}}{T_{2}^{S}}}\label{eq:tau_dIS_eff}\end{equation}
$\tau_{dIS,\, eff}$ (\ref{eq:tau_dIS_eff}) has been defined to take
account of $T_{2}$ and diffusion losses (ST b-values, $b_{a}$ and
$b_{zq}$) incurred during $\tau_{mix}$ before $\beta$ beta forms
modulated $M_{\parallel}$. $\tau_{dIS}$ is the dipolar demagnetization
time for spin $S$ defined in equation \ref{eq:tau_dIS}. 

\begin{equation}
F_{IS}(\tau_{echo})\equiv\frac{1-e^{-\tau_{echo}\,(2\pi\, q_{zq})^{2}D_{S}}}{\tau_{dIS,\, eff}(2\pi\, q_{zq})^{2}D_{S}}=\frac{1}{\tau_{dIS,\, eff}}\int_{0}^{\tau_{echo}}e^{-t\,(2\pi\, q_{zq})^{2}D_{S}}dt\label{eq:F}\end{equation}
accounts for the decay of longitudinal magnetization (and the DDF)
and can be thought of as an exponentially slowing ''unwinding''
parameter, instead of the linear time proportional unwinding parameter
as discussed in section \ref{sec:Interpreting-the-results} when diffusion
is negligible. It is the integral of the exponentially decaying DDF
during $\tau_{echo}$.

The expression for the signal amplitude in the presence of diffusion
and $T_{2}$ decay is

\begin{multline}
M_{\perp p}^{I}=-i\, M_{0}^{I}[\frac{cos(\beta)+1}{2}]\, e^{i\,\sigma\,\omega_{0}\tau_{mix}}e^{-(b_{a}+b_{zq}+b_{\perp})\, D_{I}}\\
e^{-\frac{(\tau_{mix}+\tau_{echo})}{T_{2}^{I}}}J_{1}[\frac{2}{3}\Lambda(\hat{s}_{zq})\, sin(\beta)\, F_{IS}(\tau_{echo})]\label{eq:M_tran_I_p_diffusion}\end{multline}
where $M_{\perp p}^{I}$ is the p-type cross peak amplitude. $b_{a}$
, $b_{zq}$, and $b_{\perp}$ are the ST b-values defined in equations
\ref{eq:homo_b_a}, \ref{eq:homo_b_zq}, and \ref{eq:b_tran}.

\begin{figure}
\includegraphics[%
  width=1.0\columnwidth,
  keepaspectratio]{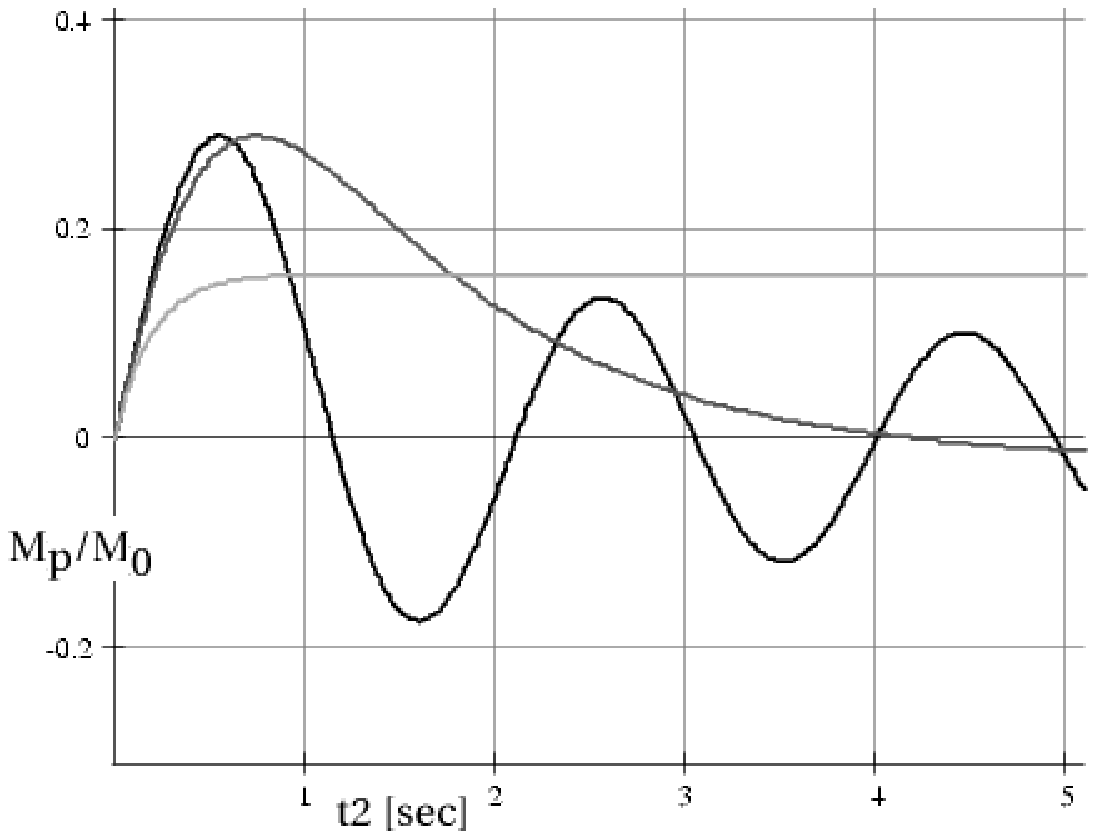}

\caption[Plot of theoretical cross peak amplitude $M_{\perp p}^{I}$ vs. $\tau_{echo}$,
for the case of negligible $T_{2}$ decay.]{\label{cap:Plot--theoretical}Plot of theoretical cross peak amplitude
$M_{\perp p}^{I}$ vs. $t2$, for the case of negligible $T_{2}$
decay. $\beta=90^{\circ}$ and $\tau_{S}=200ms$. Three situations
are shown:}

Black - negligible diffusion

\textcolor{dgray}{Dark Gray} - diffusion of $M_{\parallel}$ has
delayed the maximum and stretched the zero crossings to longer times.

\textcolor{lgray}{Light Gray} - $M_{\parallel}$ modulation has
completely diffused away before the maximum can be obtained.
\end{figure}

The effect of $F_{IS}(\tau_{echo})$ is to stretch the time axis when
diffusion weighting is significant. Equation (\ref{eq:M_tran_I_p_diffusion})
is valid as long as $S$ and $I$ are separated by $1/\tau_{S}$ in
frequency, so that only longitudinal $S$ magnetization contributes
to signal build. Steady state values $(TR<5\, T_{1}^{S}or\, T_{1}^{I})$
may be used for $\tau_{dIS}$, $M_{0}^{I}$, and $M_{0}^{S}$ as long
as diffusion has eliminated residual spatial modulation of longitudinal
magnetization\cite{CG04}. As long as the $a$ and $b$ gradient areas
are chosen correctly, radiation dampening\label{radiation damping}
is not significant. Three theoretical situations are shown in figure
\ref{cap:Plot--theoretical}.

We can also write the expressions for the other peaks of interest\begin{multline}
M_{\perp n}^{I}=i\, M_{0}^{I}[\frac{cos(\beta)-1}{2}]\, e^{i\,\sigma\,\omega_{0}\tau_{mix}}e^{-(b_{a}+b_{zq}+b_{\perp})\, D_{I}}\\
e^{-\frac{(\tau_{mix}+\tau_{echo})}{T_{2}^{I}}}J_{1}[\frac{2}{3}\Lambda(\hat{s}_{zq})\, sin(\beta)\, F_{IS}(\tau_{echo})]\label{eq:M_tran_I_n_diffusion}\end{multline}
\begin{multline}
M_{\perp p}^{S}=-i\, M_{0}^{S}[\frac{cos(\beta)+1}{2}]e^{-(b_{a}+b_{zq}+b_{\perp})\, D_{S}}e^{-\frac{(\tau_{mix}+\tau_{echo})}{T_{2}^{S}}}J_{1}[\Lambda(\hat{s}_{zq})\, sin(\beta)\, F_{SS}(\tau_{echo})]\label{eq:M_tran_S_p_diffusion}\end{multline}

\begin{multline}
M_{\perp n}^{S}=i\, M_{0}^{S}[\frac{cos(\beta)-1}{2}]e^{-(b_{a}+b_{zq}+b_{\perp})\, D_{S}}e^{-\frac{(\tau_{mix}+\tau_{echo})}{T_{2}^{S}}}J_{1}[\Lambda(\hat{s}_{zq})\, sin(\beta)\, F_{SS}(\tau_{echo})]\label{eq:M_tran_S_n_diffusion}\end{multline}
Note that the $S$ magnetization $p$ and $n$-type peaks (which appear
on the $f1=0$ axis) overlap and will cancel when $\beta=90^{\circ}$.

\begin{figure}
\includegraphics[%
  width=1.0\columnwidth,
  keepaspectratio]{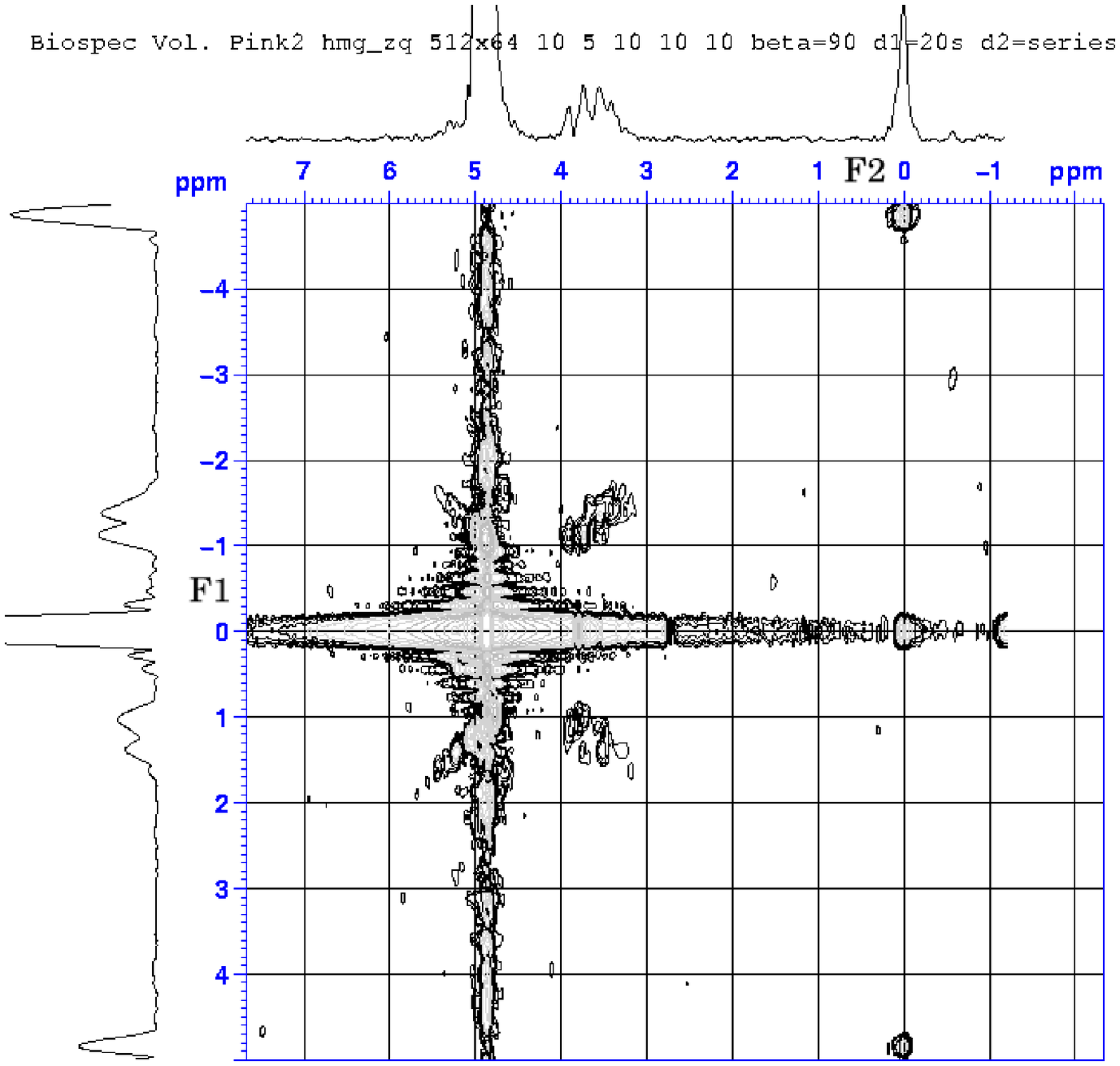}

\caption[Representative low resolution 2d HOMOGENIZED spectrum.]{\label{cap:spectrum}Representative low resolution 2d HOMOGENIZED
spectrum. TSP is referenced to -4.7ppm on F1 axis and 0.0ppm on F2
Axis. Projections are restricted to {[}0, 4{]} ppm F2 and {[}-5, -1{]}
ppm F1.}
\end{figure}

\section{Experimental Results}

\begin{figure}
\includegraphics[%
  width=1.0\columnwidth,
  keepaspectratio]{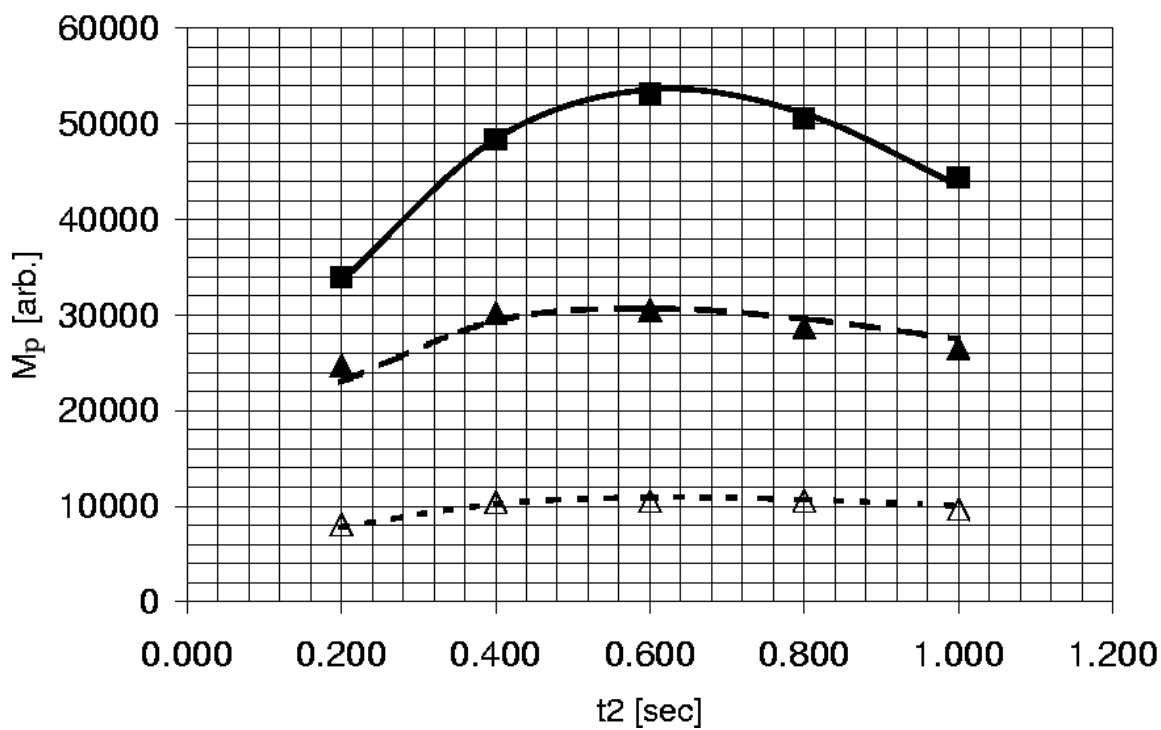}

\caption[Data points and theoretical curve of p type TSP peak for three cases.]{\label{cap:Data-points}Data points and theoretical curve of p type
TSP peak for three cases. Y axis arbitrary units.Data points and theoretical
curve of p type TSP peak for three cases. Y axis arbitrary units.}

$\alpha=\beta=90^{\circ}$, $\delta_{a}=\delta_{b}=\delta_{c}=1ms$,
$\delta_{spoil}=5ms$

$G_{a}=G_{b}=G_{c}=G_{spoil}=20\frac{mT}{m}$, $\delta_{zq}=3ms$

Upper - $TR=20s,\, G_{zq}=10\frac{mT}{m}$

Middle - $TR=20s,\, G_{zq}=40\frac{mT}{m}$

Lower - $TR=2s,\, G_{zq}=40\frac{mT}{m}$
\end{figure}

A series of low resolution (512x64) HOMOGENIZED spectra were obtained
with various strengths of $G_{zq}$ (see figure \ref{cap:spectrum}).
The solvent (S) is water at room temperature, the solute of interest
(I) was TSP at 100mM concentration. Glucose was also present in solution.
Field strength is 4.7T yielding nominal $\tau_{S}=200ms$. A best
fit, adjusting $M_{0}^{I}$and $\tau_{S}$ to account for pulse imperfections
and $B_{1}$ inhomogeneity, was obtained for the top curve, and kept
the same for the other curves. Relaxation rates were measured in separate
inversion recovery and spin-echo experiments with $T_{1}^{S}=2.57s$,
$T_{2}^{S}=140ms$ and $T_{2}^{I}=1.62s$. Effects such as $B_{1}$
inhomogeneity and RF pulse error contribute to lengthen $\tau_{Seff}$
(reduce available S magnetization). Comparison of the predicted cross
peak amplitude with experiment is shown in figure \ref{cap:Data-points}.

\chapter{SPATIALLY VARYING STEADY STATE LONGITUDINAL MAGNETIZATION\index{longitudinal magnetization}
\label{cha:JMRpaper}%
\footnote{The material in this chapter has also appeared as an arXive.org preprint
\cite{CG04d} and has been published \cite{CG04e} in the Journal
of Magnetic Resonance.%
}}

\section{Introduction}

NMR\index{NMR} and MRI\index{MRI} sequences utilizing the Distant
Dipolar Field (DDF) have the relatively unique property of preparing,
utilizing, and leaving spatially-modulated longitudinal magnetization,
$M_{z}(s)$, where \(\hat{s}\) is in the direction of an applied
gradient. In fact this is fundamental to producing the novel {}``multiple
spin-echo''\cite{DBD79,BBG90} or {}``non-linear stimulated echo''
\cite{ASDK97} of the classical picture and making the {}``intermolecular
multiple quantum coherence (iMQC\index{iMQC})'' \cite{QH93} observable
in the quantum picture.

Existing analytical signal equations for DDF/iMQC sequences depend
on $M_{z}(s)$ being sinusoidal during the signal build period\cite{alw98b,CG04b}.
Experiments that probe sample structure also require a well-defined
{}``correlation distance\index{correlation distance}'' which is
defined as the repetition distance of $M_{z}(s)$ \cite{RB96,WAM+98,ACM02}.
If the repetition time $TR$ of the DDF sequence is such that full
relaxation is not allowed to proceed $TR<5T_{1}$, or diffusion does
not average out the modulation, spatially-modulated longitudinal magnetization
will be left at the end of one iteration of the sequence. The next
repetition of the sequence will begin to establish {}``harmonics''
in what is desired to be a purely sinusoidal modulation pattern. Eventually
a steady state is established, potentially departing significantly
from a pure sinusoid. 

\begin{figure}
\includegraphics[%
  width=5.5in,
  keepaspectratio]{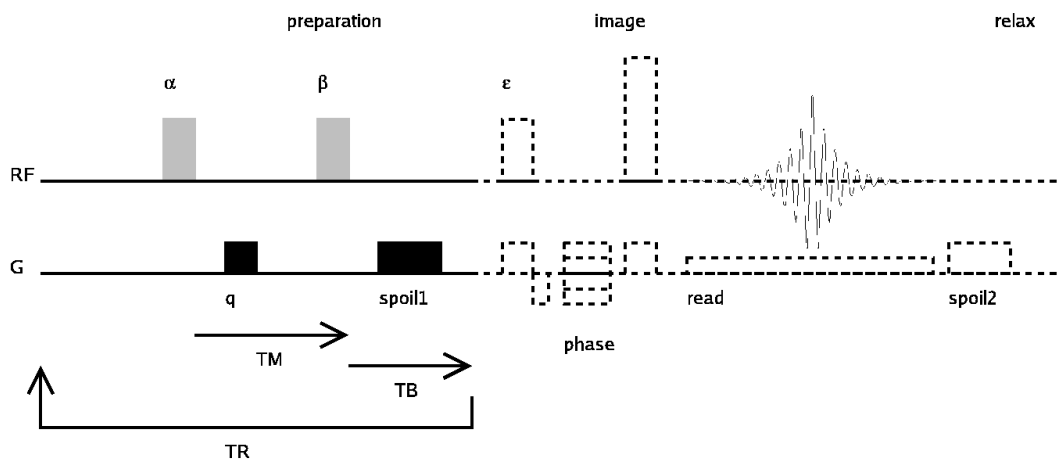}

\caption[Pulse Sequence for measuring $M_{z}^{SS}(s)$.]{\label{fig:Pulse-Sequence}Pulse Sequence for measuring $M_{z}^{SS}(s)$.
All RF pulses shown as hard for simplicity are actually Sinc3. $\alpha$
and $\beta$ are the same phase.}
\end{figure}

\section{Experimental Methods\label{sec:Experimental-Methods}}

In order to study the behavior of the steady state $M_{z}^{SS}(s)$
profile we have implemented a looped DDF preparation subsequence followed
by a standard multiple-phase encode imaging\index{imaging} sub-sequence.
(Figure \ref{fig:Pulse-Sequence}.) The $\alpha$ pulse excites the
system, the gradient $G_{q}$ twists the transverse magnetization
into a helix. $\beta$ rotates one component of the helix back into
the longitudinal direction. For simplicity we have omitted the $180^{\circ}$
pulses used to create a spin-echo during TM and/or TB sometimes present
in DDF sequences. Also, we are only interested in $M_{z}(s)$ in this
experiment, not the actual DDF-generated transverse signal. Looping
the {}``preparation'' sub-sequence thus creates the periodic $M_{z}(s)$
profile, spoils remaining transverse magnetization, and establishes
$M_{z}^{SS}(s)$. The $\varepsilon$ pulse converts $M_{z}^{SS}(s)$
into transverse magnetization, allowing it to be imaged via the subsequent
spin-echo {}``image'' sub-sequence. $M_{z}^{SS}(s)$ must be re-established
by the {}``preparation'' sub-sequence for each phase encode. After
a suitably long full relaxation delay {}``relax,'' the sequence
is repeated to acquire the next k-space line. This is clearly a slow
acquisition method because many $TR$ periods are required to reach
steady state in the preparation before each k-space line is acquired.
The sequence is intended as a tool to directly image the $M_{z}^{SS}(s)$
profile, verifying the $M_{z}^{SS}(s)$ that would occur in a steady
state DDF sequence, not as a new imaging modality. 

\begin{figure}
\includegraphics[%
  width=5.5in,
  keepaspectratio]{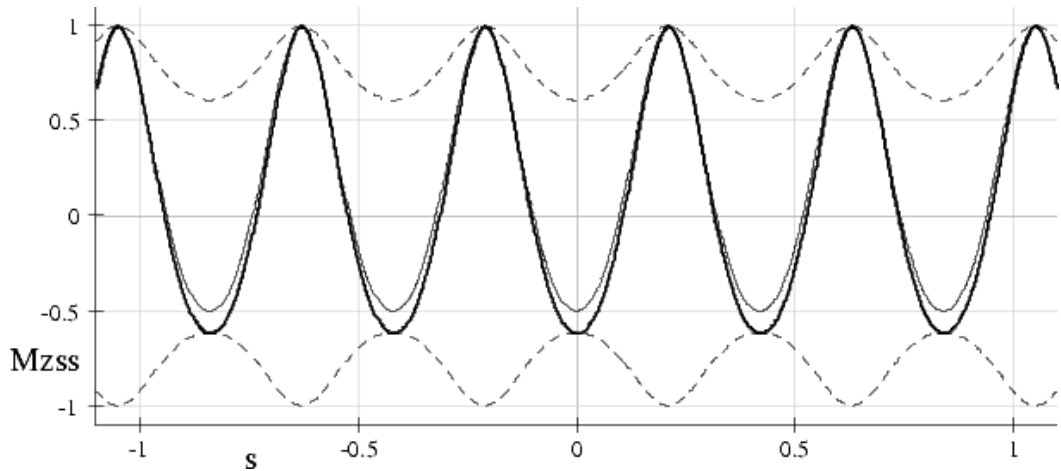}

\caption[Theoretical values of $M_{z}(s)$. ]{\label{fig:Mz_theory_image}Theoretical values of $M_{z}(s)$. $M_{z}^{SS}(s)$is
shown dashed $---$ as an envelope, $M_{z}^{SS,\,\beta}(s)$ is shown
as a heavy line, $M_{z}^{SS,\, TB}(s)$ as a normal line. $\alpha=\beta=90^{\circ},\  TR=2s,\  TM=0ms,\  TB=100ms,\ \  T_{1}=1.4s$}
\end{figure}

\section{Theory}

The effect of the ''preparation'' pulse sequence was first determined
for a single iteration. The progress along the sequence is denoted
by the the superscript.

Starting with fully relaxed equilibrium magnetization before the $\alpha$
pulse:\begin{equation}
M_{z}^{Eq}(s)=M_{0}\label{eq:initial}\end{equation}
after the $\alpha$ pulse, the mix delay $TM$ and the $\beta$ pulse
we have:\begin{align}
M_{z}^{\beta}(s) & =[A^{\beta}cos(q\, s)+B^{\beta}]\, M_{z}^{Eq}+C^{\beta}M_{0}\label{eq:beta}\end{align}
\[
A^{\beta}=-sin(\alpha)\, e^{-\frac{TM}{T_{2}}}sin(\beta)\]
\[
B^{\beta}=cos(\alpha)\, e^{-\frac{TM}{T_{1}}}cos(\beta)\]
\[
C^{\beta}=(1-e^{-\frac{TM}{T_{1}}})\, cos(\beta)\]
The parameter $q=\frac{2\pi}{\lambda}$, where $\lambda$ is the helix
pitch resulting from the applied gradient. Diffusion has been assumed
to be negligible at the scale of $\lambda$. Note that $T_{2}$ is
used in $A$ rather than $T_{2}^{*}$ when $G_{q}$ is larger than
background inhomogeneity and susceptibility gradients.

After the build delay $TB$ we have:

\begin{equation}
M_{z}^{TB}(s)=[A^{TB}cos(q\, s)+B^{TB}]\, M_{z}^{Eq}(s)+C^{TB}M_{0}\label{eq:TE}\end{equation}
\[
A^{TB}=-sin(\alpha)\, e^{-\frac{TM}{T_{2}}}sin(\beta)\, e^{-\frac{TB}{T_{1}}}\]
\[
B^{TB}=cos(\alpha)\, e^{-\frac{TM}{T_{1}}}cos(\beta)\, e^{-\frac{TB}{T_{1}}}\]
\[
C^{TB}=[(1-e^{-\frac{TM}{T_{1}}})\, cos(\beta)-1]\, e^{-\frac{TB}{T_{1}}}+1\]

At the start of the next repetition, after a $TR$ period inclusive
of $TM$ and $TB$ we have\begin{equation}
M_{z}^{TR}(s)=[A^{TR}cos(q\, s)+B^{TR}]M_{z}^{Eq}(s)+C^{TR}\, M_{0}\label{eq:one}\end{equation}
\[
A^{TR}=-sin(\alpha)\, e^{-\frac{TM}{T_{2}}}sin(\beta)\, e^{-\frac{TR-TM}{T_{1}}}\]
\[
B^{TR}=cos(\alpha)\, cos(\beta)\, e^{-\frac{TR}{T_{1}}}\]
\[
C^{TR}=[(1-e^{-\frac{TM}{T_{1}}})\, cos(\beta)-1]\, e^{-\frac{TR-TM}{T_{1}}}+1\]

If we apply the sequence $N$ times and re-arrange the terms we get
the series:\begin{equation}
M_{z}^{NxTR}(s)=M_{0}+M_{0\,}[A^{TR}cos(q\, s)+B^{TR}+C^{TR}-1]\sum\limits _{n=1}^{N}[A^{TR}cos(q\, s)+B^{TR}]^{n-1}\label{eq:sum}\end{equation}
for the starting magnetization state after $N$ repetitions of the
sequence.

Summing an infinite number of terms results in the expression for
the steady state $M_{z}^{SS}(s)$ after a large number of TR periods:\begin{equation}
M_{z}^{SS}(s)=M_{0}-M_{0\,}[\frac{A^{TR}cos(q\, s)+B^{TR}+C^{TR}-1}{A^{TR}cos(q\, s)+B^{TR}-1}]\label{eq:steady_state}\end{equation}

One can then calculate the magnetization state after the $\beta$
pulse in the steady state:\begin{equation}
M_{z}^{SS,\,\beta}(s)=[A^{\beta}cos(q\, s)+B^{\beta}]\, M_{z}^{SS}(s)+C^{\beta}M_{0}\label{eq:ss_beta}\end{equation}

and after $TB$:

\begin{equation}
M_{z}^{SS,\, TB}(s)=[A^{TB}cos(q\, s)+B^{TB}]\, M_{z}^{SS}(s)+C^{TB}M_{0}\label{eq:ss_tb}\end{equation}

We show graphs of equations {[}\ref{eq:steady_state}{]}, {[}\ref{eq:ss_beta}{]},
and {[}\ref{eq:ss_tb}{]} in Figure \ref{fig:Mz_theory_image} for
$TR=2s$.

\begin{figure}
\includegraphics[%
  width=5.5in,
  keepaspectratio]{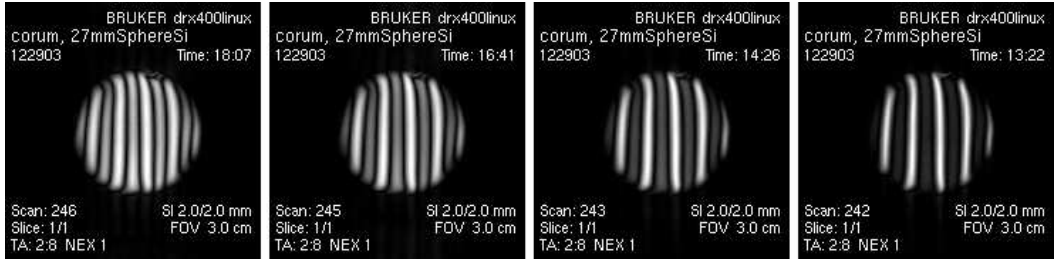}

\caption[$M_{z}^{SS}(s)$ images.]{\label{fig:Mz_images}$M_{z}^{SS}(s)$ images, $TR=5s,\ 2s,\ 1s,\ 500ms$
from left to right. $TM=TB=~7ms,\  relax=10s$.}
\end{figure}

\section{Results}

\begin{figure}
\includegraphics[%
  width=5.5in,
  keepaspectratio]{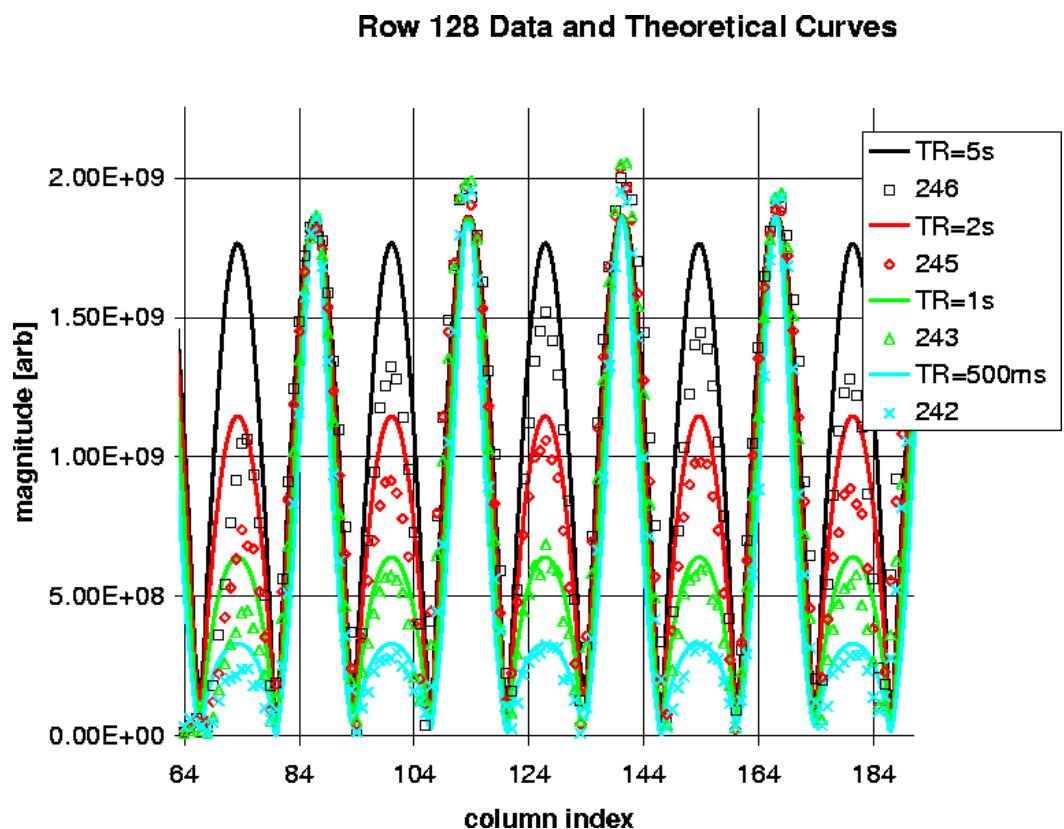}

\caption[Comparison of theory and experiment for steady state $M_{Z}$.]{\label{fig:Mz_fit}Comparison of theory and experiment for steady
state $M_{Z}$. Row 128 data (points) and Fit (lines), $\alpha=\beta=90^{\circ},\  TR=2s,\  TM=TB=7ms,\ \  T_{1}=1.4s\  relax=10s$. }
\end{figure}

We now show in Figure \ref{fig:Mz_images} representative $M_{z}^{SS}(s)$
magnitude images obtained with the sequence described in section \ref{sec:Experimental-Methods}
for four different values of $TR=5s,\ 2s,\ 1s,\ 500ms$. Figure \ref{fig:Mz_fit}
shows several cross sections through row \#128 of Figure \ref{fig:Mz_images}.
The object is an 18mm glass sphere filled with silicone oil. Data
points are superimposed with the corresponding magnitude of the theoretical
curve. The $T_{1}$ of the silicone oil (at 400MHz) was measured by
spectroscopic inversion recovery to be 1.4s. A Bruker DRX400 Micro
2.5 system was used with a custom 27mm diameter 31P/1H birdcage coil.
10 $TR$ periods were used to establish steady state. A 10s {}``relax''
delay was used between phase encodes to establish full relaxation.
$G_{q}$ was 3ms and 2.5mT/mm, with $G_{spoil1}$ of 5ms and 100mT/mm.
No attempt was made to account for $B_{1}$ inhomogeneity. A single
scaling parameter was used for all theoretical curves. We achieved
good agreement with the theoretical predictions. In the sequence as
used, $TM=TB=~7ms$. A variety of other $G_{q}$ directions and strengths
show similar agreement with theory. Better agreement in the fit between
experiment and theory can be obtained with $\alpha=\beta=75^{\circ}$than
with the nominal $90^{\circ}$. A $B_{1}$ map needs to be determined
to see if this corresponds more closely to the actual experimental
conditions.

\section{Conclusions}

The expressions developed and verified above should be useful to those
wishing to understand or utilize harmonics in the $M_{z}^{SS}(s)$
profile in DDF based sequences in the situation where the diffusion
distance during $TR$ compared with $\lambda$ in negligible. This
is especially true for those carrying out structural measurements
which depend on a well defined correlation distance. The theory should
also hold for spatially-varying magnetization density $M_{0}=M_{0}(\vec{r})$,
and longitudinal relaxation $T_{1}=T_{1}(\vec{r})$.

\chapter{THE FUTURE OF DDF\index{DDF} NMR AND HOMOGENIZED}

Research into DDF effects has only been ongoing for just over a decade.
There is still much more to be learned about the DDF and DDF based
applications in MRI\index{MRI} and MRS\index{MRS}. We can identify
several fruitful lines of research that still need more investigation.

There is still a lot to be learned about the imaging properties of
DDF based sequences, such as the point spread function, contrast mechanisms,
and whether the signal is truly {}``local'' to a voxel. There have
been a number of interesting imaging applications, beyond the initial
work of try it and see what it looks like. A very intriguing application
is {}``Multiple-Quantum Vector Imaging\index{multiple quantum vector imaging}''
which is a fancy term for utilization of the gradient direction to
detect the orientation of sub-voxel structures \cite{BW05}.

Much work has been done on using the DDF to image porous structures,
and in-vivo there has been much interest in quantifying trabecular
bone density\cite{CCA+01,SCMA+01,ACM02,SC02,BCT+03,CTB+03b}. This
work continues.

The author's (as well as at least one other research group's) work
has recently focused on adding localization to HOMOGENIZED in order
to get spectra from a voxel in-vivo. There have been some initial
successes \cite{CG05,DBCF05}. The signal equations developed in this
dissertation, and extensions, should be useful for quantification
of metabolites utilizing these new localized HOMOGENIZED sequences.

There have been recent improvements \cite{CHC+04,BF04b} to HOMOGENIZED,
utilizing selective pulses on the solvent $S$ to suppress water and
boost crosspeak signal.

Quantification of HOMOGENIZED peaks (in vitro and in vivo) is still
an active and needed research topic. Continuing the author's work
and the work of Ardelean \cite{AKK01} should help quantify the effects
of $T_{1}$ relaxation during the $\tau_{mix}$ and $\tau_{echo}$
time periods of HOMOGENIZED.

Related to the issue of quantification is determining HOMOGENIZED's
sensitivity to pulse errors, which is magnified by DDF refocusing.
Using HOMOGENIZED with adiabatic pulses should help reduce this issue,
and has recently been demonstrated \cite{CG05}.

HOMOGENIZED based Spectroscopic Imaging\index{spectroscopic maging}
is an intriguing possibility. It would have many advantages, such
as its self referencing properties which eliminate the need for frequency
shift correction (and phasing).

\part{APPENDICES}

\appendix

\chapter{SOME DERIVATIONS}

\section{Equilibrium Magnetization\label{sec:Equilibrium-Magnetization}}

\begin{figure}
\begin{center}\includegraphics[%
  width=0.50\columnwidth,
  keepaspectratio]{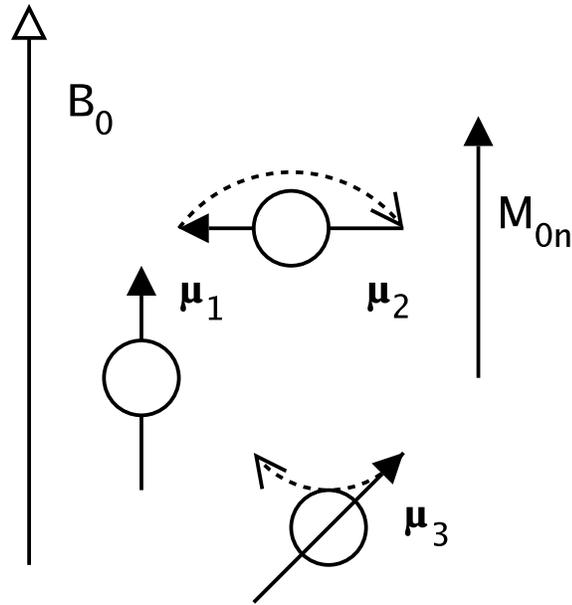}\end{center}

\caption{{\footnotesize \label{cap:moments}}Individual Nuclear Magnetic Moments
in a Magnetic Field}
\end{figure}
Any sample in an NMR experiment is composed of a large number of identical
nuclei loosely coupled to each other and the external environment
(Figure\ref{cap:moments}.) The dominant coupling is to an external
applied magnetic field $\vec{B_{0}}$. Each nucleus possesses magnetic
potential energy\begin{equation}
E_{n}=-\vec{\mu}_{n}\cdot\vec{B}_{0}.\label{eq:nucleus_energy}\end{equation}
The field will tend to cause the moments of the nuclei to align parallel
to the field, minimizing the total magnetic energy

\begin{equation}
E=-\sum_{N}\vec{\mu}_{n}\cdot\vec{B}_{0}.\label{eq:energy}\end{equation}
In competition with the field, thermal excitation will tend to randomize
the alignment. 

The net macroscopic magnetic moment per unit volume is defined as

\begin{equation}
\vec{M}\equiv\frac{1}{V}\sum_{N}\vec{\mu}_{n}.\label{eq:net_moment}\end{equation}

Since there is coupling to the environment, the sample/environment
system will eventually come to thermal equilibrium. The equilibrium
magnetization can be found by the Boltzmann law of statistical mechanics\cite[Ch I, A. p. 2]{Abr99}
$P(E_{m})\propto e^{-E_{m}/k\, T}$ where $P(E_{m})$ is the probability
of finding a nuclear moment in energy state $E_{m}$, $T$ is the
absolute temperature, and $k=1.3806505\times10^{-23}[\frac{J}{K}]$
\cite{CTGoFC02} is the Boltzmann constant. The number of energy levels
is determined by the total spin quantum number $I$ and {}``z''
component $m$ with $I\geq m\geq-I$ with $\Delta m=1$. We have

\begin{equation}
\vec{<M>_{eq}}=\frac{1}{V}\sum_{N}[\sum_{m=-I}^{I}\vec{\mu_{n}}P(E_{n})\,/\sum_{m=-I}^{I}P(E_{n})].\label{eq:M_eq1}\end{equation}
Assuming that the field $B_{0}$ is oriented in the $\hat{z}$ direction,
only the $\hat{z}$ component of $\vec{\mu}$ effects the energy,
and all orthogonal directions of $\vec{\mu}$ are equally probable,
averaging to zero. The $z$ component quantum number $m$ then determines
the potential energy, and $\vec{<M>_{eq}}=M_{0}\hat{z}$ is oriented
along the $z$ axis. We also make the substitution $\mu=\left|\vec{\mu}\right|$=$\gamma\hbar I$
where $\gamma$ is the magnetogyric ratio. Since $N$ is a very large
number we are in effect taking an ergodic average. We can assume that
$M_{0}$ does not fluctuate, and drop the expectation brackets to
get

\begin{equation}
M_{0}=\frac{\mu}{V}\sum_{N}[\sum_{m=-I}^{I}m\, e^{\frac{m\,\gamma\,\hbar\, B_{0}}{k\, T}}/\sum_{m=-I}^{I}e^{\frac{m\,\gamma\,\hbar\, B_{0}}{k\, T}}].\label{eq:M_eq2}\end{equation}
Applying the so-called high temperature approximation\begin{equation}
k\, T\gg\gamma\,\hbar\, B_{0},\label{eq:ht_condition}\end{equation}
valid for nearly all but ultra-cold temperature and ultrahigh fields,
yields\begin{equation}
e^{\frac{m\,\gamma\,\hbar\, B_{0}}{k\, T}}\approx1+\frac{m\,\gamma\,\hbar\, B_{0}}{k\, T}.\label{eq:ht_approx}\end{equation}
Substituting, we get\begin{equation}
M_{0}\approx\frac{\gamma\,\hbar}{V}\sum_{N}[\sum_{m=-I}^{I}m+\frac{m^{2}\gamma\,\hbar\, B_{0}}{k\, T}/\sum_{m=-I}^{I}1+\frac{m\,\gamma\,\hbar\, B_{0}}{k\, T}].\label{eq:M_eq3}\end{equation}
Carrying out the summation operations gives\begin{equation}
M_{0}\approx\frac{\gamma^{2}\hbar^{2}N\, B_{0}I\,(I+1)}{3\, V\, k\, T}.\label{eq:M_eq4}\end{equation}
Note since $\frac{\hbar\, B_{0}}{k\, T}\ll1$, $\vec{<M>_{eq}}\ll\frac{N\,\gamma\,\hbar\, I}{V}$,
is much less than the theoretical maximum achievable magnetization
(saturation magnetization) at low temperature or ultrahigh field.
For $^{1}H$ water $\frac{N}{V}=2[protons]\times55.56\times10^{-3}[mol\, cm^{-3}]\times N_{A}[mol^{-1}]=6.6918\times10^{22}[protons\, cm^{-3}]$,
$\gamma_{p}=2.675\times10^{8}$ at $T=310\, K$ (body temperature)
and $B_{0}=3\, T$ the ratio $\frac{M_{0}}{M_{sat}}=9.89\times10^{-6}$$\approx10^{-5}$
a very small fraction.

We can also use the above to define the nuclear magnetic susceptibility
$\chi$. From the relation

\[
M_{0}=\frac{\chi}{\mu_{0}}B_{0},\]
we get

\[
\chi=\frac{\mu_{0}\gamma^{2}\hbar^{2}N\, I\,(I+1)}{3\, V\, k\, T},\]
where $\mu_{0}=4\pi\times10^{-7}\frac{N}{A^{2}}$is the permeability
of free space. A quick check will show that $\chi$ is indeed dimensionless.

\section{Cross Product\index{cross-product} with $M_{\parallel}$ and $M_{\perp}$\label{sec:Cross-Product-with-Ml-and-Mt}}

Of most interest for the Bloch equations is the cross product of $\vec{M}\times\vec{B}$
in terms longitudinal and transverse components. In Cartesian coordinates
the cross product is defined as\begin{equation}
\vec{M}\times\vec{B}=\left|\begin{array}{ccc}
M_{x} & M_{y} & M_{z}\\
B_{x} & B_{y} & B_{z}\\
\hat{x} & \hat{y} & \hat{z}\end{array}\right|=(M_{y}B_{z}-M_{z}B_{y})\,\hat{x}-(M_{x}B_{z}-M_{z}B_{x})\,\hat{y}+(M_{x}B_{y}-M_{y}B_{x})\,\hat{z}.\label{eq:MxB}\end{equation}

In the complex representation we have\begin{equation}
\begin{array}{cc}
M_{\parallel}=M_{z} & \; B_{\parallel}=B_{z}\end{array}\label{eq:Ml_Bl}\end{equation}
and\begin{equation}
\begin{array}{cc}
M_{\perp}=M_{x}+i\, M_{y} & \; B_{\perp}=B_{x}\end{array}+i\, B_{y}.\label{eq:Mt_Bt}\end{equation}

We note that the longitudinal and transverse components of the cross
product are\begin{equation}
[\vec{M}\times\vec{B}]_{\parallel}=M_{x}B_{y}-M_{y}B_{x}\label{eq:MxB_par}\end{equation}
and\begin{equation}
[\vec{M}\times\vec{B}]_{\perp}=(M_{y}B_{z}-M_{z}B_{y})-i\,(M_{x}B_{z}-M_{z}B_{x}).\label{eq:MxB_perp}\end{equation}
By substituting\begin{equation}
\begin{array}{cc}
M_{x}=\frac{1}{2}(M_{\perp}+M_{\perp}^{*})\; & M_{y}=\frac{i}{2}(M_{\perp}-M_{\perp}^{*})\end{array},\label{eq:Mx_My}\end{equation}
and similarly for $B_{x}$and $B_{y}$ we find\begin{equation}
[\vec{M}\times\vec{B}]_{\parallel}=\frac{i}{2}(M_{\perp}B_{\perp}^{*}-M_{\perp}^{*}B_{\perp})\label{eq:Ml_result}\end{equation}
and

\begin{equation}
[\vec{M}\times\vec{B}]_{\perp}=i\,(M_{\parallel}B_{\perp}-M_{\perp}B_{\parallel}).\label{eq:Mt_result}\end{equation}

\section{Fourier Transform of $\frac{\Lambda(\vec{r})}{r^{3}}$\label{sec:Fourier-Transform-of-kernel}}

This is a derivation of fundamental importance to distant dipolar
field theory and calculation. The result in the form used was first
published in Deville et al. 1979 \cite{DBD79} who references Leggett
\cite[Appendix II]{Leg75} for the derivation. The derivation in Leggett
is somewhat terse, so we carry it out here in detail with some variation
and much enhancement for completeness%
\footnote{Thanks go to E. Clarkson for suggesting the spherical harmonic addition
theorem.%
}.

Starting with the general form of the transform\begin{equation}
\mathcal{F}_{3}\{\frac{\Lambda(\vec{r})}{r^{3}}\}\equiv\int_{\infty}d^{3}r\, e^{-i\,2\pi\,\vec{\rho}\cdot\vec{r}}\frac{\Lambda(\vec{r})}{r^{3}},\label{eq:Lambda_transform_general}\end{equation}
we put it into spherical polar coordinates\begin{equation}
\mathcal{F}_{3}\{\frac{\Lambda(\vec{r})}{r^{3}}\}=\int_{0}^{2\pi}d\phi\,\int_{0}^{\pi}d\theta\, sin(\theta)\,\int_{0}^{\infty}dr\, r^{2}e^{-i\,2\pi\,\vec{\rho}\cdot\vec{r}}\frac{\Lambda(\vec{r})}{r^{3}}.\label{eq:Lambda_transform_spherical}\end{equation}
We then constrain $\vec{\rho}=\rho\,\hat{z}$ and simplify to get\begin{equation}
\mathcal{F}_{3}\{\frac{\Lambda(\vec{r})}{r^{3}}\}=\int_{0}^{2\pi}d\phi\,\int_{0}^{\pi}d\theta\, sin(\theta)\,\int_{0}^{\infty}\frac{dr}{r}\, e^{-i\,2\pi\,\rho\, r\, cos(\theta)}\Lambda(\vec{r}).\label{eq:Lambda_transform_rho_z}\end{equation}
We note that $\Lambda(\vec{r})=P_{2}[cos(\theta)]$ where $P_{n}$
is the $n$th Legendre polynomial\index{Legendre polynomial}. We
now recognize the integral representation of the Spherical Bessel
function\index{Spherical Bessel function} \cite[10.1.14, p. 438]{AS74}
of order $n$ with $z=-2\pi\, r\,\rho$ and $n=2$, which is\begin{equation}
j_{n}(z)=\frac{(-i)^{n}}{2}\int_{0}^{\pi}d\theta\, sin(\theta)\, e^{i\, z\, cos(\theta)}P_{n}[cos(\theta)].\label{eq:Spherical_Bessel_function}\end{equation}
 Substitution leaves us with\begin{equation}
\mathcal{F}_{3}\{\frac{\Lambda(\vec{r})}{r^{3}}\}=-2\int_{0}^{2\pi}d\phi\,\int_{0}^{\infty}\frac{dr}{r}\, j_{2}(-2\pi\, r\,\rho),\label{eq:Lambda_transform_no_theta}\end{equation}
where $j_{n}$ can be generated from \cite[10.1.25, p. 439]{AS74}\begin{equation}
j_{n}(z)=z^{n}(-\frac{1}{z}\frac{\partial}{\partial z})^{n}[\frac{sin(z)}{z}].\label{eq:jn_generating_formula}\end{equation}
The first few spherical Bessel functions are\begin{equation}
j_{0}(z)=\frac{sin(z)}{z},\label{eq:j0}\end{equation}
\begin{equation}
j_{1}(z)=-\frac{cos(z)}{z}+\frac{sin(z)}{z^{2}}\label{eq:j1}\end{equation}
and\begin{equation}
j_{2}(z)=-3\frac{cos(z)}{z^{2}}+\frac{(3-z^{2})\, sin(z)}{z^{3}}.\label{eq:j2}\end{equation}
We can evaluate the integral by integrating both sides of the recurrence
relations \cite[10.1.21-22, p. 439]{AS74} obtaining the identity\begin{equation}
\int_{0}^{\infty}dz\,\frac{j_{n}(z)}{z}=\frac{n-2}{n+1}\int_{0}^{\infty}dz\,\frac{j_{n-2}(z)}{z}-\frac{1}{n+1}[j_{n-2}(z)-j_{n}(z)]_{0}^{\infty}.\label{eq:jn_over_z_recurrence}\end{equation}
Specifically we have

\begin{equation}
\int_{0}^{\infty}dz\,\frac{j_{2}(z)}{z}=\frac{0}{3}\int_{0}^{\infty}dz\,\frac{j_{0}(z)}{z}-\frac{1}{3}[j_{0}(z)-j_{2}(z)]_{0}^{\infty}=\frac{1}{3},\label{eq:j2_over_z_recurrence}\end{equation}
leading to\begin{equation}
\mathcal{F}_{3}\{\frac{\Lambda(\vec{r})}{r^{3}}\}=-\frac{4\pi}{3},\label{eq:Lambda_transform_no_theta_result}\end{equation}
and remembering the condition\begin{equation}
\vec{\rho}=\rho\,\hat{z}.\label{eq:rho_z}\end{equation}

For the case of general $\vec{\rho}$ it is easier to consider rotation
of the function being transformed, leaving $\vec{\rho}=\rho\,\hat{z}$.
We consider rotation of $\Lambda(\vec{r})$ around an arbitrary pair
of angles $\theta_{0}$ and $\phi_{0}$ leading to\begin{equation}
\Lambda_{rot}(\vec{r})=P_{2}[cos(\alpha)],\label{eq:Lambda_rot}\end{equation}
with\begin{equation}
cos(\alpha)=cos(\theta)\, cos(\theta_{0})+sin(\theta)\, sin(\theta_{0})\, cos(\phi-\phi_{0}).\label{eq:cos_alpha}\end{equation}
We need only consider $\theta_{0}$ and can set $\phi_{0}=0$ without
loss of generality due to the azimuthal symmetry of $\Lambda(\vec{r})$.

We use the spherical harmonic addition theorem \cite[8.794 1., p. 1013]{GRG+80}%
\footnote{The form in \cite[8.794 1., p. 1013]{GRG+80} can be confusing. The
series is explicitly infinite, but when one evaluates the $\Gamma$
or factorial functions in the definition of $Y_{l\, m}$ the series
is actually finite.%
}, \cite[eq. (5.83), p. 257 ]{BF92}\begin{equation}
P_{l}[cos(\alpha)]=\frac{4\pi}{2l+1}\sum_{m=-l}^{l}Y_{l\, m}^{*}(\theta_{0},\,\phi_{0})\, Y_{l,\, m}(\theta,\,\phi)\label{eq:spherical _harmonic_addition_theorem}\end{equation}
where the spherical harmonics are defined \cite[eq. (5.75), p. 255]{BF92}
as\begin{equation}
Y_{l\, m}(\theta,\,\phi)\equiv(-1)^{m}\left[\frac{2l+1}{4\pi}\frac{(l-m)!}{(l+m)!}\right]^{1/2}P_{l}^{m}[cos(\theta)]\, e^{i\, m\,\phi}\label{eq:spherical_harmonic}\end{equation}
with the condition\begin{equation}
m\geq0\label{eq:m_gr_0}\end{equation}
and further definition\begin{equation}
Y_{l,\,-m}(\theta,\,\phi)\equiv(-1)^{m}Y_{l\, m}^{*}(\theta,\,\phi).\label{eq:spherical_harmonic_neg_m}\end{equation}
After substitution of $l=2$ and $\phi_{0}=0$ and application of
the above definition we have\begin{multline}
P_{2}[cos(\alpha)]=\frac{4\pi}{5}\{ Y_{2,\,2}(\theta_{0},\,0)\,[Y_{2,\,2}(\theta,\,\phi)-Y_{2,\,2}^{*}(\theta,\,\phi)]\\
+Y_{2,\,1}(\theta_{0},\,0)\,[Y_{2,\,1}(\theta,\,\phi)-Y_{2,\,1}^{*}(\theta,\,\phi)]+Y_{2,\,0}(\theta_{0},\,0)\, Y_{2,\,0}(\theta,\,\phi)\}.\label{eq:P2_expand}\end{multline}
We note that the $\phi$ dependence of the spherical harmonic term
is\begin{equation}
Y_{l,\, m}(\theta,\,\phi)-Y_{l,\, m}^{*}(\theta,\,\phi)\sim cos(m\,\phi)\label{eq:cos_phi_depend}\end{equation}
and that when evaluated in the $\phi$ integral we have\begin{equation}
\int_{0}^{2\pi}cos(m\,\phi)\, d\phi=[_{0}^{2\pi}\frac{sin(m\,\phi)}{m}=0\label{eq:cos_phi_integral}\end{equation}
for integer $m\neq0$. The only term that survives the $\phi$ integration
is the $\phi$ independent part\begin{equation}
\frac{4\pi}{5}Y_{2,\,0}(\theta_{0},\,0)\, Y_{2,\,0}(\theta,\,\phi)=P_{2}[cos(\theta_{0})]\, P_{2}[cos(\theta)],\label{eq:only_cnst_remaining}\end{equation}
leading to the result\begin{equation}
\mathcal{F}_{3}\{\frac{\Lambda_{rot}(\vec{r})}{r^{3}}\}=-\frac{4\pi}{3}P_{2}[cos(\theta_{0})].\label{eq:Lamda_rot_r3_result}\end{equation}
Considering the rotation of $\vec{\rho}$ instead of $\Lambda$ we
have our desired result\begin{equation}
\mathcal{F}_{3}\{\frac{\Lambda(\vec{r})}{r^{3}}\}=-\frac{4\pi}{3}\Lambda(\vec{\rho}).\label{eq:Lambda_r3_transform_result}\end{equation}

\section{Secular Component of the Field of a Point Dipole\label{sec:Secular-Component-Derived}}

We start with the magnetic field of an arbitrarily oriented point
dipole $\vec{\mu}$,\begin{equation}
\vec{B}_{dip}=\frac{\mu_{0}}{4\pi}\frac{3\,(\vec{\mu}\cdot\hat{r})\,\hat{r}-\vec{\mu}}{r^{3}}.\label{eq:B_dipole}\end{equation}

The secular component is the component that is invariant to rotation
of the coordinate axes about $\vec{B}_{0}=B_{0}\hat{z}$. This definition
at first appears to suggest that the secular component is just the
$\hat{z}$ component of $\vec{B}_{dip}$. This is not so. While it
does include the $\hat{z}$ component, it can also include non $\hat{z}$
components as well. The operative definition is\begin{equation}
\vec{B}_{secular}=\frac{1}{2\pi}\int_{0}^{2\pi}\vec{B}_{dip}d\phi.\label{eq:B_secular_operative}\end{equation}

Consider $\vec{\mu}$ with arbitrary orientation\begin{equation}
\vec{\mu}=\mu\,[cos(\theta_{\mu})\,\hat{z}+sin(\theta_{\mu})\, cos(\phi_{\mu})\,\hat{x}+sin(\theta_{\mu})\, sin(\phi_{\mu})\,\hat{y}]\label{eq:Mu_vec}\end{equation}
and the definition of $\hat{r}$\begin{equation}
\hat{r}\equiv cos(\theta)\,\hat{z}+sin(\theta)\, cos(\phi)\,\hat{x}+sin(\theta)\, sin(\phi)\,\hat{y}.\label{eq:r_hat}\end{equation}
We substitute into our expression for $\vec{B}_{dip}$ and split into
Cartesian components

\begin{flushleft}\begin{multline}
B_{x}=\frac{\mu_{0}}{4\pi}\frac{\mu}{r^{3}}[3\, sin(\theta)\, cos(\theta)\, cos(\phi)\, cos(\theta_{\mu})-sin(\theta_{\mu})\, cos(\phi_{\mu})\\
+3\, sin^{2}(\theta)\, cos^{2}(\phi)\, sin(\theta_{\mu})\, cos(\phi_{\mu})+3\, sin^{2}(\theta)\, sin(\phi)\, cos(\phi)\, sin(\theta_{\mu})\, sin(\phi_{\mu})]\label{eq:Bx_dipole}\end{multline}
\begin{multline}
B_{y}=\frac{\mu_{0}}{4\pi}\frac{\mu}{r^{3}}[3\, sin(\theta)\, cos(\theta)\, sin(\phi)\, cos(\theta_{\mu})-sin(\theta_{\mu})\, sin(\phi_{\mu})\\
+3\, sin^{2}(\theta)\, sin(\phi)\, cos(\phi)\, sin(\theta_{\mu})\, cos(\phi_{\mu})+3\, sin^{2}(\theta)\, sin^{2}(\phi)\, sin(\theta_{\mu})\, sin(\phi_{\mu})]\label{eq:By_dipole}\end{multline}
\begin{multline}
B_{z}=\frac{\mu_{0}}{4\pi}\frac{\mu}{r^{3}}\{[3\, cos^{2}(\theta)-1]\, cos(\theta_{\mu})\\
+3\, sin(\theta)\, cos(\theta)\, cos(\phi)\, sin(\theta_{\mu})\, cos(\phi_{\mu})+3sin(\theta)\, cos(\theta)\, sin(\phi)\, sin(\theta_{\mu})\, sin(\phi_{\mu})\}\label{eq:Bz_dipole}\end{multline}
\end{flushleft}

We now perform the integral \ref{eq:B_secular_operative}, by Cartesian
components, yielding\begin{equation}
B_{x,\, secular}=\frac{1}{2\pi}\int_{0}^{2\pi}B_{x}d\phi=\frac{\mu_{0}}{4\pi}\frac{\mu}{r^{3}}[-sin(\theta_{\mu})\, cos(\phi_{\mu})+\frac{3}{2}\, sin^{2}(\theta)\, sin(\theta_{\mu})\, cos(\phi_{\mu})],\label{eq:Bx_secualr}\end{equation}
\begin{equation}
B_{y,\, secular}=\frac{1}{2\pi}\int_{0}^{2\pi}B_{y}d\phi=\frac{\mu_{0}}{4\pi}\frac{\mu}{r^{3}}[-sin(\theta_{\mu})\, sin(\phi_{\mu})+\frac{3}{2}\, sin^{2}(\theta)\, sin(\theta_{\mu})\, sin(\phi_{\mu})],\label{eq:By_secular}\end{equation}
and\begin{equation}
B_{z,\, secular}=\frac{1}{2\pi}\int_{0}^{2\pi}B_{z}d\phi=\frac{\mu_{0}}{4\pi}\frac{\mu}{r^{3}}[3\, cos^{2}(\theta)-1]\, cos(\theta_{\mu}).\label{eq:Bz_secular}\end{equation}
Performing the substitution $sin^{2}(\theta)\equiv1-cos^{2}(\theta)$
gives us\begin{equation}
B_{x,\, secular}=-\frac{\mu_{0}}{4\pi}\frac{\mu}{r^{3}}[\frac{3\, cos^{2}(\theta)-1}{2}]\, sin(\theta_{\mu})\, cos(\phi_{\mu}),\label{eq:Bx_secular_sub}\end{equation}
\begin{equation}
B_{y,\, secular}=-\frac{\mu_{0}}{4\pi}\frac{\mu}{r^{3}}[\frac{3\, cos^{2}(\theta)-1}{2}]\, sin(\theta_{\mu})\, sin(\phi_{\mu}).\label{eq:By_secular_sub}\end{equation}
We note\begin{equation}
B_{z,\, secular}=-(1-3)\frac{\mu_{0}}{4\pi}\frac{\mu}{r^{3}}[\frac{3\, cos^{2}(\theta)-1}{2}]\, cos(\theta_{\mu}).\label{eq:Bz_secular_note}\end{equation}

Finally, we assemble the components into vector form as\begin{equation}
\vec{B}_{secular}=\frac{\mu_{0}}{4\pi}\frac{1}{r^{3}}[\frac{3\, cos^{2}(\theta)-1}{2}]\,(3\,\mu_{z}\hat{z}-\vec{\mu}).\label{eq:B_dipole_secular_compact}\end{equation}

\chapter{The Levitt Sign Conventions\label{cha:The-Levitt-Sign-Conventions}}

\begin{figure}
\includegraphics[%
  width=1.0\columnwidth,
  keepaspectratio]{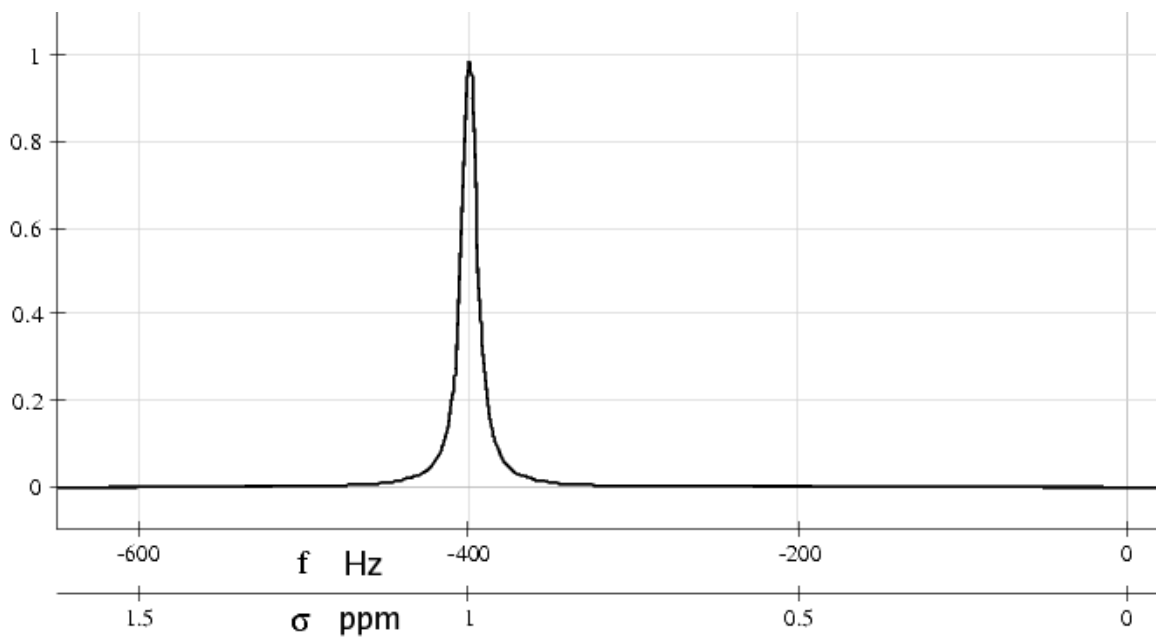}

\caption[Single peak at 1 ppm by the Levitt sign convention.]{\label{cap:Levitt_sign_convention}Single peak at 1 ppm by the Levitt
sign convention (on a $400\, MHz$ spectrometer). Note that the Levitt
sign convention correctly accounts for increasing chemical shift toward
the right, for Nuclei with positive $\gamma$ and hence negative Larmor
frequency.}
\end{figure}

We briefly review the sign conventions presented in references \cite[section 2.5]{Lev97,Lev01}
which are followed throughout this dissertation. The convention properly
accounts for the sense of rotation of net macroscopic magnetization.
First we have\[
\omega_{0}\equiv-\gamma\, B_{0}\]
and

\[
f_{0}\equiv\frac{\omega_{0}}{2\,\pi}\]

which leads to $negative$ Larmor frequency $f_{0}$ when $\gamma>0$.
Negative Larmor frequency corresponds to left handed precession about
the $B_{0}$ field. We define the chemical shift (in units of parts-per-million
or $ppm$) as\[
\delta-\delta_{ref}\equiv10^{6}\frac{\omega_{0}-\omega_{ref}}{\omega_{ref}},\]
 where $\omega_{ref}$ is the angular Larmor frequency of a reference
compound, such as $TMS$ or $DMS$ in high resolution NMR. These compounds
are often defined as $\delta_{ref}=0\, ppm$. For in-vivo spectroscopy
water is often the reference and is defined as $\delta_{ref}=4.7\, ppm$.
Note that the Levitt sign convention also properly accounts for the
{}``inverted'' axis in NMR spectroscopy where increasing $positive$
chemical shift is plotted toward the left, since it corresponds to
increasing $negative$ difference in Larmor frequency.

\chapter{Physical Constants}

From \cite{CTGoFC02}.

\begin{table}[h]
\begin{tabular}{lll}
\textbf{symbol}&
\textbf{name}&
\textbf{value (uncertainty) {[}units{]}}\tabularnewline
$\gamma_{p}$&
Proton Magnetogyric Ratio&
$-2.675\,222\,05(23)\times*10^{8}[s^{-1}T^{-1}]$\tabularnewline
$\frac{\gamma_{p}}{2\pi}$&
&
-$42.577\,481\,3(37)\,[MHz\, T^{-1}]$ \tabularnewline
$h$&
Planck Constant&
$6.626\,069\,3(11)\times10^{-34}[J\, s]$\tabularnewline
$\hbar$&
hbar, $\frac{h}{2\pi}$&
$1.054\,571\,68(18)\times10^{-34}[J\, s]$\tabularnewline
$k$&
Boltzmann Constant&
$1.380\,650\,5(24)\times10^{-23}[J\, K^{-1}]$\tabularnewline
$\mu_{0}$&
permeability of free space&
$4\pi\times10^{-7}[N\, A^{-2}]$\tabularnewline
$\mu_{p}$&
Proton Magnetic Moment&
$1.420\,606\,71(12)\times10^{-26}[J\, T^{-1}]$\tabularnewline
$N_{A}$&
Avogadro Constant&
$6.022\,141\,5(10)\times10^{23}[mol^{-1}]$\tabularnewline
$\pi$&
Pi, circle ratio&
$3.141\,592\,653\,589\,793\,238$\tabularnewline
&
&
\tabularnewline
\end{tabular}

\caption{Physical Constants\label{cap:Physical-Constants}}
\end{table}

\chapter{NMR Data\label{cha:NMR-Data}}

\begin{table}[h]
{\footnotesize }\begin{tabular}{crcrcrrr}
{\footnotesize atom}&
{\footnotesize \% abundance}&
{\footnotesize Spin}&
{\footnotesize $\gamma\ 10^{7}[\frac{rad}{T\, s}]$}&
{\footnotesize Q}&
{\footnotesize f{[}MHz{]}@11.744T}&
{\footnotesize rel. sens.}&
{\footnotesize abs. sens.}\tabularnewline
{\footnotesize $^{1}H$}&
{\footnotesize $99.980$}&
{\footnotesize $\frac{1}{2}$}&
{\footnotesize $+26.7519$}&
{\footnotesize $0$}&
{\footnotesize $-500.000$}&
{\footnotesize $1.00$}&
{\footnotesize $1.00$}\tabularnewline
{\footnotesize $^{2}H$}&
{\footnotesize $1.5\times10^{-2}$}&
{\footnotesize $1$}&
{\footnotesize $+4.1066$}&
{\footnotesize $2.8\times10^{-3}$}&
{\footnotesize $-76.753$}&
{\footnotesize $6.65\times10^{-6}$}&
{\footnotesize $1.45\times10^{-6}$}\tabularnewline
{\footnotesize $^{13}C$}&
{\footnotesize $1.108$}&
{\footnotesize $\frac{1}{2}$}&
{\footnotesize +$6.7283$}&
{\footnotesize $0$}&
{\footnotesize $-125.721$}&
{\footnotesize $1.59\times10^{-2}$}&
{\footnotesize $1.76\times10^{-4}$}\tabularnewline
{\footnotesize $^{15}N$}&
{\footnotesize $.365$}&
{\footnotesize $\frac{1}{2}$}&
{\footnotesize $-2.7120$}&
{\footnotesize $0$}&
{\footnotesize $+50.664$}&
{\footnotesize $1.04\times10^{-3}$}&
{\footnotesize $3.85\times10^{-6}$}\tabularnewline
{\footnotesize $^{19}F$}&
{\footnotesize $100.000$}&
{\footnotesize $\frac{1}{2}$}&
{\footnotesize +$25.181$}&
{\footnotesize $0$}&
{\footnotesize $-470.385$}&
{\footnotesize $.83$}&
{\footnotesize $.83$}\tabularnewline
{\footnotesize $^{31}P$}&
{\footnotesize $100.000$}&
{\footnotesize $\frac{1}{2}$}&
{\footnotesize +$10.841$}&
{\footnotesize $0$}&
{\footnotesize $-202.404$}&
{\footnotesize $6.63\times10^{-2}$}&
{\footnotesize $6.63\times10^{-2}$}\tabularnewline
{\footnotesize $^{129}Xe$}&
{\footnotesize $26.44$}&
{\footnotesize $\frac{1}{2}$}&
{\footnotesize $-7.452$}&
{\footnotesize $0$}&
{\footnotesize $+139.045$}&
{\footnotesize $2.12\times10^{-2}$}&
{\footnotesize $5.6\times10^{-3}$}\tabularnewline
\end{tabular}{\footnotesize \par}

from \cite{Par00,Lev01}

\caption{\label{cap:Some-common-Nuclei}Some common Nuclei in NMR}
\end{table}

\clearpage

\addcontentsline{toc}{chapter}{INDEX}

{\small \def\indexname{INDEX}}{\small \par}

\printindex{}

\clearpage

\addcontentsline{toc}{chapter}{REFERENCES}{\small \def\bibname{REFERENCES}\bibliographystyle{ieeetr}
\bibliography{/home/corum/references}

\begin{thebibliography}{100}

\bibitem{KRC+03}
S.~Kennedy, B.~Razavi, Z.~Chen, and J.~Zhong, ``Diffusion {M}easurements {F}ree
  of {M}otion {A}rtifacts {U}sing {I}ntermolecular {D}ipole-{D}ipole
  {I}nteractions,'' in {\em I{SMRM} {P}roceedings}, vol.~11, International
  Society for Magnetic Resonance in Medicine, July 2003.
\newblock talk 0581.

\bibitem{VLW96}
S.~Vathyam, S.~Lee, and W.~Warren, ``Homogeneous {NMR} {S}pectra in
  {I}nhomogeneous {F}ields,'' {\em Science}, vol.~272, pp.~92--96, 5 Apr. 1996.

\bibitem{GGM+93}
R.~J. Gillies, J.~P. Galons, K.~A. McGovern, P.~G. Scherer, Y.~H. Lien, C.~Job,
  R.~Ratcliff, F.~Chapa, S.~Cerdan, and B.~E. Dale, ``Design and {A}pplication
  of {NMR}-{C}ompatible {B}ioreactor {C}ircuits for {E}xtended {P}erfusion of
  {H}igh-{D}ensity {M}ammalian-{C}ell {C}ultures,'' {\em NMR In Biomedicine},
  vol.~6, pp.~95--104, Jan. 1993.

\bibitem{RZM+38}
I.~I. Rabi, J.~R. Zacharias, S.~Millman, and P.~Kusch, ``A {N}ew {M}ethod of
  {M}easuring {N}uclear {M}agnetic {M}oment,'' {\em Phys. Rev.}, vol.~53,
  p.~318, 15 Feb. 1938.

\bibitem{RMK+38}
I.~I. Rabi, S.~Millman, P.~Kusch, and J.~R. Zacharias, ``The {M}olecular {B}eam
  {R}esonance {M}ethod for {M}easuring {N}uclear {M}agnetic {M}oments {T}he
  {M}agnetic {M}oments of 3{L}i6, 3{L}i7 and 9{F}19,'' {\em Phys. Rev.},
  vol.~55, pp.~526--535, 15 Mar. 1938.

\bibitem{PTP46}
E.~Purcell, H.~Torrey, and R.~Pound, ``Resonance {A}bsorption by {N}uclear
  {M}agnetic {M}oments in a {S}olid,'' {\em Physical Review}, vol.~69,
  pp.~37--38, 1 Jan. 1946.

\bibitem{BHP46}
F.~Bloch, W.~Hansen, and M.~Packard, ``The {N}uclear {I}nduction
  {E}xperiment,'' vol.~70, pp.~474--485, 1 Oct. 1946.

\bibitem{Pfe99}
H.~Pfeifer, ``A short history of nuclear magnetic resonance spectroscopy and of
  its early years in {G}ermany,'' {\em Magnetic Resonance in Chemistry},
  vol.~37, no.~13, pp.~S154--S159, 1999.

\bibitem{Lev01}
M.~Levitt, ``Spin {D}ynamics: basics of nuclear magnetic resonance,'' 2001.
\newblock Great basic text on nmr.

\bibitem{HRG04}
L.~Hanneliusa, D.~O. Riskaa, and L.~Y. Glozmanb, ``The strangeness magnetic
  moment of the proton in the chiral quark model,'' {\em Nuclear Physics A},
  vol.~665, pp.~353--364, 28 Feb. 2004.

\bibitem{SBH+04}
D.~T. Spayde, D.~H. Beck, R.~Hasty, T.~Averett, D.~Barkhuff, G.~Dodson, K.~Dow,
  M.~Farkhondeh, W.~Franklin, E.~Tsentalovich, B.~Yang, T.~Zwart, E.~J. Beise,
  H.~Breuer, R.~Tieulent, R.~Carr, S.~Covrig, B.~W. Filippone, T.~M. Ito, R.~D.
  McKeown, W.~Korsch, S.~Kowalski, B.~Mueller, M.~L. Pitt, M.~J. Ramsey-Musolf,
  J.~Ritter, and S.~P. Wells, ``The strange quark contribution to the proton's
  magnetic moment,'' {\em Physics Letters B}, vol.~583, pp.~79--86, 11 Mar.
  2004.

\bibitem{DHK03}
C.~Durrant, M.~Hertzberg, and P.~Kuchel, ``Magnetic susceptibility: {F}urther
  insights into macroscopic and microscopic fields and the sphere of
  {L}orentz,'' {\em Concepts in Magnetic Resonance Imaging Part A}, vol.~18A,
  pp.~72--95, May 2003.

\bibitem{YH94}
D.~A. Yablonskiy and E.~M. Haacke, ``Theory of {NMR} signal behavior in
  magnetically inhomogeneous tissues: the static dephasing regime,'' {\em
  Magnetic Resonance In Medicine}, vol.~32, pp.~749--763, Dec. 1994.

\bibitem{Blo46}
F.~Bloch, ``Nuclear {I}nduction,'' {\em Physical Review}, vol.~70,
  pp.~460--474, 1 Oct. 1946.

\bibitem{Lev97}
M.~Levitt, ``The {S}igns of {F}requencies and {P}hases in {NMR},'' {\em Journal
  of Magnetic Resonance}, vol.~126, pp.~164--182, June 1997.

\bibitem{RRS54}
I.~I. Rabi, N.~F. Ramsey, and J.~Schwinger, ``Use of {R}otating {C}oordinates
  in {M}agnetic {R}esonance {P}roblems,'' {\em Rev. Mod. Phys.}, vol.~26,
  pp.~167--171, Apr. 1954.

\bibitem{DES99}
F.~Doty, J.~Entzminger, George, and J.~Staab, ``Practical {A}spects of
  {B}irdcage {C}oils,'' {\em Journal of Magnetic Resonance}, vol.~138,
  pp.~144--154, May 1999.

\bibitem{DEH99}
F.~Doty, J.~Entzminger, George, and C.~Hauck, ``Error-{T}olerant {RF} {L}itz
  {C}oils for {NMR}/{MRI},'' {\em Journal of Magnetic Resonance}, vol.~149,
  pp.~17--31, Sept. 1999.

\bibitem{HOK+00}
A.~Haase, F.~Odoj, M.~Kienlin, J.~Warnking, F.~Fidler, A.~Weisser, M.~Nittka,
  E.~Rommel, T.~Lanz, B.~Kalusche, and M.~Griswold, ``N{MR} probeheads for in
  vivo applications,'' {\em Concepts in Magnetic Resonance}, vol.~12, no.~6,
  pp.~361--388, 2000.

\bibitem{GD01}
M.~Garwood and L.~DelaBarre, ``Advances in {M}agnetic {R}esonance {T}he
  {R}eturn of the {F}requency {S}weep: {D}esigning {A}diabatic {P}ulses for
  {C}ontemporary {NMR},'' {\em Journal of Magnetic Resonance}, vol.~153,
  pp.~155--177, Dec. 2001.

\bibitem{BS40}
F.~Bloch and A.~Siegert, ``Magnetic {R}esonance for {N}onrotating {F}ields,''
  {\em Phys. Rev.}, vol.~57, pp.~522--527, 15 Mar. 1940.
\newblock cw nmr.

\bibitem{Hah50}
E.~L. Hahn, ``Nuclear {I}nduction {D}ue to {F}ree {L}armor {P}recession,'' {\em
  Phys. Rev.}, vol.~77, pp.~297--298, 15 Jan. 1950.
\newblock letter.

\bibitem{RG71}
A.~G. Redfield and R.~K. Gupta, ``Pulsed {F}ourier-{T}ransform {NMR}
  {S}pectrometer for {U}se with {H}2{O} {S}olutions,'' {\em The Journal of
  Chemical Physics}, vol.~54, pp.~1418--1419, 1 Feb. 1971.
\newblock letter.

\bibitem{Hou78}
D.~I. Hoult, ``The {NMR} receiver: {A} description and analysis of design,''
  {\em Progress in Nuclear Magnetic Resonance Spectroscopy}, vol.~12, no.~1,
  pp.~41--77, 1978.

\bibitem{EA66}
R.~R. Ernest and W.~A. Anderson, ``Application of {F}ourier {T}ransform
  {S}pectroscopy to {M}agnetic {R}esonance,'' {\em Review of Scientific
  Instruments}, vol.~37, no.~1, pp.~93--, 1966.

\bibitem{WEH71}
F.~W. Wehrli, ``Fourier {T}ransform {S}pectroscopy - {N}ew {T}echnic for
  {I}ncrease of {M}easuring {S}ensitivity in {N}uclear {M}agnetic
  {R}esonance,'' {\em Chemiker-Zeitung}, vol.~20, no.~2, pp.~58--, 1971.

\bibitem{RK75}
A.~G. Redfield and S.~D. Kunz, ``Quadrature {F}ourier {NMR} {D}etection -
  {S}imple {M}ultplex for {D}ual {D}etection and {D}iscussion,'' {\em JOURNAL
  OF MAGNETIC RESONANCE}, vol.~19, no.~2, pp.~250--254, 1975.

\bibitem{LF80}
J.~C. Lindon and A.~G. Ferrige, ``Digitisation and data processing in {F}ourier
  transform {NMR},'' {\em Progress in Nuclear Magnetic Resonance Spectroscopy},
  vol.~14, no.~1, pp.~27--66, 1980.

\bibitem{EAB+74}
R.~R. Ernst, W.~P. Aue, E.~Barthold, A.~Hohener, and A.~Schaublis,
  ``Equivalence of {F}ourier {S}pectroscopy and {S}low {P}assage in {N}uclear
  {M}agnetic-{R}esonance,'' {\em Pure and Applied Chemistry}, vol.~37, no.~1-2,
  pp.~47--60, 1974.

\bibitem{BPP48}
N.~Bloembergen, E.~Purcell, and R.~Pound, ``Relaxation {E}ffects in {N}uclear
  {M}agnetic {R}esonance {A}bsorption,'' {\em Phys. Rev.}, vol.~73, no.~7,
  pp.~679--712, 1948.

\bibitem{HB97}
D.~Hoult and B.~Bhakar, ``N{MR} signal reception: {V}irtual photons and
  coherent spontaneous emission,'' {\em Concepts in Magnetic Resonance},
  vol.~9, no.~5, pp.~277--297, 1997.

\bibitem{Dic51}
W.~C. Dickinson, ``The {T}ime {A}verage {M}agnetic {F}ield at the {N}ucleus in
  {N}uclear {M}agnetic {R}esonance {E}xperiments,'' {\em Phys. Rev.}, vol.~81,
  pp.~717--731, 1 Mar. 1951.

\bibitem{WB53}
R.~K. Wangsness and F.~Bloch, ``The {D}ynamical {T}heory of {N}uclear
  {I}nduction,'' {\em Phys. Rev.}, vol.~89, pp.~728--739, 15 Feb. 1953.

\bibitem{Blo56}
F.~Bloch, ``Dynamical {T}heory of {N}uclear {I}nduction. {II},'' {\em Phys.
  Rev.}, vol.~102, pp.~104--135, 1 Apr. 1956.

\bibitem{Bloch57}
F.~Bloch, ``Generalized {T}heory of {R}elaxation,'' {\em Phys. Rev.}, vol.~105,
  pp.~1206--1222, 15 Feb. 1957.

\bibitem{Abr99}
A.~Abragam, {\em The {P}rinciples of {N}uclear {M}agnetism}, vol.~32 of {\em
  International Series of Monographs on Physics}.
\newblock Oxford University Press, 1961.
\newblock based on corrected 1978 edition.

\bibitem{Gol01}
M.~Goldman, ``Formal theory of spin-lattice relaxation,'' {\em Journal of
  Magnetic Resonance}, vol.~149, pp.~160--187, Apr. 2001.
\newblock review.

\bibitem{Nis96}
D.~G. Nishimura, ``Principles of {M}agnetic {R}esonance {I}maging.'' Book for
  Art's 638 Class, 1996.

\bibitem{Bloem56}
N.~Bloembergen, ``Spin {R}elaxation {P}rocesses in a {T}wo-{P}roton {S}ystem,''
  {\em Phys. Rev.}, vol.~104, pp.~1542--1547, 15 Dec. 1956.

\bibitem{HBT+99}
E.~M. Haacke, R.~W. Brown, M.~R. Thompson, and R.~Venkatesan, {\em Magnetic
  {R}esonance {I}maging: {P}hysical {P}rinciples and {S}equence {D}esign}.
\newblock John Wiley \& Sons, 1~ed., 15 July 1999.

\bibitem{Hah50b}
E.~Hahn, ``Spin {E}choes,'' {\em Phys. Rev.}, vol.~80, pp.~580--594, 15 Nov.
  1950.

\bibitem{HM51}
E.~L. Hahn and D.~E. Maxwell, ``Chemical {S}hift and {F}ield {I}ndependent
  {F}requency {M}odulation of the {S}pin {E}cho {E}nvelope,'' {\em Phys. Rev.},
  vol.~84, pp.~1246--1247, 15 Dec. 1951.
\newblock letter.

\bibitem{GMS51}
H.~S. Gutowsky, D.~W. McCall, and C.~P. Slichter, ``Coupling among {N}uclear
  {M}agnetic {D}ipoles in {M}olecules,'' {\em Phys. Rev.}, vol.~84,
  pp.~589--590, 1 Nov. 1951.
\newblock letter.

\bibitem{HM52}
E.~L. Hahn and D.~E. Maxwell, ``Spin {E}cho {M}easurements of {N}uclear {S}pin
  {C}oupling in {M}olecules,'' {\em Phys. Rev.}, vol.~88, pp.~1070--1084, 1
  Dec. 1952.

\bibitem{Arn56}
J.~T. Arnold, ``Magnetic {R}esonances of {P}rotons in {E}thyl {A}lcohol,'' {\em
  Phys. Rev.}, vol.~102, pp.~136--150, 1 Apr. 1956.

\bibitem{CP54}
H.~Carr and E.~Purcell, ``Effects of {D}iffusion on {F}ree {P}recession in
  {N}uclear {M}agnetic {R}esonance {E}xperiments,'' {\em Phys. Rev.}, vol.~94,
  pp.~630--638, 1 May 1954.

\bibitem{MG58}
S.~Meiboom and D.~Gill, ``Modified {S}pin-{E}cho {M}ethodfor {M}easuring
  {N}uclear {R}elaxation {T}imes,'' {\em Review of Scientific Instruments},
  vol.~29, no.~8, pp.~688--691, 1958.

\bibitem{LCB+03}
J.~Leggett, S.~Crozier, S.~Blackband, B.~Beck, and R.~Bowtell, ``Multilayer
  transverse gradient coil design,'' {\em Concepts in Magnetic Resonance Part
  B: Magnetic Resonance Engineering}, vol.~16B, pp.~38--46, 23 Jan. 2003.

\bibitem{PS55}
D.~Pines and C.~P. Slichter, ``Relaxation {T}imes in {M}agnetic {R}esonance,''
  {\em Phys. Rev.}, vol.~100, pp.~1014--1020, 15 Nov. 1955.

\bibitem{Sli90}
C.~P. Slichter, {\em Principles of {M}agnetic {R}esonance}.
\newblock Springer Series in Solid-State Sciences, Springer Verlag, third
  enlarged and updated edition~ed., Jan. 1990.
\newblock also isbn 3540501576.

\bibitem{Pri97}
W.~Price, ``Pulsed-field gradient nuclear magnetic resonance as a tool for
  studying translational diffusion: {P}art {I}. {B}asic theory,'' {\em Concepts
  in Magnetic Resonance}, vol.~9, no.~5, pp.~299--336, 1997.

\bibitem{BML94}
P.~J. Basser, J.~Mattiello, and D.~Lebihan, ``Estimation of the {E}ffective
  {S}elf-{D}iffusion {T}ensor from the {NMR} {S}pin {E}cho,'' {\em Journal of
  Magnetic Resonance, Series B}, vol.~103, no.~3, pp.~247--254, 1994.

\bibitem{ST65}
E.~O. Stejskal and J.~E. Tanner, ``Spin {D}iffusion {M}easurements - {S}pin
  {E}cho in the {P}resence of a {T}ime {D}ependent {F}ield {G}radient,'' {\em
  Journal of Chemical Physics}, vol.~42, no.~1, p.~288, 1965.

\bibitem{Pri98}
W.~Price, ``Pulsed-field gradient nuclear magnetic resonance as a tool for
  studying translational diffusion: {P}art {II}. {E}xperimental aspects,'' {\em
  Concepts in Magnetic Resonance}, vol.~10, pp.~197--237, 7 Dec. 1998.

\bibitem{Tor56}
H.~C. Torrey, ``Bloch {E}quations with {D}iffusion {T}erms,'' {\em Phys. Rev.},
  vol.~104, pp.~563--565, 1 Nov. 1956.

\bibitem{MR87}
G.~Morrow and C.~Rosner, ``Superconducting magnets for magnetic resonance
  imaging applications,'' {\em Magnetics, IEEE Transactions on}, vol.~23,
  pp.~1294--1298, Mar. 1987.

\bibitem{Wil89}
J.~Williams, ``Superconducting magnets and their applications,'' {\em
  Proceedings of the IEEE}, vol.~77, pp.~1132--1142, Aug. 1989.

\bibitem{Fon95}
S.~Foner, ``High-field magnets and high-field superconductors,'' {\em Applied
  Superconductivity, IEEE Transactions on}, vol.~5, pp.~121--140, June 1995.

\bibitem{RH84}
F.~Romeo and D.~I. Hoult, ``Magnet field profiling: analysis and correcting
  coil design.,'' {\em Magn Reson Med.}, vol.~1, pp.~44--65, Mar. 1984.

\bibitem{CH90}
G.~Chmurny and D.~Hoult, ``The {A}ncient and {H}onourable {A}rt of
  {S}himming,'' {\em Concepts in Magnetic Resonanc}, vol.~2, pp.~131--149,
  1990.

\bibitem{Tan70}
J.~E. Tanner, ``Use of the {S}timulated {E}cho in {NMR} {D}iffusion
  {S}tudies,'' {\em The Journal of Chemical Physics}, vol.~52, pp.~2523--2526,
  1 Mar. 1970.
\newblock also see Erratum.

\bibitem{Tan72}
J.~E. Tanner, ``Erratum: {U}se of the {S}timulated {E}cho in {NMR} {D}iffusion
  {S}tudies,'' {\em The Journal of Chemical Physics}, vol.~57, p.~3586, 15 Oct.
  1972.
\newblock Original: J. Chem. Phys. 52, 2523 (1970).

\bibitem{Cul72}
B.~D. Cullity, {\em Introduction to {M}agnetic {M}aterials}.
\newblock Addison-Wesley Series in Metallurgy and Materials, Addison-Wesley,
  1972.

\bibitem{WA98}
W.~Warren and S.~Ahn, ``The boundary between liquidlike and solidlike behavior
  in magnetic resonance,'' {\em The Journal of Chemical Physics}, vol.~108,
  pp.~1313--1325, 22 Jan. 1998.

\bibitem{DBD79}
G.~Deville, M.~Bernier, and J.~Delrieux, ``N{MR} multiple echoes observed in
  solid 3{H}e,'' {\em Phys. Rev. B}, vol.~19, pp.~5666--5688, 1 June 1979.

\bibitem{DE84}
D.~Einzel, G.~Eska, Y.~Hirayoshi, T.~Kopp, and P.~W\"olfle, ``Multiple {S}pin
  {E}choes in a {N}ormal {F}ermi {L}iquid,'' {\em Phys. Rev. Lett.}, vol.~53,
  pp.~2312--2315, 10 Dec. 1984.

\bibitem{DE85}
D.~Einzel, G.~Eska, Y.~Hirayoshi, T.~Kopp, and P.~W\"olfle, ``Multiple {S}pin
  {E}choes in a {N}ormal {F}ermi {L}iquid,'' {\em Phys. Rev. Lett.}, vol.~54,
  p.~608?609, 11 Feb. 1985.
\newblock erratum.

\bibitem{BBG90}
R.~Bowtell, R.~M. Bowley, and P.~Glover, ``Multiple {S}pin {E}choes in
  {L}iquids in a {H}igh {M}agnetic {F}ield,'' {\em J. Magn. Reson.}, vol.~88,
  pp.~641--651, July 1990.

\bibitem{KDE91}
H.~K\"orber, E.~Dormann, and G.~Eska, ``Multiple spin echoes for protons in
  water,'' {\em Journal of Magnetic Resonance}, vol.~93, pp.~589--595, July
  1991.

\bibitem{WHM+92}
W.~Warren, Q.~He, M.~McCoy, and F.~Spano, ``Reply to the {C}omment on: {I}s
  multiple quantum nuclear magnetic resonance of water real?,'' {\em The
  Journal of Chemical Physics}, vol.~96, pp.~1659--1661, 15 Jan. 1992.

\bibitem{QH93}
Q.~He, W.~Richter, S.~Vathyam, and W.~Warren, ``Intermolecular multiple-quantum
  coherences and cross correlations in solution nuclear magnetic resonance,''
  {\em The Journal of Chemical Physics}, vol.~98, pp.~6779--6800, 1 May 1993.

\bibitem{ADL92}
D.~Abergel, M.~Delsuc, and J.-Y. Lallemand, ``Comment on: {I}s multiple quantum
  nuclear magnetic resonance spectroscopy of liquid water real?,'' {\em The
  Journal of Chemical Physics}, vol.~96, pp.~1657--1658, 15 Jan. 1992.

\bibitem{Jee00}
J.~Jeener, ``Equivalence between the "classical" and the "{W}arren" approaches
  for the effects of long range dipolar couplings in liquid nuclear magnetic
  resonance,'' {\em The Journal of Chemical Physics}, vol.~112, pp.~5091--5094,
  15 Mar. 2000.

\bibitem{AKK01}
I.~Ardelean, E.~Kossel, and R.~Kimmich, ``Attenuation of homo- and
  heteronuclear multiple spin echoes by diffusion,'' {\em The Journal of
  Chemical Physics}, vol.~114, pp.~8520--8529, 15 May 2001.

\bibitem{CG04b}
C.~A. Corum and A.~F. Gmitro, ``Effects of {T}2 relaxation and diffusion on
  longitudinal magnetization state and signal build for {HOMOGENIZED} cross
  peaks,'' in {\em I{SMRM} 12th {S}cientific {M}eeting}, International Society
  of Magnetic Resonance in Medicine, 15 May 2004.
\newblock poster 2323, cos(beta) should be (cos(beta)+1)/2 in abstract.

\bibitem{RB96}
R.~Bowtell and P.~Robyr, ``Structural {I}nvestigations with the {D}ipolar
  {D}emagnetizing {F}ield in {S}olution {NMR},'' {\em Phys. Rev. Lett.},
  vol.~76, pp.~4971--4974, 24 June 1996.

\bibitem{SCMA+01}
S.~Capuani, M.~Alesiani, F.~Alessandri, and B.~Maraviglia, ``Characterization
  of porous media structure by non linear {NMR} methods,'' {\em Magnetic
  Resonance Imaging}, vol.~19, pp.~319--323, Apr. 2001.

\bibitem{BRW02}
L.-S. Bouchard, R.~Rizi, and W.~Warren, ``Magnetization structure contrast
  based on intermolecular multiple-quantum coherences,'' {\em Magnetic
  Resonance in Medicine}, vol.~48, pp.~973--979, 3 Dec. 2002.

\bibitem{WAM+98}
W.~Warren, S.~Ahn, M.~Mescher, M.~Garwood, K.~Ugurbil, W.~Richter, R.~Rizi,
  J.~Hopkins, and J.~Leigh, ``M{R} imaging contrast enhancement based on
  intermolecular zero quantum coherences.,'' {\em Science}, vol.~281, no.~5374,
  pp.~247--51, 1998.

\bibitem{RAA+00}
R.~Rizi, S.~Ahn, D.~Alsop, S.~Garrett-Roe, M.~Mescher, W.~Richter, M.~Schnall,
  J.~Leigh, and W.~Warren, ``Intermolecular zero-quantum coherence imaging of
  the human brain,'' {\em Magnetic Resonance in Medicine}, vol.~43, no.~5,
  pp.~627--632, 2000.

\bibitem{SC02}
S.~Capuani, F.~Alessandri, A.~Bifone, and B.~Maraviglia, ``Multiple spin echoes
  for the evaluation of trabecular bone quality,'' {\em Magnetic Resonance
  Materials in Biology, Physics, and Medicine}, vol.~14, pp.~3--9, 1 Mar. 2002.

\bibitem{CTB+03b}
C.-L. Chin, X.~Tang, L.-S. Bouchard, P.~Saha, W.~Warren, and F.~Wehrli,
  ``Isolating quantum coherences in structural imaging using intermolecular
  double-quantum coherence {MRI},'' {\em Journal of Magnetic Resonance},
  vol.~165, pp.~309--314, 3 Nov. 2003.

\bibitem{MB04}
J.~Marques and R.~Bowtell, ``Optimizing the sequence parameters for
  double-quantum {CRAZED} imaging,'' {\em Magnetic Resonance in Medicine},
  vol.~51, pp.~148--157, Jan. 2004.

\bibitem{BW05}
L.-S. Bouchard and W.~S. Warren, ``Multiple-quantum vector imaging,'' in {\em
  46th {ENC} {C}onference}, Apr. 2005.

\bibitem{YYL00}
Y.-Y. Lin, S.~Ahn, N.~Murali, C.~Bowers, and W.~Warren, ``High-{R}esolution, >1
  {GH}z {NMR} in {U}nstable {M}agnetic {F}ields,'' {\em Physical Review
  Letters}, vol.~85, pp.~3732--3735, 23 Oct. 2000.

\bibitem{FPH03}
C.~Faber, E.~Pracht, and A.~Haase, ``Resolution enhancement in in vivo {NMR}
  spectroscopy: detection of intermolecular zero-quantum coherences,'' {\em
  Journal of Magnetic Resonance}, vol.~161, pp.~265--274, Apr. 2003.

\bibitem{CG04}
C.~A. Corum and A.~F. Gmitro, ``Experimental and {T}heoretical study of {TR}
  and {T}1 {E}ffects on {S}teady {S}tate {M}z in {D}istant {D}ipolar
  {F}ield-based {S}equences,'' in {\em 45th {ENC} {C}onference} ({Warren S.
  Warren}, ed.), Experimental Nuclear Magnetic Resonance Conference, 21 Apr.
  2004.
\newblock Time Slot/Poster Number: 016.

\bibitem{BF04b}
D.~Balla and C.~Faber, ``Solvent suppression in liquid state {NMR} with
  selective intermolecular zero-quantum coherences,'' {\em Chemical Physics
  Letters}, vol.~393, pp.~464--469, 1 Aug. 2004.

\bibitem{CHC+04}
Z.~Chen, T.~Hou, Z.-W. Chen, D.~W. Hwang, and L.-P. Hwang, ``Selective
  intermolecular zero-quantum coherence in high-resolution {NMR} under
  inhomogeneous fields,'' {\em Chemical Physics Letters}, vol.~386,
  pp.~200--205, 1 Mar. 2004.

\bibitem{ZCC+03}
J.~Zhong, Z.~Chen, Z.~Chen, and S.~Kennedy, ``High {R}esolution {NMR} {S}pectra
  in {I}nhomogeneous {F}ields via {I}ntermolecular {D}ouble {Q}uantum
  {C}oherences,'' in {\em I{SMRM} {P}roceedings}, vol.~11, International
  Society for Magnetic Resonance in Medicine, July 2003.
\newblock talk 0520.

\bibitem{RB97b}
P.~Robyr and R.~Bowtell, ``Nuclear magnetic resonance microscopy in liquids
  using the dipolar field,'' {\em The Journal of Chemical Physics}, vol.~106,
  pp.~467--476, 8 Jan. 1997.

\bibitem{LSB+04}
M.~P. Ledbetter, I.~M. Savukov, L.-S. Bouchard, and M.~V. Romalis, ``Numerical
  and experimental studies of long-range magnetic dipolar interactions,'' {\em
  Journal of Chemical Physics}, vol.~121, pp.~1454--1465, 15 July 2004.

\bibitem{BM03}
H.~H. Barrett and K.~Myers, {\em Foundations of {I}mage {S}cience}.
\newblock Wiley Series in Pure and Applied Optics, Wiley-Interscience, 1st~ed.,
  Oct. 2003.

\bibitem{CXB+90}
S.~C. Chu, Y.~Xu, J.~A. Balschi, and C.~S.~J. Springer, ``Bulk magnetic
  susceptibility shifts in {NMR} studies of compartmentalized samples: use of
  paramagnetic reagents.,'' {\em Magn Reson Med.}, vol.~13, pp.~239--262, Feb.
  1990.
\newblock paper copy only.

\bibitem{LdS+04}
P.~Loureiro~de Sousa, D.~Gounot, and D.~Grucker, ``Observation of
  diffraction-like effects in {M}ultiple {S}pin {E}choe ({MSE}) experiments in
  structured samples,'' {\em Comptes Rendus Chimie}, vol.~7, pp.~311--319, 12
  Apr. 2004.

\bibitem{TCB+04}
X.~P. Tang, C.~L. Chin, L.~S. Bouchard, F.~W. Wehrli, and W.~S. Warren,
  ``Observing {B}ragg-like diffraction via multiple coupled nuclear spins,''
  {\em Physics Letters A}, 2004.

\bibitem{BTC+03}
L.~Bouchard, X.~Tang, C.~Chin, F.~Wehrli, and W.~Warren, ``Magnetic {R}esonance
  {I}maging of the {D}istant {D}ipolar {F}ield in {S}tructured {S}amples
  {U}sing {I}ntermolecular {M}ultiple-{Q}uantum {C}oherences of {V}arious
  {O}rders,'' in {\em I{SMRM} {P}roceedings}, vol.~11, International Society
  for Magnetic Resonance in Medicine, July 2003.
\newblock poster 1110.

\bibitem{BCT+03}
L.~Bouchard, C.~Chin, X.~Tang, W.~Warren, and F.~Wehrli, ``Structural
  {C}haracterization of {T}rabecular {B}one {U}sing {B}ulk {NMR} {M}easurements
  of {I}ntermolecular {M}ultiple-{Q}uantum {C}oherences,'' in {\em I{SMRM}
  {P}roceedings}, vol.~11, International Society for Magnetic Resonance in
  Medicine, July 2003.
\newblock poster 1113.

\bibitem{CPL+02}
G.~Charles-Edwards, G.~Payne, M.~Leach, and A.~Bifone, ``Contrast mechanisms in
  {I}ntermolecular {D}ouble {Q}uantum {C}oherence {I}maging: {A} {W}arning,''
  in {\em I{SMRM} {P}roceedings}, vol.~10, International Society for Magnetic
  Resonance in Medicine, May 2002.
\newblock talk 0614.

\bibitem{ZHC+05}
B.~Zheng, D.~W. Hwang, Z.~Chen, and L.-P. Hwang, ``Rotating-frame
  intermolecular double-quantum spin-lattice relaxation {T}1rho, {DQC}-weighted
  magnetic resonance imaging,'' {\em Magnetic Resonance in Medicine}, vol.~53,
  pp.~930--936, Apr. 2005.

\bibitem{BP54}
N.~Bloembergen and R.~Pound, ``Radiation {D}amping in {M}agnetic {R}esonance
  {E}xperiments,'' {\em Phys. Rev.}, vol.~95, pp.~8--12, 1 July 1954.

\bibitem{Blo57}
S.~Bloom, ``Effects of radiation damping on spin dynamics.,'' {\em J. Appl.
  Phys.}, vol.~28, pp.~800--805, 1957.

\bibitem{SM59}
A.~Sz\"oke and S.~Meiboom, ``Radiation {D}amping in {N}uclear {M}agnetic
  {R}esonance,'' {\em Phys. Rev.}, vol.~113, pp.~585--586, 15 Jan. 1959.

\bibitem{VJB95}
A.~Vlassenbroek, J.~Jeener, and P.~Broekaert, ``Radiation damping in high
  resolution liquid {NMR}: {A} simulation study,'' {\em The Journal of Chemical
  Physics}, vol.~103, pp.~5886--5897, 8 Oct. 1995.

\bibitem{WHB89}
W.~S. Warren, S.~L. Hammes, and J.~L. Bates, ``Dynamics of radiation damping in
  nuclear magnetic resonance,'' {\em J. Chem. Phys}, vol.~91, pp.~5895--5904,
  15 Nov. 1989.

\bibitem{MW90}
M.~McCoy and W.~Warren, ``Three-quantum nuclear magnetic resonance spectroscopy
  of liquid water: {I}ntermolecular multiple-quantum coherence generated by
  spin?cavity coupling,'' {\em The Journal of Chemical Physics}, vol.~93,
  pp.~858--860, 1 July 1990.

\bibitem{BBH+96}
G.~Ball, G.~Bowden, T.~Heseltine, M.~Prandolini, and W.~Bermel, ``Radiation
  damping artifacts in 2{D} {COSY} {NMR} experiments,'' {\em Chemical Physics
  Letters}, vol.~261, pp.~421--424, 25 Oct. 1996.

\bibitem{aug02}
M.~Augustine, ``Transient properties of radiation damping,'' {\em Progress in
  Nuclear Magnetic Resonance Spectroscopy}, vol.~40, pp.~111--150, 25 Feb.
  2002.

\bibitem{LLAW00}
Y.-Y. Lin, N.~Lisitza, S.~Ahn, and W.~Warren, ``Resurrection of {C}rushed
  {M}agnetization and {C}haotic {D}ynamics in {S}olution {NMR}
  {S}pectroscopy,'' {\em Science}, vol.~290, pp.~118--121, 2000.

\bibitem{Abe02}
D.~Abergel, ``Chaotic solutions of the feedback driven {B}loch equations,''
  {\em Physics Letters A}, vol.~302, pp.~17--22, 9 Sept. 2002.

\bibitem{HL03}
S.~Huang and Y.~Lin, ``A {N}ovel {M}echanism for {MRI} {C}ontrast {E}nhancement
  based on {C}ontrol of {S}pin {C}haos,'' in {\em I{SMRM} {P}roceedings},
  vol.~11, International Society for Magnetic Resonance in Medicine, July 2003.
\newblock poster 1111.

\bibitem{SD99}
J.~Sz\'antay, Csaba and A.~Demeter, ``Radiation damping diagnostics,'' {\em
  Concepts in Magnetic Resonance}, vol.~11, no.~3, pp.~121--145, 1999.

\bibitem{BJ95}
P.~Broekaert and J.~Jeener, ``Suppression of {R}adiation {D}amping in {NMR} in
  {L}iquids by {A}ctive {E}lectronic {F}eedback,'' {\em Journal of Magnetic
  Resonance, Series A}, vol.~113, pp.~60--64, Mar. 1995.

\bibitem{RLW+95}
W.~Richter, S.~H. Lee, W.~S. Warren, and Q.~H. He, ``Imaging with
  {I}ntermolecular {M}ultiple-{Q}uantum {C}oherences in {S}olution
  {N}uclear-{M}agnetic-{R}esonance,'' {\em Science}, vol.~267, pp.~654--657, 3
  Feb. 1995.

\bibitem{CG04c}
C.~A. Corum and A.~F. Gmitro, ``Visualizing {D}istant {D}ipolar {F}ield and
  {I}ntermolecular {M}ultuple {Q}uantum {C}oherence {S}equences,'' in {\em
  I{SMRM} 12th {S}cientific {M}eeting}, International Society of Magnetic
  Resonance in Medicine, 15 May 2004.
\newblock ePoster 2711.

\bibitem{alw98b}
S.~Ahn, N.~Lisitza, and W.~Warren, ``Intermolecular {Z}ero-{Q}uantum
  {C}oherences of {M}ulti-component {S}pin {S}ystems in {S}olution {NMR},''
  {\em J. Magn. Reson.}, vol.~133, pp.~266--272, Aug. 1998.

\bibitem{GRG+80}
I.~S. Gradshteyn, I.~M. Ryzhik, Y.~V. Geronimus, M.~Y. Tseytlin, and
  A.~Jeffrey, {\em Table of {I}ntegrals, {S}eries, and {P}roducts}.
\newblock Academic Press, Inc., incorporating 4th edition~ed., 1980.
\newblock Translation of Tablitsy integralov, summ, riadov i proizvedenii.

\bibitem{AK00b}
I.~Ardelean and R.~Kimmich, ``Diffusion {M}easurements {U}sing the {N}onlinear
  {S}timulated {E}cho,'' {\em Journal of Magnetic Resonance}, vol.~143,
  pp.~101--105, Mar. 2000.

\bibitem{AK00e}
I.~Ardelean and R.~Kimmich, ``Diffusion measurements with the pulsed gradient
  nonlinear spin echo method,'' {\em The Journal of Chemical Physics},
  vol.~112, pp.~5275--5280, 22 Mar. 2000.

\bibitem{FB04}
C.~Faber and D.~Balla, ``Water suppression in 2{D} i{ZQC} spectroscopy for in
  vivo application,'' in {\em 45th {ENC} {C}onference} ({Warren S. Warren},
  ed.), Experimental Nuclear Magnetic Resonance Conference, 22 Apr. 2004.
\newblock Time Slot/Poster Number: 227.

\bibitem{JZ01}
J.~Zhong, Z.~Chen, E.~Kwok, and S.~Kennedy, ``Enhanced sensitivity to molecular
  diffusion with intermolecular double-quantum coherences: implications and
  potential applications,'' {\em Magnetic Resonance Imaging}, vol.~19,
  pp.~33--39, Jan. 2001.

\bibitem{ZCJZ01}
Z.~Chen and J.~Zhong, ``Unconventional diffusion behaviors of intermolecular
  multiple-quantum coherences in nuclear magnetic resonance,'' {\em The Journal
  of Chemical Physics}, vol.~114, pp.~5642--5653, 1 Apr. 2001.

\bibitem{ZCGL01}
Z.~Chen, G.~Lin, and J.~Zhong, ``Diffusion of intermolecular zero- and
  double-quantum coherences in two-component spin systems,'' {\em Chemical
  Physics Letters}, vol.~333, pp.~96--102, 5 Jan. 2001.

\bibitem{MBL94}
J.~Mattiello, P.~J. Basser, and D.~Lebihan, ``Analytical {E}xpressions for the
  b {M}atrix in {NMR} {D}iffusion {I}maging and {S}pectroscopy,'' {\em Journal
  of Magnetic Resonance, Series A}, vol.~108, pp.~131--141, June 1994.

\bibitem{CG04d}
C.~A. Corum and A.~F. Gmitro, ``Spatially {V}arying {S}teady {S}tate
  {L}ongitudinal {M}agnetization in {D}istant {D}ipolar {F}ield-based
  {S}equences,'' 1 Apr. 2004.
\newblock submitted to Journal of Magnetic Resonance,
  http://arxiv.org/abs/physics/0406045.

\bibitem{CG04e}
C.~A. Corum and A.~F. Gmitro, ``Spatially {V}arying {S}teady {S}tate
  {L}ongitudinal {M}agnetization in {D}istant {D}ipolar {F}ield-based
  {S}equences,'' {\em Journal of Magnetic Resonance}, vol.~171, pp.~131--134, 1
  Apr. 2004.

\bibitem{ASDK97}
I.~Ardelean, S.~Stapf, D.~Demco, and R.~Kimmich, ``The {N}onlinear {S}timulated
  {E}cho,'' {\em J. Magn. Reson.}, vol.~124, pp.~506--508, Feb. 1997.

\bibitem{ACM02}
F.~Alessandri, S.~Capuani, and B.~Maraviglia, ``Multiple {S}pin {E}choes in
  heterogeneous systems: {P}hysical origins of the observed dips,'' {\em J.
  Magn. Reson.}, vol.~156, pp.~72--78, May 2002.

\bibitem{CCA+01}
S.~Capuani, F.~Curzi, F.~Alessandri, B.~Maraviglia, and B.~Bifone,
  ``Characterization of trabecular bone by dipolar demagnetizing field {MRI},''
  {\em Magnetic Resonance in Medicine}, vol.~46, no.~4, pp.~683--689, 2001.

\bibitem{CG05}
C.~A. Corum and M.~Garwood, ``First {R}esults with "{LASER}" localized
  {HOMOGENIZED} sequence,'' Apr. 2005.

\bibitem{DBCF05}
{David Balla} and {Cornelius Faber}, ``Localized {I}ntermolecular
  {Z}ero-{Q}uantum {C}oherence {S}pectroscopy in vivo,'' in {\em 46th {ENC}
  {C}onference}, Apr. 2005.

\bibitem{CTGoFC02}
{CODATA Task Group on Fundamental Constants}, ``The {NIST} {R}eference on
  {C}onstants, {U}nits, and {U}ncertainty.'' Web, 2002.
\newblock Latest (2002) values of the constants.

\bibitem{Leg75}
A.~Leggett, ``A theoretical description of the new phases of liquid 3{H}e,''
  {\em Rev. Mod. Phys.}, vol.~47, pp.~331--414, Apr. 1975.
\newblock Review.

\bibitem{AS74}
M.~Abramowitz and I.~A. Stegun, {\em Handbook of {M}athematical {F}unctions,
  with {F}ormulas, {G}raphs, and {M}athematical {T}ables}.
\newblock Dover Publications, 1 June 1974.

\bibitem{BF92}
F.~W. Byron and R.~W. Fuller, {\em Mathematics of {C}lassical and {Q}uantum
  {P}hysics}, vol.~1-2 of {\em Addison-Wesley Series in Advanced Physics}.
\newblock Dover, reprint edition~ed., 20 Aug. 1992.
\newblock originally published as two volumes in 1969 and 1970 by
  Addison-Wesley.

\bibitem{Par00}
T.~Parella, ``{e}{NMR}.'' Web, 2000.
\newblock many nmr topics.

\end{thebibliography}
}
\end{document}